\documentclass{aa}  

\usepackage{placeins}
\usepackage{adjustbox}
\usepackage{etoolbox} 

\usepackage{graphicx}
\usepackage{pgffor}
\usepackage{txfonts}
\usepackage[table]{xcolor}
\usepackage{hyperref}
\usepackage{booktabs}
\usepackage{comment}
\usepackage{enumitem}
\usepackage{subcaption}
\usepackage{multirow}

\usepackage{makecell}
\usepackage{diagbox}  

\newcommand{\JWST}{\textit{JWST }}

\newcommand{\galfit}{\textsc{galfit }}

\usepackage{pgffor}

\usepackage{xcolor}
\usepackage{soulutf8}

\definecolor{revisionblue}{RGB}{210,235,255}

\soulregister\citep7
\soulregister\citet7
\soulregister\citealp7
\soulregister\citealt7
\soulregister\citeauthor7
\soulregister\ref7
\soulregister\pageref7

\newif\ifshowrevisions
\showrevisionsfalse   

\ifshowrevisions
    \newcommand{\rev}[1]{%
        \sethlcolor{revisionblue}%
        \hl{#1}%
    }
\else
    \newcommand{\rev}[1]{#1}
\fi

\usepackage{morefloats}     
\extrafloats{200}           
\begin{document} 

    \title{MSA-3D: Rotation Curves and Dark Matter Fractions at $z \sim 0.5$--$1.7$ with \textit{JWST}/NIRSpec}
    \titlerunning{MSA-3D: Dark Matter and Rotation Curves at $z \sim 0.5$--$1.7$}

   \author{
    Juan M. Espejo Salcedo\inst{1}\thanks{\email{jespejo@mpe.mpg.de}}
    \and
    Danail Obreschkow\inst{2}
    \and
    Karl Glazebrook\inst{3}
    \and
    Tucker Jones\inst{4}
    \and
    Ivana Bari{\v{s}}i{\'c}\inst{4}
    \and
    Natascha M. Förster Schreiber\inst{1}
    \and
    Takafumi Tsukui\inst{5}
    \and
    Xin Wang\inst{6,7,8}
    \and
    Mengting Ju\inst{6}
    \and
    Qianqiao Zhou\inst{6}
    \and
    Amit Nestor Schachar\inst{1,9}
    \and
    Ryan L. Sanders\inst{10}
    \and
    Stavros Pastras\inst{1, 15}
    \and
    Namrata Roy\inst{11}
    \and
    Alaina Henry\inst{12}
    \and
    Kyle Westfall\inst{13}
    \and
    Themiya Nanayakkara\inst{3}
    \and
    Matthew Malkan\inst{14}
    \and
    Fahmi M. Al Farisy\inst{3}
    \and 
    Isaac Kanowski\inst{16}
}

\institute{
    Max-Planck-Institut für extraterrestrische Physik (MPE), Giessenbachstraße 1., 85748 Garching, Germany
    \and
    International Centre for Radio Astronomy Research, University of Western Australia, 7 Fairway, Crawley, WA 6009, Australia
    \and
    Centre for Astrophysics \& Supercomputing, Swinburne University of Technology, PO Box 218, Hawthorn, VIC 3122, Australia
    \and
    Department of Physics and Astronomy, University of California, Davis, 1 Shields Avenue, Davis, CA 95616, USA
    \and
    Kavli Institute for the Physics and Mathematics of the Universe (WPI), The University of Tokyo Institutes for Advanced Study, The University of Tokyo, Kashiwa, Chiba 277-8583, Japan
    \and
    School of Astronomy and Space Science, University of Chinese Academy of Sciences (UCAS), Beijing 100049, China
    \and
    National Astronomical Observatories, Chinese Academy of Sciences, Beijing 100101, China
    \and
    Institute for Frontiers in Astronomy and Astrophysics, Beijing Normal University, Beijing 102206, China    
    \and
    School of Physics and Astronomy, Tel Aviv University, Tel Aviv 69978, Israel
    \and
    Department of Physics and Astronomy, University of Kentucky, 505 Rose Street, Lexington, KY 40506, USA
    \and
    School of Earth and Space Exploration, Arizona State University, Tempe, AZ 85287
    \and
    Space Telescope Science Institute; 3700 San Martin Drive, Baltimore, MD, 21218, USA
    \and
    University of California Observatories, University of California, Santa Cruz, 1156 High St., Santa Cruz, CA 95064, USA
    \and 
    Department of Physics and Astronomy, University of California, Los Angeles, CA 90095-1547, USA
    \and 
    Max-Planck-Institut für Astrophysik (MPA), Karl-Schwarzschild-Str. 1, 85748, Garching, Germany
    \and
    Research School of Astronomy and Astrophysics, Australian National University, Canberra, ACT 2611, Australia
}

   \date{Received ...; accepted ...}

\abstract{We present rotation curves and inner mass distributions for 30 star-forming galaxies at $0.5<z<1.7$, \rev{down to stellar masses of $\sim10^9\,M_\odot$}, observed with \textit{JWST}/NIRSpec in the MSA-3D Cycle 1 survey. \rev{We combine} spatially resolved ionised-gas kinematics and \textit{JWST}/NIRCam imaging in forward dynamical models to constrain baryonic and dark matter contributions. \rev{The 24 golden-sample galaxies exhibit ordered rotation}, with intrinsic dispersions of \rev{$\sigma_0\simeq34$--63~km\,s$^{-1}$} and median \rev{$V_{\rm rot}/\sigma_0=2.9$ at $2.2R_{\rm s}$. Their stellar Tully--Fisher relation has a slope consistent with previous $z\sim1$ studies ($a=3.8\pm0.5$) but substantial inter-survey differences in normalisation. Within the disk effective radius, we obtain a median dark matter fraction $f_{\rm DM}(R_{\rm e})=0.73$ (68\% range: 0.49--0.84), with systematic dependence on the adopted baryonic mass scale. Across comparison surveys, $f_{\rm DM}(R_{\rm e})$ correlates more strongly with baryonic surface density than stellar mass, although neither trend is significant within MSA-3D. Five galaxies favour inner halo slopes shallower than NFW (versus $\approx2$ expected by chance), three significantly in model-matched NFW injection tests. Among 16 galaxies with sufficient radial coverage, five have rising, eight flat, and three falling rotation curves (although individual slopes are mostly not significant), with no clear dependence on other properties. Together, these results indicate that these low-mass disks at $z\sim1$ are dark-matter dominated within $R_{\rm e}$ for the baryonic mass scale favoured by the kinematics, although adopting the catalogue-based masses lowers the median to $\approx0.5$. Their inner halo profiles are generally consistent with or moderately shallower than NFW. These observations demonstrate the efficiency of \textit{JWST} for extending resolved dynamical studies to lower-mass galaxies.}}


\maketitle

\section{Introduction}
\label{section:Introduction}

{\looseness=-1
Rotation curves encode the dynamical mass distribution of galaxies and, combined with imaging data, provide one of the most direct constraints on the relative contributions of baryonic matter and dark matter. This offers key insights into how galaxies assemble and evolve. In particular, the persistence of flat rotation curves at large radii cannot be explained by the Newtonian gravity of the observed baryonic components alone, providing some of the strongest evidence for extended dark matter halos \citep{Rubin_1980, Bosma_1981, Persic_1996, Sofue_2001, de_Blok_2001, de_Blok_2008}. The apparent coordination between the radial distributions of baryons and dark matter implied by these observations is sometimes referred to as the ``disk-halo conspiracy'' (e.g., \citealp{van_Albada_1986, Hoekstra_2001}).\par}

In the local Universe, high-quality extended rotation curves show that dark matter dominates galaxy mass budgets at large radii, while revealing diversity in inner kinematic profiles \citep[e.g.,][]{Courteau_2015, Lelli_2016}. Massive spirals typically exhibit steeply rising inner rotation curves, whereas low-surface-brightness and dwarf galaxies show more gradually rising profiles and shallow central density slopes \citep{de_Blok_1997, de_Blok_2001}. This diversity challenges the cuspy inner density profiles predicted by collisionless $\Lambda$CDM simulations, commonly described by the Navarro--Frenk--White profile \citep{Navarro_1997}. Baryonic feedback provides a plausible resolution to this ``core--cusp problem'', as repeated star formation and outflows can redistribute matter and lower central densities \citep[e.g.,][]{Governato_2010, Pontzen_2012, Lazar_2020}. The same feedback processes are expected to influence the inner dark matter content of high-redshift disks, motivating observations of dark matter fractions across cosmic time. Extending such studies to earlier epochs, however, first requires establishing whether dynamically mature, rotationally supported disks are already in place at high redshift.

Integral-field spectroscopy has established that rotating, albeit dynamically turbulent, disks are already common at $1<z<3$ \citep[see reviews by][]{Glazebrook_2013, Forster_review_2020, Tacconi_review_2020}. Studies of massive star-forming galaxies at $z\sim1$--2 have often inferred low inner dark matter fractions and, in some cases, falling rotation curves \citep[e.g.,][]{Genzel_2017, Lang_2017, Ubler_2018, Genzel_2020, Price_2021, Nestor_2023}. Larger samples have subsequently shown that much of the diversity in $f_{\rm DM}(R_{\rm e})$ is associated with galaxy properties, particularly central baryonic surface density, with weaker evidence for evolution with redshift \citep[e.g.,][]{Wuyts_2016, Genzel_2020, Price_2021, Nestor_2023, Puglisi_2023, Sharma_2025, Ciocan_2026}.

In light of these trends, $z\sim1$ emerges as a particularly informative epoch for studying the interplay of baryons, dark matter, and dynamical evolution: rotationally supported disks are already widespread, yet their internal mass distributions have not fully converged to those of local systems.

Significant observational progress has already been made at $z\sim1$ with state-of-the-art ground-based facilities to probe galaxy kinematics and mass distributions. Spatially resolved spectroscopic surveys targeting ionised gas, mostly with integral-field spectroscopy, such as KMOS$^{\rm 3D}$ \citep{KMOS3D, Wisnioski_2019, Uebler_2019}, KROSS \citep{Stott_2016, Harrison_2017, Sharma_2021}, KGES \citep{Tiley_2021}, MOSDEF \citep{Kriek_2015, Price_2016, Price_2020}, MASSIV \citep{Contini_2012, Epinat_2012}, MUSE \citep[e.g.,][]{Contini_2016, Bouche_2022}, and KURVS \citep{Puglisi_2023}, have established statistical constraints on rotation-curve shapes and baryonic and dark matter fractions beyond the local Universe. Complementary constraints come from stellar absorption-line kinematics, with LEGA-C broadly supporting gas-based dynamical mass estimates at this epoch \citep{Ubler_2024}, and from spatially resolved CO observations, which trace molecular-gas kinematics and, with an appropriate CO-to-H$_2$ conversion factor \citep[e.g.,][]{Bolatto_2013}, constrain molecular-gas mass distributions \citep[e.g., PHIBSS and NOEMA-3D;][]{Tacconi_2018, Chen_2026, Jolly_2026}.

\rev{High-resolution observations provide complementary constraints on the inner regions of galaxies. Adaptive-optics observations have provided kiloparsec- and sub-kiloparsec-scale kinematic measurements for smaller samples of star-forming galaxies \citep{Law_2, Mieda_IROCKS, Forster_2018, Molina_2019, Sweet, Espejo_Salcedo_2022, Espejo_Salcedo_2025}, while strong gravitational lensing has enabled resolved spectroscopy of intrinsically faint galaxies at effective resolutions of a few hundred parsecs (e.g., \citealp{Jones_2010, Yuan_2011, Yuan_2012, Livermore_2015, Leethochawalit_2016, Jeanneau_2026}). These approaches provide important high-resolution benchmarks, but they each have limitations: lensing samples are intrinsically rare and subject to lens-modelling uncertainties, while AO observations at $z\sim1$ are affected by the modest Strehl ratios in the $J$ band, where H$\alpha$ is observed, and by complex point-spread functions with extended wings that can complicate beam-smearing corrections.}

Building on these approaches, the \textit{James Webb Space Telescope} (\textit{JWST}) provides a complementary advance for spatially resolved galaxy kinematics at intermediate redshift. Its stable $\sim0\farcs1$--$0\farcs2$ spatial resolution enables $\sim1$--2~kpc-scale measurements at $z\sim1$ with substantially reduced beam smearing compared with seeing-limited observations. In particular, slit stepping with the NIRSpec micro-shutter assembly (MSA) enables efficient two-dimensional spectroscopy of multiple galaxies simultaneously, producing pseudo-cubes well suited for kinematic modelling at a fraction of the cost of traditional IFU observations.

The MSA-3D survey \citep{Barisic_2025, Ju_2025} exploits this capability to obtain spatially resolved spectroscopy of star-forming galaxies at $z\sim0.5$--1.7 across a broad stellar-mass range. Its Cycle 1 observations provide one of the largest samples of near-main-sequence galaxies with resolved ionised-gas kinematics at $\sim1$--2 kpc resolution at this epoch.

\begin{figure*}
    \centering
    \includegraphics[width=\linewidth]{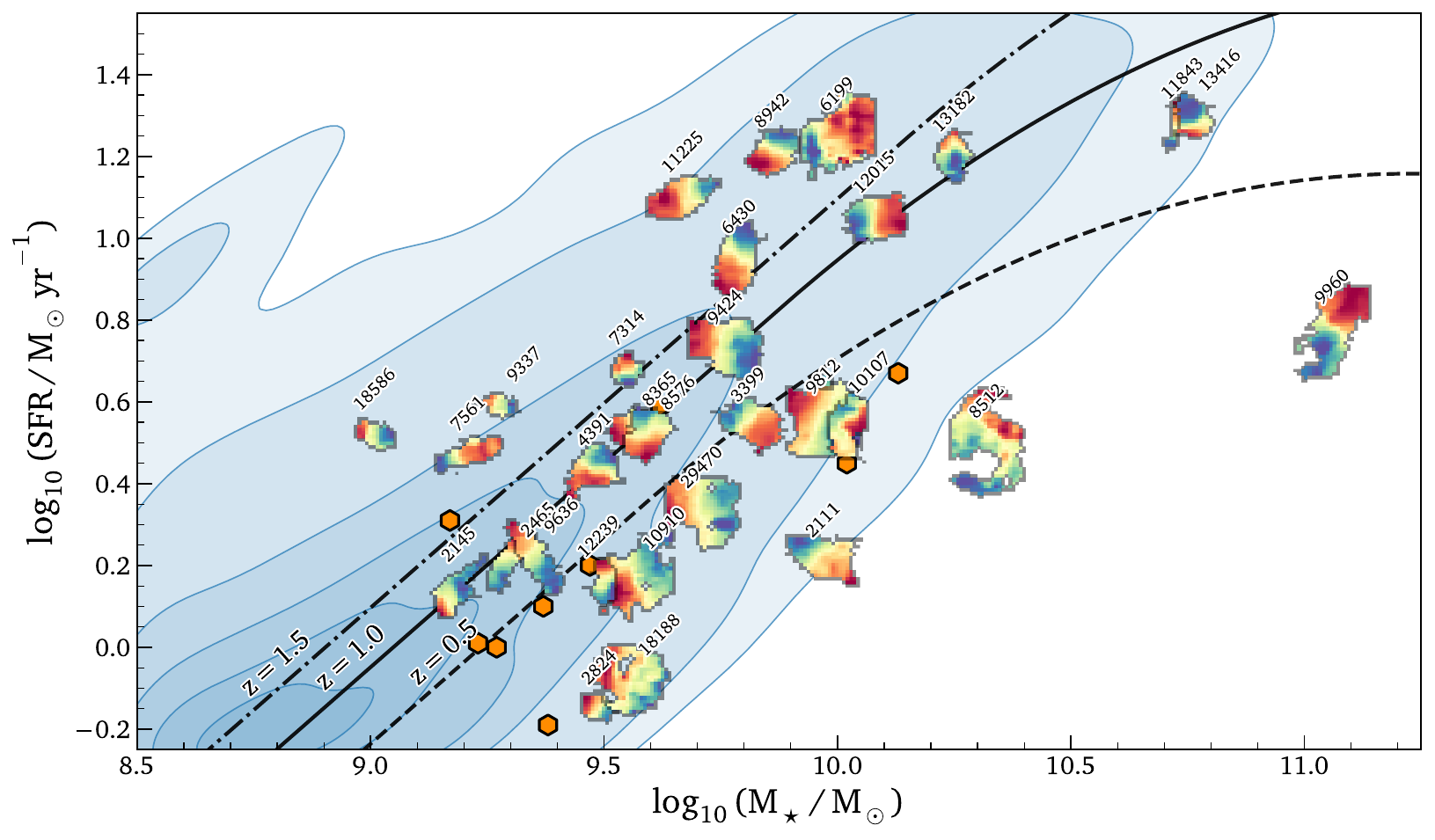}
    \caption{Star-formation rate as a function of stellar mass for our kinematic sample ($0.5 \lesssim z \lesssim 1.7$). The 30 galaxies used in this work are represented by their observed ionised-gas velocity fields rather than by single data points, while the orange hexagons indicate galaxies from the MSA-3D Cycle 1 programme that were excluded from the kinematic analysis. Black curves show the star-forming main sequence from \citet{Schreiber_2015} (converted to the \citet{Chabrier_2003} IMF of the 3D-HST catalogues) evaluated at $z=0.5$ (dashed), $z=1.0$ (solid), and $z=1.5$ (dash-dotted). Blue contours show the density distribution of star-forming galaxies from the 3D-HST survey \citep{Brammer_3DHST, Momcheva_3DHST}. Our sample lies along the locus of typical star-forming galaxies across the probed redshift range, mostly on or slightly below the main sequence (median offset $-0.16$~dex when each galaxy is compared with the relation at its own redshift, with 16 of the 30 galaxies within $\pm0.3$~dex).}
    \label{fig: main sequence}
\end{figure*}

In this work, we use the MSA-3D Cycle 1 sample to investigate the resolved dynamics and inner mass distributions of star-forming galaxies at $z\sim0.5$--1.7. We combine \textit{JWST}/NIRCam imaging with forward modelling of the NIRSpec ionised-gas kinematics using \texttt{DysmalPy} (\citealp{Davies_2004b, Davies_2004a, Cresci_2009, Davies_2011, Wuyts_2016, Lang_2017, Price_2021, Lee_2025}) to constrain rotation curves, baryonic and dark matter contributions, and intrinsic velocity dispersions. We then investigate the stellar Tully--Fisher relation, inner dark matter fractions and halo profiles, and the diversity of outer rotation-curve shapes.

\begin{figure*}
    \centering
    \includegraphics[width=\linewidth]{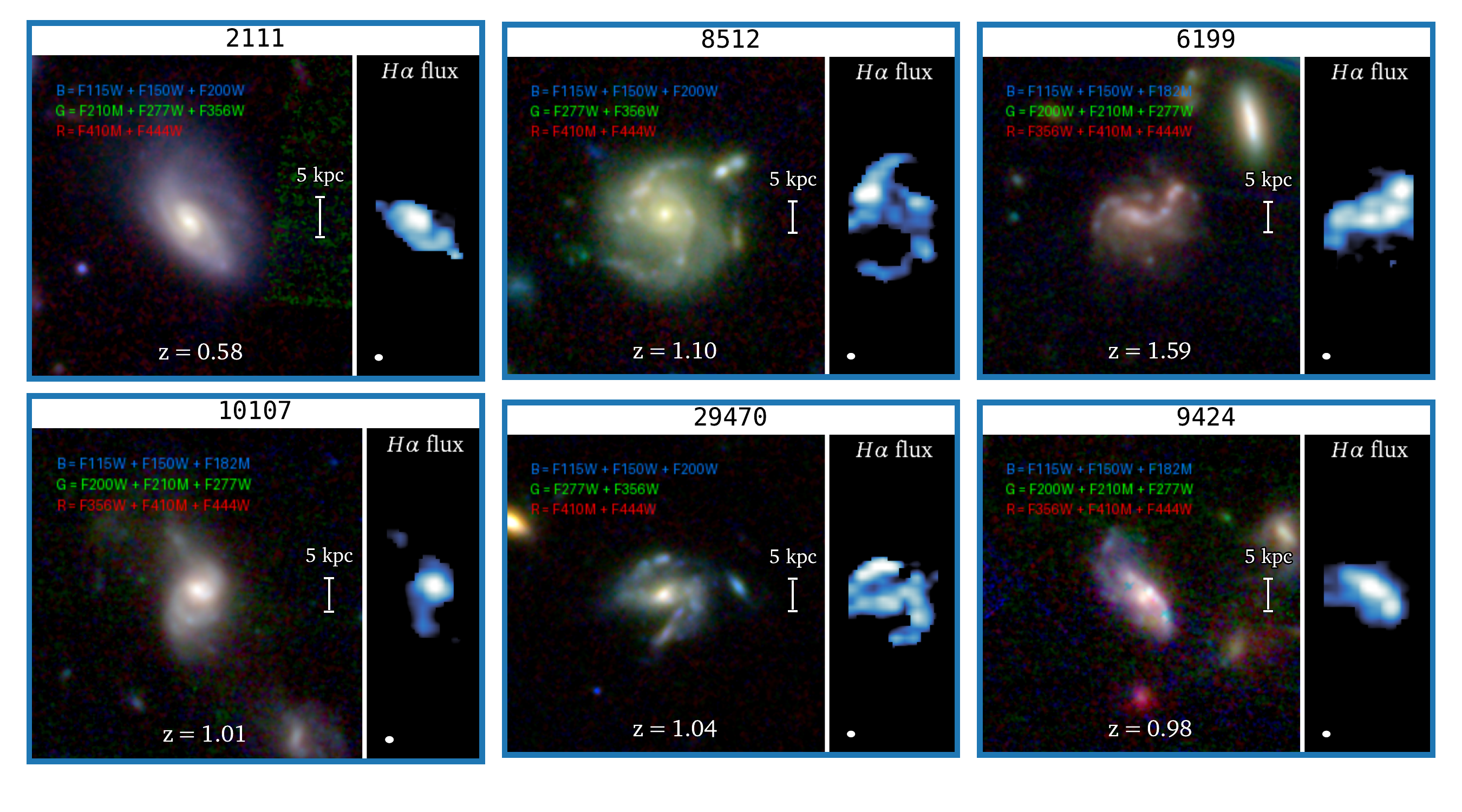}
    \caption{Colour composite images of a representative subsample of galaxies from the MSA-3D Cycle 1 sample. For each galaxy, the left panel shows the \textit{JWST}/NIRCam colour composite, while the right panel displays the corresponding H$\alpha$ flux map derived from MSA-3D spectra. The colour scheme used for the NIRCam images is indicated in the upper left corner of each panel, listing the filters assigned to the blue, green, and red channels. All images are shown at the same angular scale, corresponding to a field of view of 6" × 6". The galaxy ID is given at the top of each panel, and the spectroscopic redshift is indicated in the lower part of the NIRCam image. \rev{The white scale bar in each panel corresponds to a physical size of 5 kpc at the galaxy redshift and the white ellipse in the lower-left corner of each H$\alpha$ panel shows the FWHM of the point-spread function of the slit-stepping data ($0\farcs19\times0\farcs16$)}. These sources illustrate the range of disk-like morphologies in the sample, with H$\alpha$ mapped spectroscopically across a large region of each galaxy.}
    \label{fig: Colour images}
\end{figure*}

The paper is organised as follows. Sections~\ref{section: Data and sample selection}--\ref{section: Forward modelling with DYSMALPY} describe the sample, kinematic measurements, and dynamical modelling; Sections~\ref{section: Disk classification and rotational support}--\ref{section: Rotation curve shapes} present the disk kinematics, Tully--Fisher relation, dark matter constraints, and rotation-curve shapes; and Section~\ref{section: Conclusions} summarises our conclusions. Additional tests, including the complementary baryonic rotation-curve analysis and modelling systematics, are presented in the Appendices, while diagnostic figures for all individual galaxies are provided in the \href{https://www.mpe.mpg.de/resources/IR/MSA-3D_cycle_1/MSA_3D_rotation_curves_supplementary_material.pdf}{supplementary material}.

Throughout this paper, we adopt a $\Lambda$CDM cosmology with $\Omega_\mathrm{m}=0.3$, $\Omega_\Lambda=0.7$, and $H_0=70~\mathrm{km\, s^{-1}\, Mpc^{-1}}$. At $z=1.0$, one arcsecond corresponds to $\approx8$ kpc.

\section{Data and sample selection}
\label{section: Data and sample selection}

\subsection{MSA-3D sample properties}
\label{subsection: MSA-3D data}

The parent MSA-3D sample (\citealp{Barisic_2025}) consists of 43 star-forming galaxies spanning the redshift range $0.5 < z < 1.7$, stellar masses of $9.0 < \log(M_\star/M_\odot) < 11.2$ and star formation rates $\mathrm{SFR} \gtrsim 0.6\, M_\odot \,\mathrm{yr}^{-1}$. As shown in Figure~\ref{fig: main sequence}, the majority of targets lie close to the star-forming main sequence, providing broad coverage of main-sequence star-forming galaxies across the probed mass range at the end of the cosmic noon epoch. All targets are located in the Extended Groth Strip (EGS), a field with extensive ancillary multi-wavelength photometric and spectroscopic coverage. In particular, 40/43 have complementary \textit{JWST}/NIRCam \rev{(\citealp{Rieke_2023})} and/or MIRI imaging\footnote{The \textit{JWST} imaging filters available for each galaxy are listed in the colour-composite panels provided in the \href{https://www.mpe.mpg.de/resources/IR/MSA-3D_cycle_1/MSA_3D_rotation_curves_supplementary_material.pdf}{supplementary material}.} obtained from CEERS JWST Early Release Science (\citealp{Finkelstein_2023}).

In addition, we utilise ancillary multi-wavelength photometric and spectroscopic data from the 3D\textit{-HST} \citep{Momcheva_3DHST} and CANDELS \citep{Koekemoer_2011} surveys. \rev{We adopt the stellar masses and star formation rates from these catalogues, derived from homogeneous integrated SED fitting, as the fiducial global values used throughout this work.} These measurements also provide consistency with previous large-scale studies of the EGS field and allow us to place the MSA-3D galaxies within the broader population of star-forming systems at similar redshifts. \rev{Spectroscopic redshifts are measured directly from the NIRSpec data by fitting H$\alpha$ and [N\,\textsc{ii}] (or [O\,\textsc{iii}]$\lambda5007$ for ID 13182) to the spectrum integrated over the emission-line maps; these redshifts are used for all distance-dependent quantities.}

For each galaxy we define a reference image, the reddest available high-resolution image covering the galaxy, which we use for all structural measurements and visual classifications in this work: NIRCam F444W for 18 galaxies and F356W for eight; F115W for \texttt{12239}, which lies just outside the footprint of the redder NIRCam filters; MIRI F770W for \texttt{13182}, which lies outside the NIRCam footprint; and \textit{HST}/WFC3 F160W from CANDELS for \texttt{13416} and \texttt{18188}, which have no \textit{JWST} imaging.

\rev{Spatially resolved stellar mass maps were derived from pixel-by-pixel spectral energy distribution (SED) fitting of the \textit{JWST}/NIRCam (F115W, F150W, F200W, F277W, F356W, F410M and F444W) and \textit{HST} (F606W, F814W, F105W, F125W, F140W and F160W) imaging, all resampled to a common $0\farcs03$ pixel scale and matched to the F444W PSF, following \citet{Jiang_2024}. The fits used \texttt{Bagpipes} \citep{Carnall_2018} with a \citet{Chabrier_2003} IMF, an exponentially declining star-formation history ($0.3\le\tau\le10$~Gyr), stellar metallicities $0\le Z/Z_\odot\le2.5$, and the \citet{Calzetti_2000} attenuation law ($0\le A_V\le2$). Maps with complete filter coverage and without artefacts are available for only ten galaxies of the kinematic sample (\texttt{2145}, \texttt{2465}, \texttt{2824}, \texttt{3399}, \texttt{4391}, \texttt{6199}, \texttt{7561}, \texttt{8365}, \texttt{8512} and \texttt{8576}). We therefore use these maps only to characterise the spatial distribution of the stellar mass (Section~\ref{subsection - Characterisation of geometrical parameters}), while retaining the catalogue stellar masses as the fiducial integrated values for the whole sample.}

Figure~\ref{fig: Colour images} shows colour images generated with \texttt{MORPHANG} (Pastras, 2026, in prep.), which uses the \texttt{Trilogy} image-combination algorithm \citep{Coe_2012}, together with their corresponding H$\alpha$ flux distributions for six representative galaxies in the sample.

\subsection{NIRSpec slit-stepping strategy}
\label{subsection: slit-stepping strategy}

Observations were carried out with \textit{JWST}/NIRSpec \rev{(\citealp{Jakobsen_2022, Ferruit_2022})} using the high-resolution G140H/F100LP grating–filter combination ($R \sim 2700$), providing wavelength coverage over 0.97–1.82~$\mu$m. The exact spectral coverage varies slightly from target to target, depending on slit position within the micro-shutter assembly. In addition to H$\alpha$, the observations typically include other rest-frame optical emission lines such as H$\beta$, [O\textsc{iii}], [N\textsc{ii}], [S\textsc{ii}], and [S\textsc{iii}], depending on redshift, enabling complementary diagnostics of ionised gas conditions \citep{Barisic_2025,Ju_2025}.

Each galaxy was observed using the NIRSpec multi-slit configuration with seven dither positions arranged in a $7\times9$ slit-stepping dither pattern (see \citealp{Barisic_2025} for details). This strategy yields a total spatial coverage of approximately $1\farcs8 \times (2\farcs0{-}3\farcs0)$. The slit width is $0\farcs2$, and the stepped pattern of nine slitlets ensures full two-dimensional coverage across the galaxy disks, enabling the reconstruction of resolved velocity fields at kiloparsec-scale resolution.

The spatial sampling and angular coverage were designed to trace galaxy kinematics out to radii of $\simeq 3 R_{\mathrm{s}}$, where $R_{\mathrm{s}}\simeq1.0$--2.5~kpc is the range of exponential scale lengths expected for the stellar masses of our targets from the average H$\alpha$ profiles of star-forming galaxies at $z\simeq0.7$--1.5 \citep{Nelson_2016}, with $R_{\mathrm{s}}\simeq2.5$~kpc at the upper-mass end ($M_\star\approx10^{11}\,M_\odot$). This configuration allows us to probe well beyond the turnover radius ($R \gtrsim 2.2 R_{\mathrm{s}}$), where the rotation curve of a self-gravitating exponential disk reaches its maximum \citep{Freeman}. With a total effective integration time of $\sim$2 hours per target, the observations were designed, using the NIRSpec exposure time calculator, to reach ${\rm SNR}=10$ per $0\farcs1\times0\farcs2$ element for an H$\alpha$ surface brightness of $7\times10^{-17}$~erg\,s$^{-1}$\,cm$^{-2}$\,arcsec$^{-2}$ at $\lambda\approx1.3\,\mu$m, assuming an intrinsic line width of $\sigma=50$~km\,s$^{-1}$. This is the average H$\alpha$ surface brightness at $2.2R_{\rm s}$ for these galaxies \citep{Nelson_2016}, so that the average profiles are detected at ${\rm SNR}\approx5$ out to $R\simeq3R_{\rm s}$, although the achieved radial extent varies from target to target depending on line brightness, size, and surface-brightness distribution.
This depth is sufficient to map both the rising and outer regions of the rotation curves for a substantial fraction of the sample.

\rev{We measured the effective point-spread function (PSF) of the reconstructed slit-stepping cubes empirically for two galaxies with bright, extended continuum emission (\texttt{8942} and \texttt{2145}). For each, we compared a line-free continuum image extracted from the cube (median over $\simeq0.2\,\mu$m near $1.2\,\mu$m) with the \textit{JWST}/NIRCam F115W image, rotated to the orientation of the cube and resampled to a common pixel scale, and fitted for the elliptical Gaussian kernel that, convolved with the NIRCam image, best reproduces the NIRSpec image. Adding the F115W PSF in quadrature, we obtain FWHMs of $0\farcs15$--$0\farcs23$ along the stepping direction and $0\farcs16$--$0\farcs17$ along the slit. We adopt their averages, $0\farcs19\times0\farcs16$, in the kinematic modelling (Section~\ref{section: Forward modelling with DYSMALPY}) and include alternative PSF shapes in the systematic uncertainties (Section~\ref{subsection: Fit quality and uncertainties}).}

\begin{table*}[ht]
\centering
\caption{Properties of the MSA-3D galaxies used in this study. The upper section corresponds to the golden sample, while the lower section corresponds to the good sample. The table lists the galaxy ID, spectroscopic redshift ($z$), right ascension (RA), declination (Dec), stellar mass ($M_\star$), and star formation rate (SFR) as derived from the 3D\textit{-HST} and CANDELS catalogues \citep{Koekemoer_2011, Momcheva_3DHST}, and effective radius ($R_{\rm e,S\acute{e}rsic}$). The reported $R_{\rm e,S\acute{e}rsic}$ corresponds to the photometric effective radius measured from a single-component \texttt{GALFIT} (\citealp{Peng_2002}) fit to the reference image of each galaxy (Section~\ref{subsection: MSA-3D data}).}
\begin{tabular}{ccccccc}
\hline
ID     & $z$    & RA            & Dec          & $\log(M_\star/M_\odot)$      & SFR & $R_{\rm e,\,S\acute{e}rsic}$\\
       &        & deg           & deg          &  & $M_\odot/\mathrm{year}$ & arcsec \\
\hline
\multicolumn{7}{c}{\textbf{Golden sample}}\\
\hline
\texttt{2111}   & 0.58   & 215.0627824	&	52.9070766 & 9.97  & 1.6  & 0.66 \\
\texttt{2145}   & 1.17   & 215.0694675	&	52.9108516 & 9.19  & 1.4  & 0.2 \\
\texttt{2465}   & 1.25   & 215.0704381	&	52.9137241 & 9.30  & 1.6  & 0.43 \\
\texttt{2824}   & 0.98   & 215.0685025	&	52.914333  & 9.49  & 0.7  & 0.32 \\
\texttt{3399}   & 1.34   & 215.042511	&	52.8996031 & 9.81  & 3.5  & 0.4 \\
\texttt{4391}   & 1.08   & 215.0676175	&	52.9232437 & 9.48  & 2.7 & 0.18 \\
\texttt{6199}   & 1.59   & 215.0450158	&	52.9194652 & 10.00 & 17.8 & 1.21 \\
\texttt{6430}   & 1.17   & 215.0131444	&	52.8980378 & 9.79  & 8.9 & 0.32 \\
\texttt{7314}   & 1.28   & 214.9989097	&	52.8925151 & 9.55  & 4.8 & 0.19 \\
\texttt{7561}   & 1.03   & 215.0609094	&	52.9383828 & 9.21  & 3.0 & 0.4 \\
\texttt{8365}   & 1.68   & 215.0599904	&	52.9422373 & 9.56  & 3.4 & 0.26 \\
\texttt{8576}   & 1.57   & 215.0595679	&	52.9434335 & 9.60  & 3.3 & 0.3 \\
\texttt{8942}   & 1.18   & 215.0094032	&	52.9100655 & 9.86  & 16.2 & 0.22 \\
\texttt{9424}   & 0.98   & 214.9926533	&	52.900911  & 9.76  & 5.4 & 0.58 \\
\texttt{9636}   & 0.74   & 215.0365985	&	52.9328783 & 9.35  & 1.7 & 0.7 \\
\texttt{9812}   & 0.74   & 215.0402843	&	52.9376004 & 9.97  & 3.6 & 0.88 \\
\texttt{9960}   & 1.51   & 215.031894	&	52.9331513 & 11.06 & 6.0 & 1.93 \\
\texttt{10910}  & 0.74   & 215.0561932	&	52.9553701 & 9.57  & 1.5 & 1.3 \\
\texttt{11843}  & 1.46   & 215.0390468	&	52.9471037 & 10.74 & 19.0 & 0.59 \\
\texttt{11225}  & 1.05   & 215.0415788	&	52.9454655 & 9.67  & 12.6 & 0.19 \\
\texttt{12015}  & 1.24   & 215.0323151	&	52.9431798 & 10.08 & 11.2 & 0.34 \\
\texttt{12239}  & 0.89   & 215.0495343	&	52.9560252 & 9.49  & 1.5 & 0.22 \\
\texttt{18586}  & 0.76   & 214.9712282	&	52.9337862 & 9.01  & 3.3 & 0.22 \\
\texttt{29470}  & 1.04   & 214.9689505	&	52.9453733 & 9.71  & 2.1 & 1.39 \\
\hline
\multicolumn{7}{c}{\textbf{Good sample}}\\
\hline
\texttt{8512}   & 1.1    & 215.0497806	&	52.9380795 & 10.32 & 3.2 & 1.81 \\
\texttt{9337}   & 1.17   & 214.9957065	&	52.9019407 & 9.27  & 3.9 & 1.3 \\
\texttt{10107}  & 1.01   & 214.9817759	&	52.8975743 & 10.01 & 3.6 & 0.31 \\

\texttt{13182$^{(a)}$} & 1.54   & 214.99983	&	52.9268186 & 10.25 & 15.8 & 0.34 \\ 
\texttt{13416}  & 1.54   & 215.0252815	&	52.9456859 & 10.76 & 20.0 & 0.6 \\ 
\texttt{18188}  & 0.82   & 214.9839579	&	52.9413562 & 9.56  & 0.8 & 0.78 \\
\hline
\end{tabular}
\tablefoot{
        \tablefoottext{$a$}{For this system, H$\alpha$ falls outside the wavelength coverage, so we used [O\textsc{iii}]$\lambda5007$ as the kinematic tracer.}\\
        }
\label{tab: basic parameters}
\end{table*}

\subsection{Sample selection}
\label{subsection: Sample selection}

For the present analysis, we focus on a subset of 30 galaxies selected from the parent sample of 43 targets based on the quality and spatial extent of their emission-line kinematics. \rev{Specifically, all galaxies included in the modelling have emission detected at ${\rm S/N}\geq5$ out to radial extents of $\gtrsim2$ PSF FWHM and exhibit an ordered velocity gradient consistent with rotation. Galaxies that do not meet these criteria, either because of insufficient emission-line signal-to-noise, limited spatial coverage, or the absence of a clear ordered velocity gradient, are excluded from the dynamical modelling.}

\rev{We divide the final kinematic sample of 30 galaxies into a \textit{golden} and a \textit{good} sample using three quantitative criteria, applied uniformly and without per-galaxy exceptions. All three are based on the imaging or on the spatial extent of the data: \textit{i)} an imaging inclination $25^\circ < i < 75^\circ$, excluding nearly face-on systems, whose deprojected velocities are dominated by inclination uncertainties \citep{Harrison_2017}, and nearly edge-on systems, in which extinction flattens the inner rotation curve \citep{Sharma_2025}; \textit{ii)} a disk effective radius (from a disk+bulge decomposition of the imaging; Section~\ref{subsection - Characterisation of geometrical parameters}) at least as large as the PSF FWHM, $R_{\rm e,\,disk}/{\rm FWHM}_{\rm PSF} \geq 1$; and \textit{iii)} modelled data extending to at least $1.5\,R_{\rm e,\,disk}$, so that the dark matter fraction at $R_{\rm e,\,disk}$ and the shape of the rotation curve beyond it are constrained by the data. Twenty-four galaxies satisfy all three criteria and form the golden sample. The remaining six galaxies form the complementary \textit{good sample}: four have low inclinations (8512, 10107, 13182, 18188), one is close to edge-on (13416), and one has insufficient radial coverage (9337). We model them with the same two-dimensional approach and show them in figures and qualitative comparisons for completeness, but their inferred mass distributions and dark matter fractions are less well constrained, and we exclude them from all statistical analyses and regression fits, which are based exclusively on the golden sample.}

In the full sample, nearly all galaxies have both H$\alpha$ and [O\textsc{iii}]$\lambda5007$ covered by the NIRSpec spectra. Throughout this study, H$\alpha$ is adopted as the primary kinematic tracer, as it typically provides the highest S/N and is the most robustly detected emission line. In one case (\texttt{ID:13182}), H$\alpha$ falls outside the wavelength coverage, and we therefore use [O\textsc{iii}]$\lambda5007$ as the kinematic tracer, as it is the strongest available emission line. Aside from the choice of emission line, this system was analysed using the same reduction and kinematic fitting procedures as the rest of the sample.

The resulting kinematic sample remains broadly representative of the parent sample in the SFR--$M_\star$ plane, but is intentionally biased toward systems with regular, spatially resolved disk-like kinematics best suited for detailed disk-based dynamical analysis. Its basic properties are listed in Table~\ref{tab: basic parameters}.

\section{Data processing and extraction of kinematic maps}
\label{section: Data Processing and maps extraction}

\subsection{Emission line fitting}
\label{subsection: Emission line fitting}

Spatially resolved velocity and dispersion maps were derived from the \textit{JWST}/NIRSpec spectra following a uniform Gaussian line-fitting procedure applied on a pixel-by-pixel basis on the reconstructed cube. \rev{For each spaxel, we identified emission-free spectral regions around H$\alpha$ (or [O\textsc{iii}]$\lambda5007$ for \texttt{ID:13182}), estimated the continuum level as the median flux density in these regions, and subtracted this constant level from the spectrum.} For a small subset of sources exhibiting artificial band-like features, these regions were manually masked prior to fitting. \rev{The same maps and masks are used for all the dynamical modelling and tests presented in this work; none of the detection or clipping thresholds were changed after the kinematic fits.}

A single Gaussian profile was fit to each spaxel \rev{with emission exceeding $2\sigma$ in at least two consecutive spectral channels. The channel width is 37.1--64.4~km\,s$^{-1}$ (median 47.5~km\,s$^{-1}$), such that the FWHM of the line-spread function (LSF; Section~\ref{subsection: Priors and modelling assumptions}) spans $\approx1.9$ channels and even an unresolved line extends over at least two adjacent channels. For independent Gaussian noise, the probability that two given consecutive channels both exceed $2\sigma$ is $5.2\times10^{-4}$, corresponding to a false-detection probability of $\approx1$\% per spaxel for the at most 27 channels of the cropped spectra. This relatively low threshold avoids biasing the maps against regions with broad emission and lower peak signal-to-noise ratios; similar multi-channel criteria have been adopted in previous studies \citep{Cortese_2026}. The intensity, velocity, and velocity dispersion maps correspond to the integrated flux, centroid, and width of the fitted Gaussians, rather than moments of the data. The noise was measured from adjacent emission-free regions and propagated into the kinematic uncertainties. We excluded spaxels with velocity or dispersion uncertainties exceeding 30~km\,s$^{-1}$, an empirical compromise between measurement precision and spatial coverage, since a stricter cut would preferentially remove regions with faint or broad emission (e.g., section 3.2.6 in \citealp{van_de_Sande_2017}). We also removed isolated ``islands'' of spurious detections arising from residual noise peaks or detector artefacts, identified by visual inspection aided by a criterion on the minimum number of connected detected spaxels. Finally, we applied a $3.5\sigma$ clipping to the velocity map, removing spaxels whose velocities deviate from those of the surrounding spaxels by more than 3.5 times the scatter among them; such isolated outliers arise from residual instrumental artefacts such as cosmic rays or hot pixels. Genuine non-circular motions, which are generally spatially coherent over several spaxels, should be largely unaffected by this criterion; in practice, this step removed only a negligible number of spaxels.}

\begin{figure*}
    \centering
    \includegraphics[width=\linewidth]{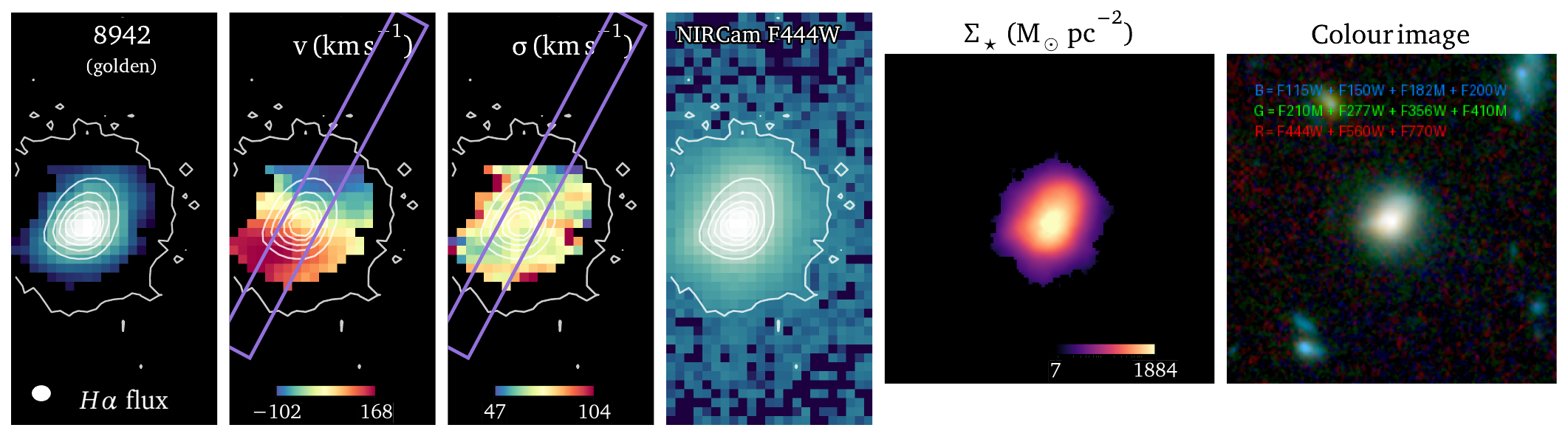}
    \caption{Summary of imaging and kinematic properties for galaxy \texttt{8942}. From left to right: \textbf{(1)} H$\alpha$ flux, \textbf{(2)} H$\alpha$ velocity field with white continuum contours (from NIRCam imaging); \textbf{(3)} H$\alpha$ velocity dispersion map; \textbf{(4)} \textit{JWST}/NIRCam F444W imaging with the same contours; \textbf{(5)} Stellar mass surface density map $\Sigma_\star$ derived from spatially resolved SED fitting; and \textbf{(6)} RGB colour composite image, constructed using NIRCam medium and wide-band filters as labelled. All maps share the same orientation, and colour bars indicate the range of each quantity in physical units. The white ellipse in the H$\alpha$ flux panel shows the FWHM of the PSF ($0\farcs19\times0\farcs16$). The corresponding diagnostic figures for all galaxies in the sample are provided in the \href{https://www.mpe.mpg.de/resources/IR/MSA-3D_cycle_1/MSA_3D_rotation_curves_supplementary_material.pdf}{supplementary material}.}
    \label{fig: 8942}
\end{figure*}

The continuum level was not refitted simultaneously with the emission line, as its residual uncertainty after subtraction was much smaller than the spectral noise level and would not contribute meaningfully to the error budget. The full set of continuum-subtraction and Gaussian-fit validation plots for all spaxels was visually inspected to ensure the robustness of the resulting maps. See Figs.~\ref{fig: Colour images} and \ref{fig: 8942} for some examples of the extracted intensity, velocity, and dispersion maps.

\rev{Because H$\alpha$ traces star-forming regions, its surface brightness is often enhanced along spiral arms, bars and clumps. The kinematic fits, however, constrain the gravitational potential through the line-of-sight velocities and dispersions, which are measured wherever emission is detected; in our sample, fainter emission between these structures is also detected, so that the maps cover the disks contiguously, with a median of 34 resolution elements per map and 97\% of the pixels within $1.5R_{\rm e}$ detected. Non-circular motions associated with bars, arms or clumps appear as residual structure in the velocity maps, which we propagate into the parameter uncertainties through the block-resampling calibration (Section~\ref{subsection: Fit quality and uncertainties}). The geometry is also not inferred from the H$\alpha$ morphology: the inclinations are measured from the outer continuum isophotes, avoiding bars and spiral arms, and the kinematic and photometric position angles agree to within $30^\circ$ for 78\% of the golden sample. The uneven H$\alpha$ surface brightness mainly affects the flux weighting of the beam-smeared central velocities, which the PSF limits to the innermost $\sim1$--1.5~kpc.}

\subsection{Characterisation of geometrical parameters}
\label{subsection - Characterisation of geometrical parameters}

The geometrical parameters (inclination and position angle) are among the most critical inputs for dynamical modelling, as they strongly influence the inferred circular velocities and derived dark matter fractions. Due to the intrinsic degeneracies between inclination, rotation velocity, and disk thickness in kinematic forward modelling, we determined these quantities independently (prior to the dynamical fits) from the imaging.

For each galaxy, we derived the inclination from its reference image (Section~\ref{subsection: MSA-3D data}). We performed isophotal ellipse fitting on the reference image of each galaxy, using the outer isophotes and avoiding the inner regions where bulges, bars, spiral arms or clumpy star formation could bias the shape measurement. The mean axis ratio ($b/a$) of the outer isophotes was then converted to an inclination assuming a thick-disk geometry, adopting the standard relation

\begin{equation}
\cos^2 i = \frac{(b/a)^2 - q_0^2}{1 - q_0^2},
\end{equation}

where $q_0$ represents the intrinsic flattening of the disk as seen edge-on, with a commonly adopted value of $q_0 = 0.2$ for disk galaxies (e.g., \citealp{Tully_fisher, Giovanelli_1997}). \rev{This value is close to the intrinsic thickness adopted in the dynamical models ($q=0.25$; Section~\ref{section: Forward modelling with DYSMALPY}); adopting $q_0=0.25$ instead changes the inclinations by less than $3^\circ$ for $i\le75^\circ$.}

\rev{The structural parameters of the baryonic components entering the dynamical models were determined independently of the kinematic modelling. We fitted the reference image with a PSF-convolved double-component model consisting of an exponential disk (S\'ersic index $n=1$) and a de Vaucouleurs bulge ($n=4$). This provides the disk effective radius, $R_{\rm e,\,disk}$, and the bulge-to-total light ratio, $B/T$.}

For the ten galaxies with reliable stellar mass maps (Section~\ref{subsection: MSA-3D data}), we applied the same double-component fit to the stellar mass surface-density maps, adopting an exponential disk and a de Vaucouleurs bulge with a common centre and requiring the bulge to be more compact than the disk. The models were convolved with the F444W PSF to which the mass maps are matched. This provides a mass-weighted bulge-to-total ratio, which better represents the relative distribution of stellar mass than the corresponding light-weighted value and can reveal central mass concentrations that are weak in the rest-frame optical light (e.g. $B/T=0.32$ for \texttt{3399}, compared with 0.01 from the F444W decomposition). The disk effective radii inferred from the mass maps agree closely with those measured from the imaging (median ratio $R_{\rm e,\,disk}^{\rm mass}/R_{\rm e,\,disk}^{\rm light}=1.02$), and we therefore adopt the imaging-based $R_{\rm e,\,disk}$ for all galaxies. For the ten galaxies with reliable mass maps, we adopt the mass-weighted $B/T$ to describe the relative stellar mass distribution; the overall stellar mass normalisation remains fixed by the catalogue stellar mass described in Section~\ref{subsection: MSA-3D data}.

In addition to the morphological position angles derived from the reference images, we estimated the kinematic position angles using the \texttt{PAFIT} algorithm (\citealp{Kinemetry}) applied to the observed velocity fields. The two estimates are broadly consistent, allowing for the non-zero offsets expected between stellar-continuum morphologies and ionised-gas velocity fields: for the golden sample, the median absolute offset between the kinematic and photometric position angles is $18^\circ$, with 78\% of galaxies agreeing within $30^\circ$ (e.g., \citealp{KMOS3D}). \rev{In the dynamical modelling, the position angle is a free parameter with a flat prior of $\pm30^\circ$ centred on the \texttt{PAFIT} value, i.e. at least as wide as the typical disagreement between the two estimates.}

\rev{The kinematic centre was measured independently and held fixed in the dynamical modelling. We adopt the centroid in the reddest available NIRCam image (F444W or F356W) for the 26 galaxies with suitable coverage, and the centroid of the line-free continuum in the NIRSpec cube for the remaining four. Allowing the centre to vary freely produces large offsets, exceeding one PSF FWHM from the nearest light centroid in eleven galaxies and reaching the prior bounds in five, suggesting that the additional freedom can absorb asymmetric or non-circular velocity structure rather than robustly trace the gravitational centre. Nevertheless, the impact on the inferred parameters is moderate: relative to the free-centre fits, the fiducial models change $f_{\rm DM}(R_{\rm e})$ by a median of $+0.01$ (median absolute difference 0.03) and $\alpha$ by $-0.10$, while $\sigma_0$ is unchanged. We propagate the uncertainty in the adopted centre into the systematic error budget by refitting after displacements of $0\farcs07$ and $0\farcs05$ along the stepping and slit directions, respectively. The centre selection and full free-centre tests are presented in Appendix~\ref{section: free centre}.}

\section{Kinematic Forward Modelling with \texttt{DysmalPy}}
\label{section: Forward modelling with DYSMALPY}

We perform the full kinematic forward modelling using \texttt{DysmalPy}\footnote{\url{https://github.com/dysmalpy/dysmalpy}}, a publicly available Python package designed to model spatially resolved galaxy kinematics. In short, the code starts from parametrised mass and light distributions, constructs three-dimensional model datacubes, convolves them with instrumental and observational effects (PSF, LSF, projection, pixel sampling), and extracts model velocity and velocity-dispersion maps that are fitted directly to the observed two-dimensional velocity and velocity-dispersion maps. \rev{The baryonic and dark matter components are constrained through the forward modelling of the observed kinematics, with the structural parameters of the baryons (effective radii and, where available, bulge fractions from the spatially resolved stellar mass maps) fixed or constrained by the imaging.} This approach provides a physically motivated framework for jointly constraining the baryonic and dark matter distributions and the intrinsic kinematic properties of the disk, while self-consistently accounting for beam smearing (see \citealp{Price_2021} for details of the code). \rev{An alternative approach is three-dimensional tilted-ring modelling of the full data cube, as implemented in \texttt{3D-Barolo} \citep{DiTeodoro2015}, \texttt{TiRiFiC} \citep{Tirific} and \texttt{FAT} \citep{Kamphuis}, which has been extensively validated on H\,{\sc i}, CO and optical/near-infrared IFU data. We adopt forward parametric modelling because it constrains the baryonic and dark matter components simultaneously with the geometry and kinematics in a single self-consistent model, rather than through a separate mass decomposition of an empirically derived rotation curve, which is central to our focus on inner dark matter fractions.} We now describe the specific mass components, priors and parameter choices, posterior summaries, fit quality assessment and robustness tests.

\subsection{Mass model}
\label{subsection: Mass model}

We model each galaxy with a three-component mass distribution comprising an exponential stellar disk (Sérsic index $n=1$), a de Vaucouleurs bulge ($n=4$) (\citealp{de_Vaucouleurs_1953}), and a \rev{generalised NFW profile \citep{Zhao_1996} given by}

\begin{equation}
  \rho(r) = \frac{\rho_s}{(r/r_s)^{\alpha}\,(1 + r/r_s)^{3-\alpha}},
\end{equation}

\rev{where $\alpha$ is the inner logarithmic slope ($\alpha=1$ recovers the NFW profile of \citealt{Navarro_1997}, while $\alpha=0$ corresponds to a constant-density core) and $r_s = R_{\rm vir}/c$. We fit $\alpha$ with a flat prior over $0\leq\alpha\leq1.5$; the adopted parametrisation retains the NFW outer asymptotic behaviour, $\rho \propto r^{-3}$}. The concentration is fixed using the redshift-dependent relation $c = 10.9(1+z)^{-0.83}$ adopted by \citet{Genzel_2020} from \citet{Dutton_Maccio_2014}; \rev{its uncertainty is included in the systematic error budget (Section~\ref{subsection: Fit quality and uncertainties}). The virial mass $M_{\rm vir}$ is a free parameter with a Gaussian prior of width 0.5~dex centred on the stellar-to-halo mass relation of \citet{Moster_2013}, and the dark matter fraction within the disk effective radius, $f_{\rm DM}(R_{\rm e}) = V_{\rm DM}^2(R_{\rm e})/V_{\rm c}^2(R_{\rm e})$, is derived from the model for every posterior sample. Because $M_{\rm vir}$ is fitted, the prior on $f_{\rm DM}$ is induced by the priors on the mass components rather than chosen directly. Given that our kinematic data extend to only about $3\,R_{\rm e}$, they constrain the dark matter contribution over the observed radial range much more directly than the total halo virial mass. We therefore do not interpret the inferred $M_{\rm vir}$ values as direct measurements of the total halo mass. As a matched reference, we repeat every fit with $\alpha$ fixed to 1 (an NFW halo with otherwise identical parametrisation and priors), thereby isolating the effect of allowing the inner halo slope to vary.}

\rev{The structural parameters of the baryonic components are fixed from the imaging, following \citet{Price_2021} and \citet{Nestor_2023}: the disk effective radius is fixed to the value from the photometric bulge-disk decomposition (Section~\ref{subsection - Characterisation of geometrical parameters}), while the bulge effective radius is fixed to 1~kpc. This value is representative of the compact bulges observed in massive star-forming galaxies at $z\sim0.5-2.5$ (e.g. \citealp{Lang_2014, Tacchella_2015_Bulges}). Because the bulge is compact and its effective radius is not robustly constrained for individual galaxies, we fix its size rather than introducing an additional poorly constrained structural parameter.}

\rev{The bulge-to-total ratio ($B/T$) is taken from the resolved stellar mass maps for the ten galaxies with reliable maps and from the reference image otherwise. It is fixed for 19 galaxies and left free for the 11 in which a bulge component was required to describe the kinematics, with a Gaussian prior of width 0.1 centred on the measured value. Fixing $B/T$ at its measured value instead worsens the fit substantially ($\Delta\chi^2_{\rm b}>2$) for seven of these 11 galaxies, including five of the seven golden ones (median $\Delta\chi^2_{\rm b}=+20$ for those seven, where $\Delta\chi^2_{\rm b}$ is the change in the beam-weighted $\chi^2$ defined in Sect.~\ref{subsection: Priors and modelling assumptions}; Table~\ref{tab: sensitivity fit quality}), and the kinematics prefer bulge fractions that differ from the photometric estimates by 0.1--0.2, which would otherwise bias $f_{\rm DM}(R_{\rm e})$ by up to $\sim$0.2; for the remaining four the fitted $B/T$ agrees with the photometric value. As in \citet{Price_2021}, the bulge fraction is applied to the total baryonic mass, which implicitly assumes that the gas follows the stellar distribution, including the bulge. If the gas resided entirely in the disk, the baryonic bulge fraction would be lower by a factor $M_\star/M_{\rm bar}$ (median 0.4); this is relevant only for the seven golden galaxies with $B/T\geq0.1$, and for those with free $B/T$ the bulge fraction is in any case adjusted by the kinematics. The sensitivity to the fixed structural parameters is part of the systematic error budget (Section~\ref{subsection: Fit quality and uncertainties}).}

\subsection{Priors and modelling assumptions}
\label{subsection: Priors and modelling assumptions}

\rev{The total baryonic mass, $M_{\rm bar}$, is a free parameter describing the combined mass of the disk and bulge. We adopt a Gaussian prior with a width of 0.2~dex, truncated at $\pm0.5$~dex, and centred on the expected baryonic mass $M_{\rm bar,0}=M_\star+M_{\rm mol}+M_{\rm HI}(<2R_{\rm e})$ where $M_\star$ is the catalogue stellar mass and $M_{\rm mol}$ is estimated from the molecular gas-fraction scaling relation of \citet{Tacconi_2018} (their Eq.~6 with the best-fit coefficients of their Table~3b), which depends on redshift, stellar mass, and offset from the star-forming main sequence. The inferred molecular gas fractions span $M_{\rm mol}/M_\star\simeq0.1$--1.8, with a median of 0.9 for the golden sample. The highest gas fractions occur in the lowest-mass galaxies, as expected from the stellar-mass dependence of the scaling relation at fixed redshift and $\Delta{\rm MS}$, and are consistent with empirical measurements of main-sequence galaxies at $z\sim1$ \citep{Tacconi_review_2020}.}

\rev{We also include the contribution from atomic gas within $2R_{\rm e}$. We first estimate the total H\,\textsc{i} mass using the mean values measured for star-forming galaxies at $z\sim1$ by \citet{Chowdhury_2022}, and multiply this by 1.38 to account for helium. We then use the H\,\textsc{i} size--mass relation and radial surface-density profile of \citet{Wang_2016} to estimate the fraction of this gas enclosed within $2R_{\rm e}$. This contribution is modest: the atomic gas raises the centre of the $M_{\rm bar}$ prior by a median of 0.08~dex, and by up to 0.3--0.4~dex for the three most extended, lower-mass disks. Excluding it changes neither the fit quality nor $f_{\rm DM}(R_{\rm e})$ appreciably (Table~\ref{tab: sensitivity fit quality}).}

The model includes pressure support from the turbulent gas motions. For a disk with radially constant intrinsic dispersion $\sigma_0$, the observable rotation velocity is related to the circular velocity by

\begin{equation}
  V_{\rm rot}^2(R) = V_{\rm c}^2(R) + \sigma_0^2\,\frac{{\rm d}\ln\rho}{{\rm d}\ln R},
  \label{eq: adc}
\end{equation}

\rev{where $\rho$ is the gas density in the disk mid-plane \citep{Burkert_2010}. We evaluate the radial density gradient directly from the three-dimensional density profile of the adopted disk model, an exponential ($n=1$) S\'ersic disk with constant intrinsic axis ratio $q=0.25$ \citep{Noordermeer_2008} (distinct from the $q_0 = 0.2$ used to convert axis ratios into inclinations, Section~\ref{subsection - Characterisation of geometrical parameters}), rather than adopting the self-gravitating thin-disk approximation, $V_{\rm c}^2 - V_{\rm rot}^2 = 2\sigma_0^2 R/R_{\rm s}$ \citep{Burkert_2010}, where $R_{\rm s}=R_{\rm e}/1.678$ is the exponential scale length. This ensures that the pressure-support correction is evaluated consistently with the finite-thickness disk adopted in the dynamical model. Over the radial range probed by our data, this corresponds approximately to $V_{\rm c}^2 - V_{\rm rot}^2 \simeq 1.2\,\sigma_0^2 R/R_{\rm s}$. Adopting the thin-disk approximation instead would increase $f_{\rm DM}(R_{\rm e})$ by a median of only 0.03.}

\rev{The model cubes are convolved with the empirical point-spread function of the reconstructed slit-stepping cubes, an elliptical Gaussian with FWHM $0\farcs19$ along the stepping direction and $0\farcs16$ along the slit (Section~\ref{subsection: slit-stepping strategy}),} and with the line-spread function of G140H/F100LP evaluated at the observed wavelength of each galaxy's emission line using the wavelength-dependent model of \citet{Shajib_2025}, 

\begin{equation}
\sigma_{\rm LSF}(\lambda) = \sqrt{\left(\frac{\sigma'_{\rm piv}}{1 + \alpha_\lambda\,(\lambda - \lambda_{\rm piv})/10000}\right)^{\!2} - \sigma_{\rm PN}^2\,},
\label{eq:sigma_lsf}
\end{equation}

where $\lambda_{\rm piv}$, $\sigma'_{\rm piv}$, and $\alpha_\lambda$ are grating-dependent coefficients, and $\sigma_{\rm PN}$ is the intrinsic broadening associated with the planetary nebula used for calibration. All 30 galaxies in our sample were observed with the G140H/F100LP grating--filter combination, for which $(\lambda_{\rm piv},\,\sigma'_{\rm piv},\,\alpha_\lambda,\,\sigma_{\rm PN}) = (13347~\text{\AA},\ 40.63~\text{km\,s}^{-1},\ 0.703,\ 6.90~\text{km\,s}^{-1})$, see Table 1 in \citealp{Shajib_2025}. The observed wavelength $\lambda = \lambda_{\rm rest}(1+z)$ was computed using H$\alpha$ ($\lambda_{\rm rest} = 6562.8$~\AA) as the kinematic tracer for all galaxies (except \texttt{ID:13182}, for which [O\textsc{iii}]~$\lambda5007$ falls in the observed bandpass). The resulting $\sigma_{\rm LSF}$ values span $\approx$[30.5--51.1]~km~s$^{-1}$ across the sample, reflecting the increase in spectral resolving power of G140H toward longer wavelengths.

\begin{figure}
    \centering
    \includegraphics[width=0.98\linewidth]{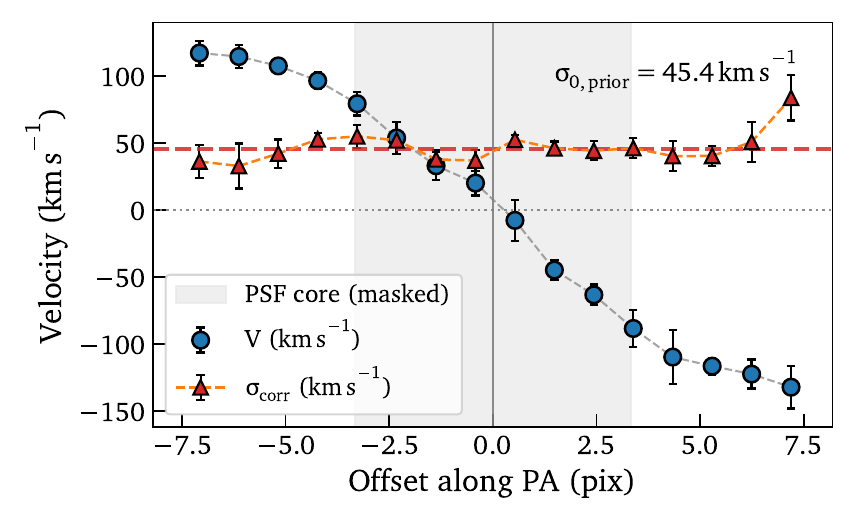}
    \caption{Line-of-sight velocity (blue circles) and velocity-dispersion (orange triangles) profiles of \texttt{8365} along the kinematic major axis, extracted in a slit of width $1.4\times$FWHM. The dispersion is corrected for the instrumental line spread function. The grey region marks the beam-smeared centre, and the dashed line the LSF-corrected outer dispersion adopted as the starting value and the centre of the flat prior on $\sigma_0$.}
    \label{fig: LOSV 8365}
\end{figure}

\rev{The observed dispersion maps are therefore fitted directly, without any prior instrumental correction. The LSF-corrected dispersion, $\sigma_{\rm corr} = (\sigma_{\rm obs}^2 - \sigma_{\rm LSF}^2)^{1/2}$, averaged over the outer disk to avoid the beam-smeared centre, is used only as the starting value and the centre of the flat prior on $\sigma_0$; an example of the extracted line-of-sight velocity and dispersion profiles is shown in Fig.~\ref{fig: LOSV 8365}.} 

\rev{The reconstructed maps oversample the PSF, so neighbouring pixels are not independent. We account for this by dividing the $\chi^2$ of the velocity and dispersion maps by the number of pixels per resolution element, so that the log-likelihood is $\ln\mathcal{L} = -\chi^2/(2N_{\rm beam})$, with $N_{\rm beam} = \pi\,{\rm FWHM}_{\rm maj}\, {\rm FWHM}_{\rm min}/(4\ln2\,\Delta^2)\simeq5.4$ the area of the elliptical PSF in pixels of size $\Delta=0\farcs08$. Each resolution element then contributes once to the likelihood, rather than once per pixel. This scalar correction does not capture the full spatial correlation of the noise and residuals; we calibrate the remaining effect on the parameter uncertainties with resampling tests (Section~\ref{subsection: Fit quality and uncertainties}). When comparing the fit quality of alternative models, we use the corresponding change in this beam-weighted statistic, $\Delta\chi^2_{\rm b}=\Delta\chi^2/N_{\rm beam}=-2\Delta\ln\mathcal{L}$, evaluated relative to the fiducial model; positive values indicate a worse fit (Appendix~\ref{section: error budget}).}

\rev{The inclination is fixed to the imaging value (Section~\ref{subsection - Characterisation of geometrical parameters}) for 20 galaxies for which the photometric estimate is considered sufficiently reliable. For the remaining ten galaxies, the imaging morphology makes the inclination more uncertain, and we therefore leave it free. In six cases, the galaxies are viewed at relatively low inclination and their outer isophotes are less well defined; in this regime the inclination is particularly sensitive to the measured axis ratio ($\Delta(b/a)=0.05$ corresponds to $\Delta i=5$--$13^\circ$, compared with $\approx4^\circ$ for the more inclined galaxies), so that the photometric estimates are not sufficiently reliable to be held fixed. One galaxy (\texttt{13416}) is close to edge-on, where the inferred inclination depends strongly on the assumed intrinsic thickness ($i=80$--$90^\circ$ for $q_0=0.15$--0.25). In the remaining three galaxies, irregular outer isophotes make the axis-ratio-based inclination estimates unreliable. For these ten galaxies, we adopt a flat inclination prior of $\pm15^\circ$ around the imaging value, truncated to $5^\circ$--$89^\circ$. For five of these galaxies (\texttt{6199}, \texttt{8942}, \texttt{9960}, \texttt{11225} and \texttt{29470}), the inclination posterior reaches the prior bound; their inclinations are reported as limits in Table~\ref{tab: Fitted parameters}, and their derived quantities are conditional on this bound.}

\rev{For galaxies with fixed inclinations, we propagate the uncertainty corresponding to $\Delta(b/a)=0.05$ into the systematic error budget, together with a 10\% uncertainty in the fixed disk effective radius (Section~\ref{subsection: Fit quality and uncertainties}). This includes four galaxies with relatively low inclinations ($i\lesssim36^\circ$; \texttt{8576}, \texttt{9812}, \texttt{10107}, and \texttt{18586}) for which the outer isophotes are regular and the imaging-based estimates are therefore considered sufficiently reliable to remain fixed. For the three belonging to the golden sample, varying the inclination within this uncertainty changes $f_{\rm DM}(R_{\rm e})$ by at most 0.02 and $\alpha$ by at most 0.06. The position angle is free with a flat prior of $\pm30^\circ$ around the \texttt{PAFIT} value, the systemic velocity with a flat prior of $\pm35$~km\,s$^{-1}$, and $\sigma_0$ with a flat prior of $\pm30$~km\,s$^{-1}$ around the LSF-corrected outer dispersion. The kinematic centre is held fixed (Section~\ref{subsection - Characterisation of geometrical parameters}). The disk scale height is tied to the disk effective radius through the fixed intrinsic axis ratio, $\sigma_z = q\,R_{\rm e}/\sqrt{2\ln2}$ with $q = 0.25$. We use Gaussian priors only for quantities with an independent external measurement (the baryonic mass, the virial mass through the stellar-to-halo mass relation, and the bulge fraction where it is free), with widths set by the uncertainty of that measurement, and flat priors otherwise. The parameters and priors are listed in Table~\ref{tab: dysmalpy parameters}.}

\subsection{Sampling and posterior summaries}
\label{subsection: Sampling and posterior summaries}

\rev{We sample the posterior using the dynamic nested sampler \texttt{dynesty} \citep{Speagle_2020} with slice sampling, adopting 250 initial live points and 200 live points per subsequent batch. We impose no maximum number of iterations, and all fits achieve effective sample sizes greater than $10^4$. For each fitted parameter, we report the median of the marginalised posterior together with the 16th--84th percentile interval, calculated from the importance-weighted posterior samples. Derived quantities, including $f_{\rm DM}$, $V_{\rm rot}$, $V_{\rm c}$, and $V_{\rm rot}/\sigma_0$, are calculated for every posterior sample and summarised in the same way.}

\rev{For the displayed model maps and fiducial residual statistics, we use the sample with the highest data likelihood from each nested-sampling run. The sensitivity comparisons in Table~\ref{tab: sensitivity fit quality} instead compare the data-only $\chi^2$ evaluated at re-optimised maximum-a-posteriori (MAP) solutions for both the fiducial and alternative models, as described in Section~\ref{subsection: Fit quality and uncertainties}. Parameters whose posteriors extend to a prior boundary are reported as limits. The posterior intervals reflect the statistical uncertainties within the adopted model and noise assumptions; the final uncertainties quoted in this work are calibrated as described in Section~\ref{subsection: Fit quality and uncertainties}. The resulting fitted and derived parameters are listed in Table~\ref{tab: Fitted parameters}.}

\rev{Unless stated otherwise, all quantities quoted at one effective radius, including $f_{\rm DM}(R_{\rm e})$, enclosed masses and rotation velocities, are evaluated at the disk effective radius, $R_{\rm e,\,disk}$, from the double-component fit (Section~\ref{subsection - Characterisation of geometrical parameters}), rather than at the single-component S\'ersic radius $R_{\rm e,\,S\acute{e}rsic}$ listed in Table~\ref{tab: basic parameters}, which describes the total light distribution.}

\subsection{Fit quality and uncertainties}
\label{subsection: Fit quality and uncertainties}

\rev{We assess the quality of each fit quantitatively at the maximum-likelihood solution. The velocity residuals have a median root-mean-square of 17.6~km\,s$^{-1}$ (11.0--37.8~km\,s$^{-1}$ across the sample), while the dispersion residuals have a median of 10.0~km\,s$^{-1}$. We compare these residuals with 999 realisations of the cube-measured, spatially correlated noise. The velocity residuals exceed the noise expectation in every galaxy ($p<0.001$), with a median RMS about three times that expected from noise, as do the dispersion residuals in 18 of the 24 golden galaxies. However, in 21 of the 24 golden galaxies the residuals are correlated on scales no larger than the noise (median 2 versus 3 pixels), indicating small-scale structure, such as non-circular motions, rather than a large-scale mismatch of the rotation curve. We therefore propagate this residual structure into the parameter uncertainties rather than treating the models as complete descriptions of the maps  (see below). To identify unusually large departures, we flag galaxies whose residual excess is an outlier relative to the sample, using a robust $z$-score threshold of 2 defined prior to the fits. This flags \texttt{6430} and \texttt{9424}, whose inferred parameters we interpret with caution. We also note \texttt{6199}: its imaging inclination ($33^\circ$) satisfies our criterion and its regular velocity field is well reproduced by the model, but its kinematic inclination and baryonic mass reach their lower prior bounds. It also has a possible companion with only a grism-based redshift. We retain \texttt{6199} according to the declared criteria; excluding it does not change our population results, and we discuss it further in Appendix~\ref{section: interaction strengths}.}

\rev{The resolved maps also provide substantial information relative to the number of fitted parameters. Each galaxy has six to eight free parameters, depending on whether the inclination and bulge fraction are allowed to vary. The golden sample has a median $R_{\rm e}/{\rm FWHM}_{\rm PSF}=2.1$ (range 1.0--4.0) and 34 independent spatial resolution elements per map (13--77), corresponding to approximately ten independent velocity and dispersion measurements per free parameter. The kinematic data extend to a median radius of $2.8R_{\rm e}$ (1.6--5.5), providing leverage on both the inner and outer mass distributions. Consistent with this information content, the nominal posterior widths are substantially narrower than the adopted priors, with median reductions of 90\% for $\alpha$, 96\% for $\sigma_0$, and 63\% for the baryonic mass. Injection--recovery tests spanning different halo slopes and intrinsic dispersions further assess parameter recoverability, including the frequency of spurious inferences of $\alpha<1$ and the recovery of dispersions below the instrumental resolution (Appendix~\ref{section: injection recovery}).}

\rev{The nominal posterior intervals account for the statistical uncertainties assumed in the likelihood, but not for spatially correlated residuals or uncertainties associated with the modelling assumptions. We therefore calibrate the statistical uncertainties using synthetic-map refits that reproduce the spatially correlated residual structure of each galaxy, and estimate systematic uncertainties by repeating the fits under alternative assumptions for the LSF, PSF, halo concentration, bulge size, atomic-gas contribution, and kinematic centre, and by displacing the fixed inclinations and disk effective radii by their uncertainties ($\Delta(b/a)=\pm0.05$ and $\pm10\%$, respectively). The calibrated statistical and systematic terms are added in quadrature, separately on each side. The full calibration procedure and fallback rules are described in Appendix~\ref{section: error budget}.}

\rev{The alternative models generally support the fiducial assumptions (Table~\ref{tab: sensitivity fit quality}). Alternative PSFs, broader LSFs, fixed larger bulge fractions, and most tested centre displacements worsen the fits, whereas changing the bulge effective radius or omitting atomic gas has little effect. Alternative halo-concentration prescriptions modestly improve the fits and lower $\alpha$ by 0.10--0.11 while leaving $f_{\rm DM}(R_{\rm e})$ nearly unchanged; we therefore retain the fiducial concentration as the more conservative choice for assessing departures from $\alpha=1$.}

\rev{The absolute baryonic mass scale is treated separately because it represents a coherent uncertainty rather than an independent modelling perturbation. Shifting the baryonic-mass prior by +0.5~dex changes $f_{\rm DM}(R_{\rm e})$ by a median of $-0.19$ for galaxies whose baryonic mass is constrained by the data, and by $-0.40$ for those whose posterior reaches the prior boundary. We therefore report this sensitivity separately from the statistical and systematic uncertainties.}

\rev{As an independent check of the map-based fits, we extract position--velocity (PV) diagrams directly from the observed cubes along the kinematic major axis, in a pseudo-slit of four pixels, and compare them with the same slice of the maximum-likelihood model cube (supplementary material). Before extraction, the velocity zero point of each cube is aligned with that of the velocity map using the data alone. The PV diagrams are not used in the fits, which are performed on the two-dimensional velocity and dispersion maps, but they test whether the model also reproduces the spectral data. For all galaxies, the model contours follow the observed emission ridge, and we find no departures beyond the small-scale residual structure already present in the maps.}


\rev{Throughout this paper, parameter values are the medians of the marginalised posteriors, and quoted uncertainties combine the calibrated statistical and systematic terms in quadrature, separately on each side. These total uncertainties are listed in Table~\ref{tab: Fitted parameters}, shown in the main-text figures, and used in subsequent analyses; their statistical and systematic components are given separately in Table~\ref{tab: stat sys}. The corner plots in the supplementary material instead show the nominal posterior intervals, and posteriors that reach a prior bound are reported as limits. The calibration increases the nominal posterior widths by median factors of 2--3, and the total uncertainties are typically three to seven times larger than the nominal intervals. For the golden sample, the median total uncertainties are $-0.15/+0.10$ in $f_{\rm DM}(R_{\rm e})$, $-0.26/+0.15$ in $\alpha$, $-7/+3$~km\,s$^{-1}$ in $\sigma_0$, and $-11/+15$~km\,s$^{-1}$ in $V_{\rm rot}(2.2R_{\rm s})$. The systematic term is dominated by the LSF for $\sigma_0$, by the halo concentration for $\alpha$, and by the kinematic centre, disk size and inclination for the rotation velocities. The sensitivity to the baryonic mass scale is not included in these uncertainties and is discussed separately where relevant.}

\subsection{Robustness}
\label{subsection: Robustness}

\rev{The modelling choices that can affect the inferred quantities are propagated into the systematic uncertainties (Section~\ref{subsection: Fit quality and uncertainties}; Appendices~\ref{section: concentration impact} and~\ref{section: B/T impact}). In addition, we tested the choice of halo profile and the fixed kinematic centre. The generalised NFW (gNFW) model is preferred over the matched NFW model at $p<0.05$ for 15 of the 30 galaxies, based on a likelihood-ratio test in which the statistic $2\Delta\ln\mathcal{L}$ of each galaxy is divided by the square of its bootstrap calibration factor for $\alpha$ (Appendix~\ref{section: error budget}), to approximately account for the residual correlations not captured by the likelihood, and compared with a $\chi^2$ distribution with one degree of freedom. Because this is an approximate rescaling rather than an empirically calibrated null distribution, we regard the resulting $p$-values as indicative. The Bayes factors favour the gNFW model ($\ln B>1$) for 20 galaxies and the NFW model ($\ln B<-1$) for four; since they are computed from the same likelihood without this correction, their magnitudes are likely overestimated, and we use them only as qualitative support. The rate of spurious detections of $\alpha<1$ is instead quantified empirically with injection--recovery tests (Appendix~\ref{section: injection recovery}). Refitting with a free kinematic centre improves the fits substantially (median $\Delta\chi^2_{\rm b}\approx-100$ for the golden sample, from the maximum-likelihood solutions of the full fits), as expected if the additional freedom absorbs asymmetries and non-circular motions (Section~\ref{subsection - Characterisation of geometrical parameters}), but changes the dark matter fractions by a median of only 0.01 (Appendix~\ref{section: free centre}).}

\begin{table*}[ht]
\centering
\caption{Parameters and priors in the \texttt{DysmalPy} dynamical forward modelling.}
\begin{tabular}{c p{3.8cm} p{1.9cm} p{3.3cm} p{4.2cm} p{1.5cm}}
\toprule
Parameter & Description & Fixed / Free & Initial value & Prior ($\sigma=$ standard deviation) & Bounds \\
\midrule

$M_{\mathrm{bar}}$
  & \parbox[t]{4.0cm}{Total baryonic mass\\ \small(log$_{10} M_\odot$)}
  & Free
  & $M_\star + M_{\rm mol} + M_{\rm HI}^{(c)}$
  & Gaussian, $\sigma = 0.2$
  & $\pm 0.5$ \\[6pt]

$R_{\mathrm{e,disk}}$
  & \parbox[t]{4.0cm}{Disk effective radius\\ \small(kpc)}
  & Fixed
  & Reference image$^{(b)}$
  & --
  & -- \\[6pt]

$R_{\mathrm{e,bulge}}$
  & \parbox[t]{4.0cm}{Bulge effective radius\\ \small(kpc)}
  & Fixed
  & 1.0
  & --
  & -- \\[6pt]

$n_{\mathrm{disk}},\,n_{\mathrm{bulge}}$
  & \parbox[t]{4.0cm}{S\'ersic indices}
  & Fixed
  & 1, 4
  & --
  & -- \\[6pt]

$B/T$
  & \parbox[t]{4.0cm}{Bulge-to-total ratio}
  & Fixed/Free$^{(d)}$
  & Stellar-mass maps or reference image$^{(d)}$
  & Gaussian, $\sigma = 0.1$
  & $[0.02,\ B/T+0.2]$ \\[6pt]

$M_{\mathrm{vir}}$
  & \parbox[t]{4.0cm}{Virial mass\\ \small(log$_{10} M_\odot$)}
  & Free
  & Abundance matching$^{(e)}$
  & Gaussian, $\sigma = 0.5$
  & $9$--$17$ \\[6pt]

$\alpha$
  & \parbox[t]{4.0cm}{Inner slope of the dark-matter density profile}
  & Free
  & 1 (NFW)
  & Uniform
  & $0$--$1.5$ \\[6pt]

$\beta$
  & \parbox[t]{4.0cm}{Outer slope of the dark-matter density profile}
  & Fixed
  & 3
  & --
  & -- \\[6pt]

$c$
  & \parbox[t]{4.0cm}{Halo concentration}
  & Fixed
  & $10.9\,(1+z)^{-0.83}$$^{(f)}$
  & --
  & -- \\[6pt]

$f_{\mathrm{DM}}(R_{\mathrm{e,disk}})$
  & \parbox[t]{4.0cm}{Dark-matter fraction within $R_{\mathrm{e,disk}}$}
  & Derived$^{(g)}$
  & --
  & Induced by the $M_{\rm vir}$ and $M_{\rm bar}$ priors
  & -- \\[6pt]

$\sigma_0$
  & \parbox[t]{4.0cm}{Intrinsic velocity dispersion\\ \small(km\,s$^{-1}$)}
  & Free
  & $\langle\sigma_{\rm corr}\rangle$
  & Uniform
  & $\pm 30$$^{(h)}$ \\[6pt]

$h_z$
  & \parbox[t]{4.0cm}{Vertical scale height}
  & Tied to $R_{\mathrm{e,disk}}$$^{(i)}$
  & --
  & --
  & -- \\[6pt]

$i$
  & \parbox[t]{4.0cm}{Inclination\\ \small(degrees)}
  & Fixed/Free$^{(a)}$
  & Outer isophotes (reference image)
  & Uniform
  & $\pm 15$$^{(a)}$ \\[6pt]

$\theta_{\rm PA}$
  & \parbox[t]{4.0cm}{Kinematic position angle\\ \small(degrees)}
  & Free
  & \texttt{PAFIT}
  & Uniform
  & $\pm 30$ \\[6pt]

$x_{0},\,y_{0}$
  & \parbox[t]{4.0cm}{Dynamical centre\\ \small(pixels)}
  & Fixed$^{(j)}$
  & Measured centre$^{(j)}$
  & --
  & -- \\[6pt]

$V_{\mathrm{sys}}$
  & \parbox[t]{4.0cm}{Systemic velocity\\ \small(km\,s$^{-1}$)}
  & Free
  & Integrated line centroid
  & Uniform
  & $\pm 35$ \\[6pt]

\bottomrule
\end{tabular}

\tablefoot{
  \tablefoottext{a}{Fixed to the imaging value for 20 galaxies; free with a flat prior of $\pm15^\circ$ around it (truncated to $5^\circ$--$89^\circ$) for the ten galaxies with poorly constrained photometric inclinations (\texttt{4391}, \texttt{6199}, \texttt{8512}, \texttt{8942}, \texttt{9960}, \texttt{11225}, \texttt{13182}, \texttt{13416}, \texttt{18188}, \texttt{29470}; Section~\ref{subsection: Priors and modelling assumptions}).}\\ 
  \tablefoottext{b}{From the double-component (disk + bulge) fit to the reference image (Section~\ref{subsection - Characterisation of geometrical parameters}).}\\
  \tablefoottext{c}{$M_\star$ from the 3D-HST/CANDELS catalogues; $M_{\rm mol}$ from the gas-fraction scaling relation of \citet{Tacconi_2018}; $M_{\rm HI}$ is the atomic gas expected within $2R_{\rm e}$ \citep{Chowdhury_2022, Wang_2016}.}\\
  \tablefoottext{d}{Free for the 11 galaxies whose kinematics require a bulge, fixed otherwise; measured from the stellar mass maps for the ten galaxies with reliable maps and from the reference image otherwise. The lower bound is 0.02.}\\
  \tablefoottext{e}{Prior centred on the stellar-to-halo mass relation of \citet{Moster_2013} at the catalogue $M_\star$.}\\
  \tablefoottext{f}{Redshift-dependent relation adopted by \citet{Genzel_2020} from \citet{Dutton_Maccio_2014}.}\\
  \tablefoottext{g}{Derived for each posterior sample, $f_{\rm DM}=V_{\rm DM}^2/V_{\rm c}^2$ at $R_{\rm e,disk}$; its prior is induced by those on $M_{\rm vir}$ and $M_{\rm bar}$.}\\
  \tablefoottext{h}{Centred on the LSF-corrected outer dispersion; lower bound 5~km\,s$^{-1}$.}\\
  \tablefoottext{i}{Gaussian vertical profile tied to the disk radius through the intrinsic axis ratio $q=0.25$.}\\
  \tablefoottext{j}{Fixed at the measured kinematic centre (Section~\ref{subsection - Characterisation of geometrical parameters}); its uncertainty is included in the systematic error budget.}
}
\label{tab: dysmalpy parameters}
\end{table*}

\section{Disk classification and rotational support}
\label{section: Disk classification and rotational support}

A visual inspection of the \rev{reference images} (Section~\ref{subsection: MSA-3D data}) and of the colour composites provided in the supplementary material, indicates that the majority of galaxies from the MSA-3D sample used in this work exhibit disk-like morphologies, characterised by flattened light distributions and, in several cases, visible spiral arms. Overall, 24/30 systems are morphologically consistent with disk galaxies. For the remaining objects, the disk morphology is less obvious in the imaging, primarily because of face-on orientations, compact sizes, or limited angular extent. For these systems, the resolved ionised-gas kinematics provide the stronger diagnostic, showing ordered velocity gradients consistent with rotating disks even when the stellar morphology is less visually distinctive.

The six galaxies with less certain visual classifications illustrate this point \rev{(see their diagnostic figures in the supplementary material)}. Galaxy~\texttt{4391} appears close to face-on and somewhat elliptical in projection, although its kinematic maps show well-ordered rotation. Galaxy~\texttt{7314} is compact, and the spatial coverage of the kinematic data is limited, but the observed velocity field is nonetheless consistent with a rotating disk. Similarly, galaxy~\texttt{8576} is viewed nearly face-on, making the disk morphology less evident in the imaging despite displaying regular rotational kinematics. Galaxy~\texttt{10107} shows signs of possible disturbance, potentially related to a companion to the south-west, but still exhibits broadly ordered rotation. \rev{Finally, galaxy 12239 is covered only by F115W imaging, which at its redshift traces rest-frame $\approx0.6\,\mu$m light that is more affected by clumpy star formation and dust than the redder bands used for the rest of the sample; in this image it shows no clear disk structure, and its moderate axis ratio ($b/a\simeq0.77$, $i=41^\circ$) does not by itself establish a disk geometry}. Finally, 13182 is close to face-on, which makes its morphological classification more ambiguous. The velocity fields of both galaxies nevertheless remain compatible with rotating systems.

\begin{figure}
    \centering
    \includegraphics[width=0.98\linewidth]{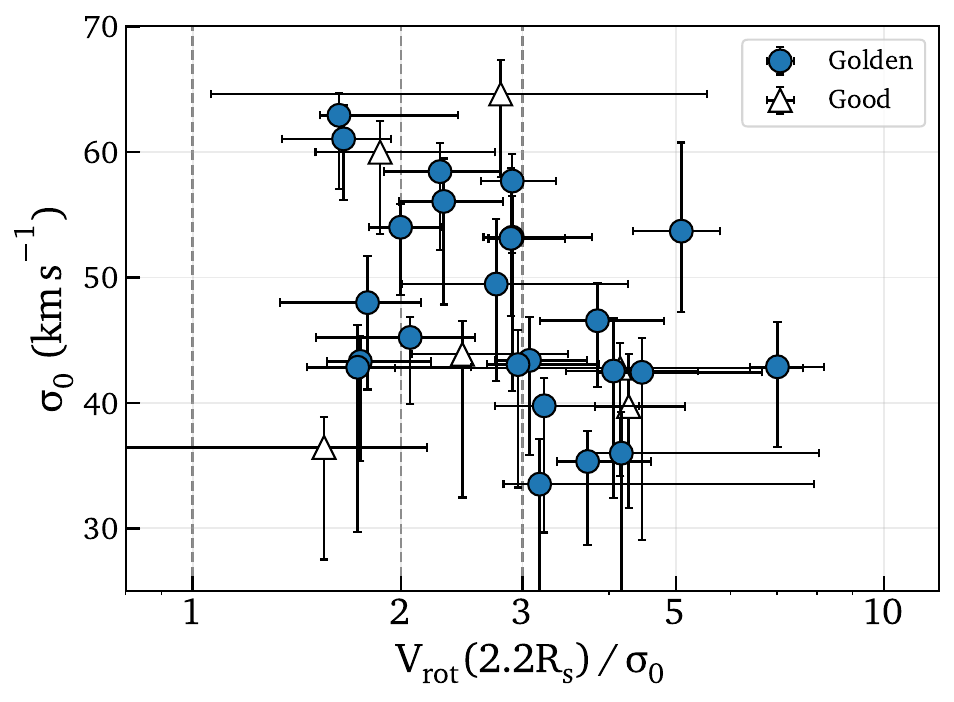}
    \caption{Intrinsic velocity dispersion $\sigma_0$ as a function of rotational support, $V_{\rm rot}/\sigma_0$, with $V_{\rm rot}$ evaluated at $2.2R_{\rm s}$. Filled circles show the golden sample and open triangles the good sample; this convention is used throughout the paper. Error bars are total (statistical and systematic) uncertainties; the large lower uncertainties of $\sigma_0$ are dominated by the line-spread-function systematic. Dashed lines mark $V_{\rm rot}/\sigma_0=1$, 2 and 3.}
    \label{fig: sigma vs vsig}
\end{figure}

Beyond the qualitative assessment of the imaging and kinematic maps, we quantify the degree of rotational support using the ratio of the intrinsic rotation velocity to the intrinsic velocity dispersion, $V_{\rm rot}/\sigma_0$, evaluated \rev{for each posterior sample of the dynamical model at $2.2R_{\rm s}$ ($=1.31R_{\rm e}$), the radius at which the rotation curve of an exponential disk peaks (cf. $1.38R_{\rm e}$ in \citealp{Uebler_2019})}. This quantity provides a useful diagnostic of the dynamical state of the systems and allows us to distinguish rotation-dominated disks from more dispersion-supported systems. We note that this ratio refers specifically to the ionised-gas component and may differ from measurements based on molecular gas, which traces a dynamically colder phase of the interstellar medium \rev{(e.g. \citealp{Ubler_2018, Girard_2021, Wisnioski_2025, Barfety_2026})}. The intrinsic velocity dispersions obtained from the kinematic modelling span $\sigma_0 \sim 34$--65~km\,s$^{-1}$ (see Table~\ref{tab: Fitted parameters}), consistent with expectations for turbulent star-forming disks at $z\sim1$ (e.g., \citealp{forster_schreiber_sins_2009, Johnson_2018, Tiley_2021}). \rev{All galaxies have $V_{\rm rot}/\sigma_0>1$ (median 2.9; Fig.~\ref{fig: sigma vs vsig}): of the 24 golden galaxies, ten have $V_{\rm rot}/\sigma_0>3$, eight have $2<V_{\rm rot}/\sigma_0\le3$ and six have $1<V_{\rm rot}/\sigma_0\le2$.} These values are commonly observed in turbulent star-forming disks at cosmic noon and do not, by themselves, imply dispersion-dominated dynamics (e.g., \citealp{Forster_review_2020, Tiley_2021}).

\rev{The rotation velocity increases with stellar mass (Spearman\footnote{Throughout, we assess correlations with the Spearman rank coefficient $\rho$ (and its partial form when controlling for a third quantity), differences between groups with the Mann--Whitney (two groups) and Kruskal--Wallis (three or more groups) tests, and offsets of a distribution from zero with the Wilcoxon signed-rank test.} $\rho=0.57$, $p=0.003$), as expected from the Tully--Fisher relation. In contrast, $\sigma_0$ shows no significant dependence on stellar mass ($\rho=0.25$, $p=0.24$) or redshift, but nominal associations with SFR ($\rho=0.42$, $p=0.04$) and with the star-formation-rate surface density, $\Sigma_{\rm SFR}=\mathrm{SFR}/(2\pi R_{\rm e}^2)$ ($\rho=0.47$, $p=0.02$); neither remains significant after a Holm correction for the four relations tested ($p\approx0.13$ and 0.08). The association with SFR is of similar strength at fixed stellar mass (partial $\rho=0.35$, $p=0.09$), whereas the dependence on stellar mass vanishes at fixed SFR, suggesting that $\sigma_0$ may be linked more directly to star-formation activity than to the underlying SFR--$M_\star$ relation. This behaviour is qualitatively consistent with a connection between gas turbulence and star formation, although it does not distinguish its physical origin. Both stellar feedback and gravitational instabilities sustained by gas accretion can produce higher dispersions in gas-rich, actively star-forming disks (e.g., \citealp{Krumholz_2018, Uebler_2019}), and the limited dynamic range of our sample ($\approx1.4$~dex in SFR) prevents us from discriminating between these scenarios. In absolute terms, the median $\sigma_0\approx46$~km\,s$^{-1}$ at $z\approx1.1$ is $\approx12$~km\,s$^{-1}$ above the relation of \citet{Uebler_2019} at the same redshift, but this difference is comparable to their intrinsic scatter and may partly reflect the uncertainty in the LSF, which dominates our systematic error budget on $\sigma_0$ (see below). Together with the morphological evidence, these kinematic properties support treating the MSA-3D sample as a population of rotating disk galaxies.}

\rev{Because $\sigma_{\rm LSF}\simeq30$--51~km\,s$^{-1}$ is comparable to the measured intrinsic dispersions, we tested the sensitivity of $\sigma_0$ to instrumental broadening. The inferred dispersions show no correlation with $\sigma_{\rm LSF}$, and both LSF-variation and injection--recovery tests demonstrate that $\sigma_0$ can be constrained even below the instrumental width (Appendix~\ref{section: sigma0 LSF}). Uncertainty in the LSF nevertheless remains the dominant systematic: adopting an LSF broadened by 20\%, appropriate to the maximum expected effect of spatially extended emission, lowers $\sigma_0$ by a median of $\approx7$~km\,s$^{-1}$ and by up to 20~km\,s$^{-1}$ for galaxies with $\sigma_0\lesssim\sigma_{\rm LSF}$. This broadening is mildly disfavoured by the fits (median $\Delta\chi^2_{\rm b}=+4.0$), so we treat it as a conservative systematic uncertainty.}

\begin{figure}
    \centering
    \includegraphics[width=0.98\linewidth]{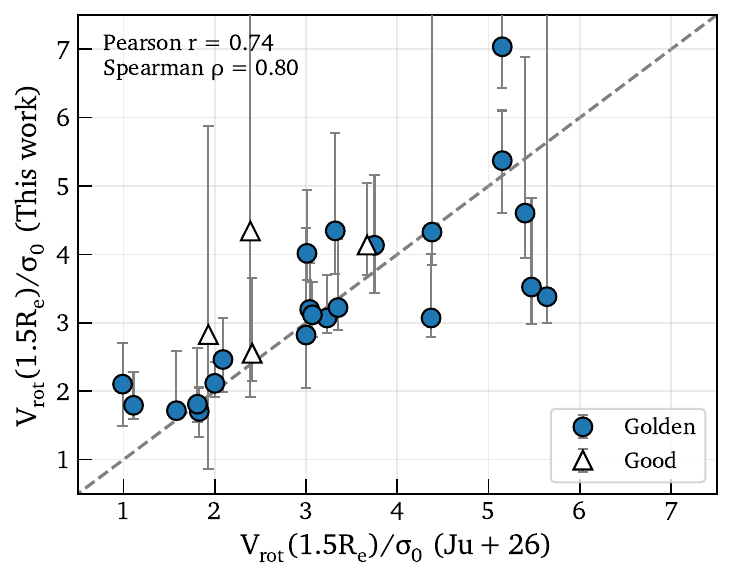}
    \caption{Comparison of the rotational support $V_{\rm rot}/\sigma_0$ at $1.5R_{\rm e}$ measured in this work and by \citet{Ju_2026} for the 26 galaxies in common. Circles and triangles indicate the golden and good samples, respectively; error bars are total uncertainties. The dashed grey line shows the 1:1 relation. The two measurements are strongly correlated (Spearman $\rho=0.80$), with a median ratio of 1.05 and a 16th--84th percentile range of 0.93--1.37.}
    \label{fig: sigma vs vsig comparison}
\end{figure}

Our measurements are in broad agreement with those of \citet{Ju_2026}, who analysed the same sample using an arctangent \rev{rotation-curve model and measured $V_{\rm rot}/\sigma_0$ at $1.5R_{\rm e}$. Evaluating our models at the same radius, the median ratio for the 26 galaxies in common is $(V_{\rm rot}/\sigma_0)_{\rm this\,work}/(V_{\rm rot}/\sigma_0)_{\rm Ju+26}=1.05$, with a 16th--84th percentile range of 0.93--1.37, and the two measurements are strongly correlated (Spearman $\rho=0.80$, $p=1\times10^{-6}$; Fig.~\ref{fig: sigma vs vsig comparison}), indicating that both methods trace the same kinematic structure despite their different modelling assumptions.}

\begin{table*}[ht]
\centering
\setlength{\tabcolsep}{3.6pt}
\caption{Fitted and derived parameters of the MSA-3D galaxies: spectroscopic redshift $z$; inclination $i$ (imaging value,
or posterior for the ten galaxies with free inclination); kinematic position angle $\theta_{\rm PA}$; disk effective
radius $R_{\rm e,disk}$ (fixed to the bulge--disk decomposition); intrinsic velocity dispersion $\sigma_0$; rotation velocity $V_{\rm rot}(2.2R_{\rm s})$ and rotational support $V_{\rm rot}/\sigma_0$, \textbf{both at $2.2R_{\rm s}=1.31R_{\rm e}$}; dark matter fraction
$f_{\rm DM}(R_{\rm e})$; inner slope $\alpha$ of the dark matter halo; bulge-to-total ratio $B/T$; and outer
rotation-curve class (Section~\ref{section: Rotation curve shapes}). Values are posterior medians with total
(statistical and systematic) uncertainties (Section~\ref{subsection: Fit quality and uncertainties}); values without
uncertainties are limits. Statistical and systematic uncertainties are listed separately in Table~\ref{tab: stat sys}.}
\begin{tabular}{cccccccccccc}
\toprule
Galaxy ID & $z$ & $i$ & $\theta_{\rm PA}$ & $R_{\rm e,disk}$ & $\sigma_0$ & $V_{\rm rot}(2.2R_{\rm s})$ & $V_{\rm rot}/\sigma_0$ & $f_{\rm DM}(R_{\rm e})$ & $\alpha$ & $B/T$ & RC shape \\
 &  & deg & deg & kpc & km\,s$^{-1}$ & km\,s$^{-1}$ &  &  &  &  &  \\
\midrule
\multicolumn{12}{c}{\textbf{Golden sample}}\\
\hline
\texttt{2111} & 0.575 & $57.2^{a}$ & $41.4^{+1.7}_{-1.7}$ & 3.97 & $53.2^{+3.3}_{-12.3}$ & $154^{+12}_{-11}$ & $2.90^{+0.89}_{-0.27}$ & $0.71^{+0.08}_{-0.13}$ & $0.74^{+0.11}_{-0.43}$ & $0.26^{+0.04}_{-0.06}$ & ---$^{c}$ \\
\texttt{2145} & 1.173 & $51.7^{a}$ & $299.1^{+4.8}_{-4.7}$ & 3.16 & $43.3^{+2.0}_{-7.9}$ & $76^{+12}_{-8}$ & $1.75^{+0.46}_{-0.18}$ & $0.47^{+0.27}_{-0.18}$ & $0.59^{+0.22}_{-0.42}$ & $<0.02$ & Flat \\
\texttt{2465} & 1.245 & $70.1^{a}$ & $168.9^{+2.3}_{-2.3}$ & 3.31 & $43.4^{+3.5}_{-7.5}$ & $133^{+7}_{-8}$ & $3.07^{+0.65}_{-0.34}$ & $0.76^{+0.17}_{-0.25}$ & $0.99^{+0.14}_{-0.21}$ & $0.02^{a}$ & Flat \\
\texttt{2824} & 0.980 & $53.9^{a}$ & $241.0^{+4.1}_{-4.0}$ & 2.47 & $42.5^{+4.2}_{-10.1}$ & $173^{+17}_{-17}$ & $4.06^{+1.31}_{-0.59}$ & $0.92^{+0.03}_{-0.06}$ & $0.89^{+0.15}_{-0.23}$ & $0.21^{a}$ & ---$^{c}$ \\
\texttt{3399} & 1.337 & $64.2^{a}$ & $37.5^{+3.1}_{-3.0}$ & 2.65 & $54.0^{+1.9}_{-5.4}$ & $108^{+11}_{-10}$ & $2.00^{+0.29}_{-0.20}$ & $0.76^{+0.05}_{-0.08}$ & $0.82^{+0.15}_{-0.35}$ & $0.32^{a}$ & ---$^{c}$ \\
\texttt{4391} & 1.079 & $36.0^{+16.4}_{-21.1}$ & $343.5^{+3.0}_{-2.9}$ & 1.43 & $49.5^{+5.2}_{-7.7}$ & $136^{+69}_{-34}$ & $2.75^{+1.52}_{-0.74}$ & $0.50^{+0.46}_{-0.21}$ & $1.01^{+0.49}_{-0.34}$$^{b}$ & $<0.05$ & Flat \\
\texttt{6199} & 1.592 & $<18.1$ & $115.1^{+10.6}_{-10.8}$ & 5.86 & $45.2^{+1.6}_{-5.3}$ & $93^{+20}_{-25}$ & $2.06^{+0.50}_{-0.56}$ & $0.66^{+0.09}_{-0.09}$ & $0.68^{+0.62}_{-0.68}$$^{b}$ & $0.18^{a}$ & ---$^{c}$ \\
\texttt{6430} & 1.172 & $64.3^{a}$ & $314.6^{+6.4}_{-6.1}$ & 4.38 & $58.4^{+2.3}_{-6.3}$ & $133^{+24}_{-24}$ & $2.28^{+0.51}_{-0.39}$ & $0.83^{+0.17}_{-0.40}$$^{b}$ & $0.64^{+0.23}_{-0.39}$ & $0.04^{a}$ & Rising \\
\texttt{7314} & 1.283 & $38.7^{a}$ & $188.6^{+3.4}_{-3.3}$ & 1.60 & $61.0^{+2.7}_{-4.9}$ & $101^{+16}_{-17}$ & $1.65^{+0.29}_{-0.31}$ & $0.62^{+0.38}_{-0.33}$ & $0.92^{+0.16}_{-0.25}$ & $<0.09$ & Flat \\
\texttt{7561} & 1.029 & $67.2^{a}$ & $105.0^{+7.8}_{-8.0}$ & 2.84 & $48.0^{+3.7}_{-6.9}$ & $86^{+12}_{-19}$ & $1.79^{+0.35}_{-0.45}$ & $0.13^{+0.87}_{-0.13}$$^{b}$ & $0.90^{+0.60}_{-0.90}$$^{b}$ & $0.07^{a}$ & Falling \\
\texttt{8365} & 1.685 & $51.8^{a}$ & $291.7^{+1.8}_{-1.8}$ & 2.09 & $35.3^{+2.4}_{-6.6}$ & $132^{+10}_{-10}$ & $3.73^{+0.88}_{-0.36}$ & $0.79^{+0.06}_{-0.07}$ & $0.68^{+0.12}_{-0.13}$ & $0.02^{a}$ & Rising \\
\texttt{8576} & 1.567 & $32.3^{a}$ & $12.5^{+1.9}_{-1.9}$ & 2.38 & $46.5^{+3.0}_{-5.3}$ & $179^{+38}_{-27}$ & $3.85^{+0.96}_{-0.67}$ & $0.87^{+0.06}_{-0.08}$ & $0.82^{+0.12}_{-0.23}$ & $0.02^{a}$ & Rising \\
\texttt{8942} & 1.176 & $>48.6$ & $332.4^{+2.3}_{-2.0}$ & 1.70 & $53.1^{+5.6}_{-6.2}$ & $153^{+26}_{-6}$ & $2.89^{+0.57}_{-0.21}$ & $0.69^{+0.08}_{-0.06}$ & $0.77^{+0.19}_{-0.24}$ & $0.02^{a}$ & Flat \\
\texttt{9424} & 0.976 & $61.2^{a}$ & $256.1^{+3.1}_{-2.3}$ & 3.86 & $43.1^{+2.8}_{-9.8}$ & $127^{+12}_{-11}$ & $2.95^{+0.91}_{-0.29}$ & $0.71^{+0.09}_{-0.20}$ & $0.42^{+0.07}_{-0.33}$ & $>0.23$ & Flat \\
\texttt{9636} & 0.737 & $68.5^{a}$ & $232.1^{+2.5}_{-2.6}$ & 3.28 & $33.5^{+3.6}_{-20.3}$ & $106^{+7}_{-6}$ & $3.17^{+4.75}_{-0.36}$ & $0.75^{+0.13}_{-0.15}$ & $0.69^{+0.12}_{-0.32}$ & $0.02^{a}$ & Falling \\
\texttt{9812} & 0.738 & $34.4^{a}$ & $239.6^{+1.4}_{-1.4}$ & 4.51 & $42.4^{+2.8}_{-13.4}$ & $189^{+34}_{-23}$ & $4.46^{+2.20}_{-0.61}$ & $0.44^{+0.28}_{-0.25}$ & $0.79^{+0.26}_{-0.39}$ & $0.02^{a}$ & ---$^{c}$ \\
\texttt{9960} & 1.514 & $>59.6$ & $147.6^{+0.9}_{-0.9}$ & 2.69 & $42.9^{+3.6}_{-6.4}$ & $300^{+9}_{-9}$ & $7.01^{+1.18}_{-0.61}$ & $0.33^{+0.19}_{-0.16}$ & $1.28^{+0.22}_{-0.27}$$^{b}$ & $<0.30$ & Flat \\
\texttt{10910} & 0.742 & $46.0^{a}$ & $308.4^{+1.8}_{-1.8}$ & 4.98 & $36.0^{+3.3}_{-17.4}$ & $150^{+13}_{-11}$ & $4.17^{+3.88}_{-0.46}$ & $0.87^{+0.10}_{-0.15}$ & $1.06^{+0.15}_{-0.23}$ & $0.02^{a}$ & ---$^{c}$ \\
\texttt{11225} & 1.052 & $>63.1$ & $278.4^{+3.3}_{-3.3}$ & 3.39 & $62.9^{+1.8}_{-5.9}$ & $102^{+49}_{-6}$ & $1.63^{+0.79}_{-0.10}$ & $0.79^{+0.09}_{-0.15}$ & $0.85^{+0.12}_{-0.17}$ & $0.02^{a}$ & Falling \\
\texttt{11843} & 1.465 & $74.3^{a}$ & $161.8^{+3.2}_{-3.2}$ & 3.17 & $53.7^{+7.1}_{-6.5}$ & $273^{+15}_{-17}$ & $5.09^{+0.70}_{-0.75}$ & $0.69^{+0.07}_{-0.14}$ & $0.87^{+0.13}_{-0.20}$ & $0.02^{a}$ & ---$^{c}$ \\
\texttt{12015} & 1.235 & $46.4^{a}$ & $88.5^{+4.6}_{-5.1}$ & 2.39 & $57.7^{+2.2}_{-5.7}$ & $167^{+18}_{-16}$ & $2.90^{+0.45}_{-0.29}$ & $0.75^{+0.10}_{-0.14}$ & $0.71^{+0.23}_{-0.20}$ & $0.02^{a}$ & Rising \\
\texttt{12239} & 0.894 & $40.6^{a}$ & $210.5^{+7.3}_{-7.6}$ & 2.10 & $56.1^{+3.4}_{-8.2}$ & $129^{+18}_{-16}$ & $2.31^{+0.51}_{-0.32}$ & $0.85^{+0.09}_{-0.13}$ & $0.84^{+0.15}_{-0.23}$ & $0.02^{a}$ & ---$^{c}$ \\
\texttt{18586} & 0.756 & $35.9^{a}$ & $264.5^{+5.2}_{-5.5}$ & 1.49 & $42.8^{+3.3}_{-13.1}$ & $74^{+12}_{-9}$ & $1.73^{+0.79}_{-0.27}$ & $0.67^{+0.19}_{-0.23}$ & $0.87^{+0.19}_{-0.26}$ & $0.02^{a}$ & Flat \\
\texttt{29470} & 1.042 & $>47.6$ & $276.9^{+1.8}_{-1.7}$ & 3.02 & $39.8^{+2.2}_{-10.1}$ & $128^{+15}_{-12}$ & $3.22^{+1.19}_{-0.48}$ & $0.83^{+0.05}_{-0.05}$ & $0.56^{+0.12}_{-0.30}$ & $<0.11$ & Rising \\
\hline
\multicolumn{12}{c}{\textbf{Good sample}}\\
\hline
\texttt{8512} & 1.104 & $19.2^{+21.5}_{-19.3}$ & $138.3^{+3.1}_{-3.1}$ & 5.45 & $42.8^{+2.0}_{-8.6}$ & $177^{+163}_{-106}$ & $4.15^{+3.48}_{-2.19}$ & $0.71^{+0.29}_{-0.33}$$^{b}$ & $0.83^{+0.61}_{-0.68}$ & $0.22^{a}$ & ---$^{c}$ \\
\texttt{9337} & 1.172 & $51.6^{a}$ & $269.8^{+14.9}_{-16.9}$ & 5.56 & $36.5^{+2.4}_{-8.9}$ & $57^{+27}_{-48}$ & $1.55^{+0.63}_{-1.21}$ & $0.69^{+0.19}_{-0.23}$ & $<0.16$ & $<0.11$ & ---$^{c}$ \\
\texttt{10107} & 1.010 & $25.0^{a}$ & $35.4^{+4.1}_{-4.1}$ & 3.24 & $60.0^{+2.5}_{-6.5}$ & $112^{+51}_{-22}$ & $1.87^{+0.87}_{-0.36}$ & $0.35^{+0.42}_{-0.35}$$^{b}$ & $>0.28$ & $0.10^{+0.09}_{-0.09}$ & Rising \\
\texttt{13182} & 1.540 & $20.1^{+31.8}_{-22.7}$ & $209.8^{+8.7}_{-9.1}$ & 4.17 & $64.6^{+2.7}_{-6.6}$ & $179^{+206}_{-128}$ & $2.79^{+2.76}_{-1.73}$ & $0.60^{+0.40}_{-0.60}$$^{b}$ & $1.03^{+0.47}_{-0.98}$$^{b}$ & $0.21^{+0.16}_{-0.10}$ & ---$^{c}$ \\
\texttt{13416} & 1.540 & $81.1^{+1.6}_{-1.6}$ & $12.5^{+2.9}_{-2.7}$ & 2.30 & $39.7^{+4.2}_{-8.1}$ & $169^{+15}_{-13}$ & $4.27^{+0.87}_{-0.46}$ & $0.03^{+0.09}_{-0.03}$$^{b}$ & $<0.18$ & $0.60^{a}$ & ---$^{c}$ \\
\texttt{18188} & 0.822 & $28.7^{+15.3}_{-24.1}$ & $245.8^{+2.0}_{-2.0}$ & 6.24 & $43.9^{+2.6}_{-11.4}$ & $108^{+29}_{-15}$ & $2.46^{+1.03}_{-0.38}$ & $0.69^{+0.31}_{-0.25}$ & $0.60^{+0.47}_{-0.60}$$^{b}$ & $<0.07$ & ---$^{c}$ \\
\bottomrule
\end{tabular}

\tablefoot{
\tablefoottext{a}{Fixed in the modelling (inclination from the imaging; $B/T$ from the stellar mass maps or the reference image).}\\
\tablefoottext{b}{Uncertainties truncated at the physical range of $f_{\rm DM}$ (0--1) or the prior range of $\alpha$ (0--1.5).}\\
\tablefoottext{c}{Rotation curve not classified (insufficient radial coverage or too few pixels in the outer window).}
}
\label{tab: Fitted parameters}
\end{table*}

\section{Stellar Tully--Fisher relation}
\label{section: Tully--Fisher}

The stellar Tully--Fisher relation (sTFR) links the circular velocity of disk galaxies to their stellar mass and provides a fundamental empirical connection between galaxy dynamics and mass assembly \citep{Tully_fisher}. \rev{In the local Universe, the sTFR is remarkably tight for rotation-dominated disk galaxies (e.g., \citealp{Reyes_2011})}, whereas disturbed morphologies and non-equilibrium kinematics are associated with substantially increased scatter (e.g., \citealp{Flores_2006, Weiner2006a, Weiner2006b, Kassin_2007, Simons2019, Abril-Melgarejo_2021}). Its redshift evolution remains debated: some studies report mild or negligible evolution out to $z\sim2$ (e.g., \citealp{Kassin_2007, Harrison_2017, Tiley}), while others find significant offsets in normalisation or slope, consistent with evolving mass-to-light ratios or internal dynamical structure (e.g., \citealp{Cresci_2009, Turner_2017, Ubler_2017}). Recent measurements at $z\sim1$ suggest that the stellar TFR is already in place but exhibits systematic offsets relative to the local relation, indicative of evolving disk-halo mass relations over the past $\sim8$~Gyr (e.g., \citealp{Mancera_Pina_2026}).

\begin{figure}
    \centering
    \includegraphics[width=0.99\linewidth]{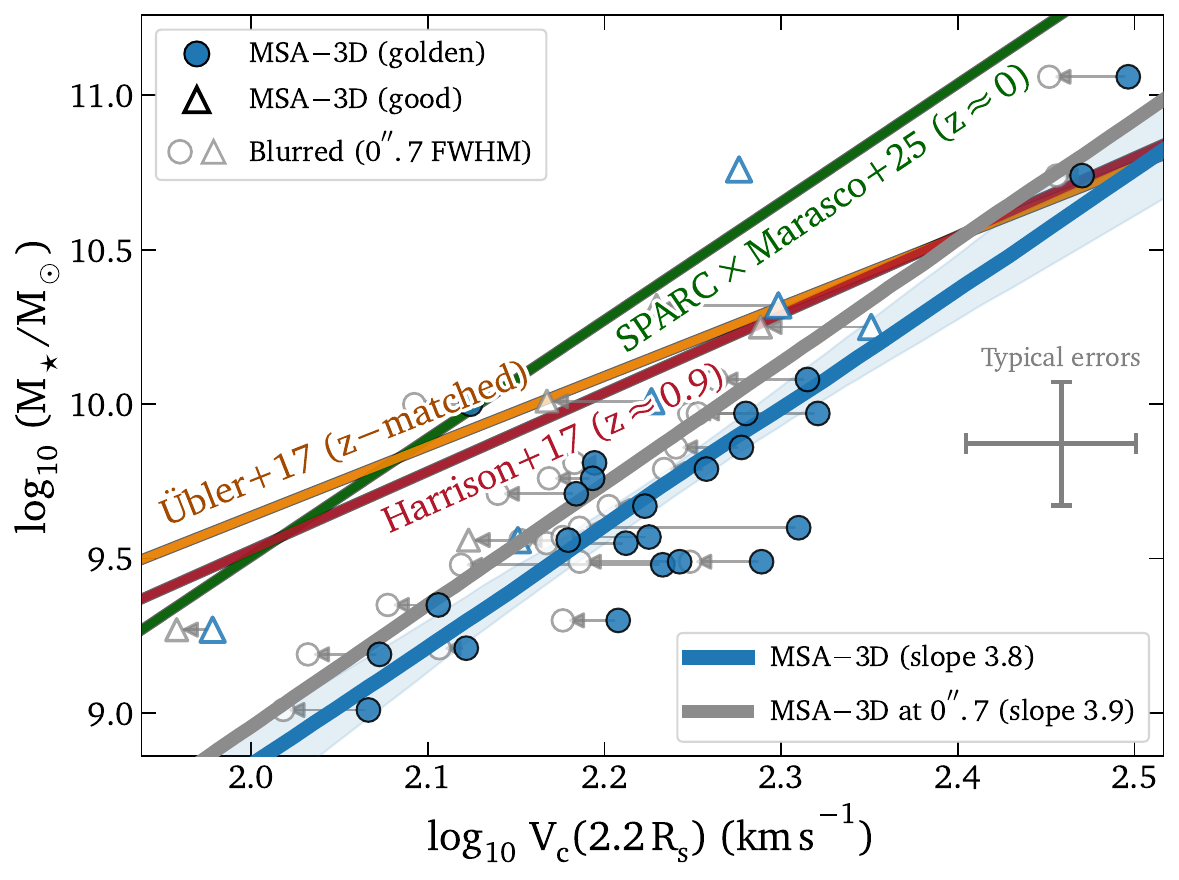}
    \caption{Stellar Tully--Fisher relation of MSA-3D with a uniform velocity definition, $V_{\rm c}^2(2.2R_{\rm s}) = V_{\rm rot}^2 + 4.4\,\sigma_0^2$, applied to all samples. Filled circles and open triangles show the golden and good samples; the inset shows a representative total uncertainty. Grey symbols show the same galaxies after degrading the best-fitting models to $0\farcs7$ seeing and re-extracting the velocity as a seeing-limited analysis would (with the dispersion measured in the degraded maps); arrows connect each galaxy to its degraded counterpart. The blue line and band show the free-slope fit to the golden sample and its uncertainty, and the grey line shows the fit to the degraded measurements. Coloured lines show the free-slope relations refitted on the same velocity definition for the $z$-matched KMOS$^{\rm 3D}$ sample of \citet{Ubler_2017}, disk-like KROSS galaxies at $z\sim0.9$ \citep{Harrison_2017} and local SPARC galaxies \citep{Lelli_2016_SPARC} with the stellar masses of \citet{Marasco_2025}. The offsets quoted in the text are derived at a common slope ($a=3.7$).}
    \label{fig: Tully--Fisher}
\end{figure}

We construct the stellar Tully--Fisher relation using the circular velocity $V_{\rm c}$ \rev{at $2.2R_{\rm s}$, where the rotation curve of a self-gravitating exponential disk peaks ($\simeq1.31R_{\rm e}$). All \textit{golden}-sample galaxies are reliably traced to at least this radius, so no extrapolation is required. Figure~\ref{fig: Tully--Fisher} compares MSA-3D with literature samples for which $V_{\rm c}(2.2R_{\rm s})$ can be reconstructed consistently: the $z$-matched KMOS$^{\rm 3D}$ disks of \citet{Ubler_2017}, based on \texttt{DYSMAL} forward modelling and the same 3D-HST photometry used here; the disk-like KROSS galaxies at $z\simeq0.9$ \citep{Harrison_2017}; and, as a local baseline, the SPARC galaxies \citep{Lelli_2016_SPARC}, with rotation velocities at 2.2 disk scale lengths \citep{Lelli_2019} and stellar masses from \citet{Marasco_2025}. Samples with velocities reported only at other radii are instead considered in Section~\ref{subsection: Comparison with a self-similar halo-scaling reference}.}

\rev{For MSA-3D, KMOS$^{\rm 3D}$, and KROSS, we adopt a common pressure-support correction, $V_{\rm c}^2(2.2R_{\rm s}) = V_{\rm rot}^2(2.2R_{\rm s}) + 4.4\,\sigma_0^2$, corresponding to the thin-exponential-disk prescription of \citet{Burkert_2010}; for \citet{Ubler_2017}, we first remove their native correction. For SPARC, whose cold-gas dispersions are $\sim10$~km\,s$^{-1}$, the correction is below 0.015~dex and is neglected. For MSA-3D, $V_{\rm rot}$ and $\sigma_0$ are evaluated for each posterior sample. This estimator differs from the native circular velocity of our models (Section~\ref{subsection: Priors and modelling assumptions}) by only 0.03~dex in the median, and adopting the effective coefficient of our models ($2.65$ instead of $4.4$) changes the offsets below by only 0.01--0.03~dex. The same radius is used for $V_{\rm rot}/\sigma_0$ in Section~\ref{section: Disk classification and rotational support}.}

\rev{We fit each sample using the same pivoted relation,}

\begin{equation}
    \log_{10}(M_\star/M_\odot) = a\,\bigl[\log_{10}(V_{\rm c}/{\rm km\,s^{-1}}) - \log_{10}150\bigr] + b,
\label{eq: sTFR}
\end{equation}

\rev{with the slope $a$, zero point $b$, and intrinsic scatter in $\log M_\star$ as free parameters, accounting for uncertainties in both $V_{\rm c}$ and $M_\star$ (0.2~dex). The pivot at 150~km\,s$^{-1}$, close to the golden-sample median (168~km\,s$^{-1}$), minimises the covariance between slope and zero point (e.g., \citealp{Tremaine_2002, Ubler_2017}) and allows the zero points of the samples to be compared directly. For the golden sample, we obtain $a=3.8\pm0.5$ and $b=9.51\pm0.06$, with an intrinsic scatter of $0.08\pm0.07$. Fixing the slope to $a=3.7$, representative of measurements at $z\sim1$ (e.g., \citealp{DiTeodoro_2016, Harrison_2017, Pelliccia_2017, Ubler_2017, Tiley, Abril-Melgarejo_2021, Sharma_2024, Mancera_Pina_2026, Jeanneau_2026}), gives an indistinguishable zero point of $b=9.52\pm0.06$.}

\rev{Because the comparison relations have different free slopes over different velocity and mass ranges, their offsets from MSA-3D depend on where and in which direction they are evaluated. Following common practice (e.g., \citealp{Ubler_2017, Tiley}), we therefore compare the zero points of all samples fitted with the same slope, $a=3.7$. The offset is then a single number, $\Delta b$, independent of the velocity at which it is evaluated, and corresponds to $\Delta\log V_{\rm c}=-\Delta b/a$ at fixed stellar mass. At this slope, MSA-3D ($b=9.52\pm0.06$) lies $0.23\pm0.07$, $0.38\pm0.07$, and $0.65\pm0.06$~dex below the $z$-matched KMOS$^{\rm 3D}$, KROSS, and SPARC relations in stellar mass at fixed $V_{\rm c}$, i.e. MSA-3D galaxies rotate faster at fixed stellar mass by $0.06\pm0.02$, $0.10\pm0.02$, and $0.17\pm0.02$~dex, respectively. These offsets change by at most 0.025~dex for common slopes between 3.3 and 4.1, and free-slope fits in either regression direction give consistent results for KROSS and SPARC. The offset from the $z$-matched KMOS$^{\rm 3D}$ sample is the least certain because that sample lies at higher stellar masses (median $\log(M_\star/M_\odot)=10.6$, compared with 9.6 for our golden sample), making its comparison with MSA-3D more sensitive to the adopted slope.}

\rev{Because spatial resolution differs substantially among these surveys, we next test whether it could contribute to the observed offsets. We degrade the best-fitting \texttt{DysmalPy} model of each MSA-3D galaxy to a representative seeing-limited resolution of $0\farcs7$ FWHM. We compute the maximum-likelihood model cube on a wide field, convolve each spectral channel with the seeing PSF, and measure the rotation velocity at $2.2R_{\rm s}$ along the major axis as a seeing-limited analysis would, using the true geometry and no beam-smearing correction. We then apply the same pressure-support correction as above, using the dispersion measured in the outer parts of the degraded maps. At this resolution the recovered rotation velocity is strongly suppressed (by a median of 0.31~dex), but beam smearing also inflates the outer dispersion (by a median of 29\%), so the pressure-support correction partly compensates: the recovered $V_{\rm c}$ is lower by a median of 0.040~dex (from $-0.12$ to $-0.01$~dex), or $\sim$9\%. Had the intrinsic $\sigma_0$ been used, the suppression would be 0.11~dex. The degraded measurements move the MSA-3D zero point by $+0.14$~dex in $\log M_\star$ at fixed $V_{\rm c}$, toward the seeing-limited relations (Fig.~\ref{fig: Tully--Fisher}), about 60\% of the offset from the $z$-matched KMOS$^{\rm 3D}$ sample and 40\% of that from KROSS.}

The magnitude of this effect depends on how the velocity dispersion entering the pressure-support correction is measured. Beam smearing reduces the apparent rotation velocity by smoothing unresolved velocity gradients, but it also broadens the observed line profiles; if this inflated dispersion is used as $\sigma_0$, the pressure-support correction partially compensates for the suppressed $V_{\rm rot}$. This test should therefore be interpreted as a controlled experiment rather than a full mock observation of seeing-limited data.

\begin{figure}
    \centering
    \includegraphics[width=0.99\linewidth]{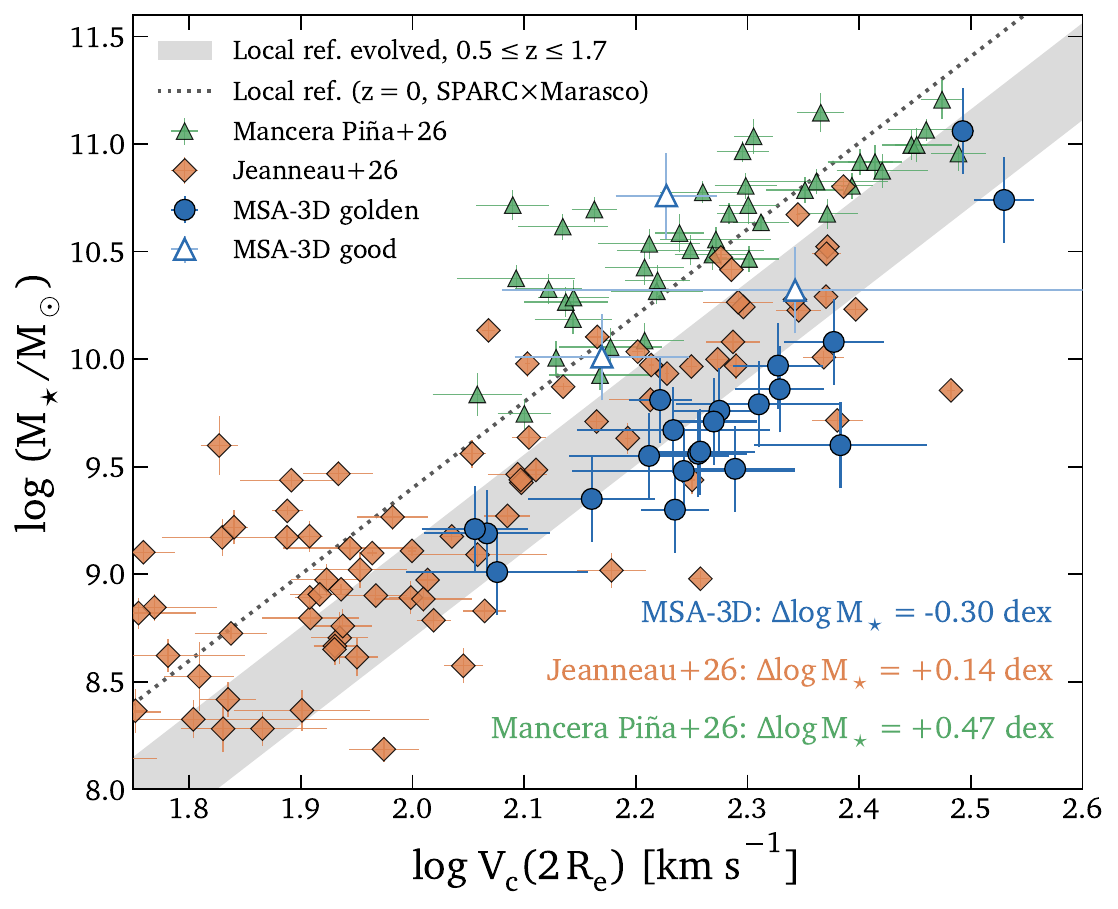}
    \caption{Stellar Tully--Fisher relation at a common radius, $V_{\rm c}(2R_{\rm e})$, compared with the local relation fitted to the SPARC galaxies at $2R_{\rm e}$ \citep{Lelli_2016_SPARC}, with the stellar masses of \citet{Marasco_2025}, evolved with the self-similar halo scaling of Eq.~\ref{eq: halo scaling} over $0.5\le z\le1.7$ (grey band; dotted line: $z=0$). Filled blue circles show the MSA-3D golden sample and open blue triangles the good sample; orange diamonds show the lensed galaxies of \citet{Jeanneau_2026} at $2R_{\rm e}$ and green triangles \citet{Mancera_Pina_2026}. The labels give the median stellar-mass offset of each sample from the evolved reference, each galaxy evaluated at its own redshift.}
    \label{fig: TFR halo scaling}
\end{figure}

\subsection{Comparison with a self-similar halo-scaling reference}
\label{subsection: Comparison with a self-similar halo-scaling reference}

\rev{Two recent studies provide particularly relevant comparisons for interpreting the sTFR in the context of self-similar halo evolution. Both \citet{Mancera_Pina_2026} and \citet{Jeanneau_2026} study galaxies at $z\sim1$ and compare their measured sTFR with the evolution expected from self-similar halo scaling, making them natural benchmarks for MSA-3D. However, unlike the samples considered above, their measurements cannot be placed uniformly at $2.2R_{\rm s}$ using the same velocity definition. \citet{Mancera_Pina_2026} derive $V_{\rm c}(2R_{\rm e})$ for 43 KROSS and KMOS$^{\rm 3D}$ disks at $0.8\lesssim z\lesssim1.0$ from three-dimensional forward modelling of seeing-limited data, while \citet{Jeanneau_2026} model 95 gravitationally lensed [O\,{\sc ii}] emitters at $0.6\lesssim z\lesssim1.4$ with \texttt{GalPaK$^{\rm 3D}$} \citep{Galpak} and report pressure-support-corrected circular velocities at $1.8R_{\rm e}$. We therefore compare the three samples separately through their offsets from a common local relation evolved according to the same self-similar halo scaling, evaluated at the redshift of each galaxy.}

\rev{As a local reference, we use the SPARC galaxies \citep{Lelli_2016_SPARC} with the stellar masses of \citet{Marasco_2025}, for which pressure support is negligible. We use the SPARC rotation velocities at $2R_{\rm e}$ and refit the relation with the same model as above, obtaining $a=4.01$ and $b=10.11$ (pivot 150~km\,s$^{-1}$; 113 galaxies).} We evolve this $z=0$ relation following the self-similar halo scaling adopted by \citet{Mancera_Pina_2026}, assuming redshift-independent ratios $f_V=V_{\rm c}/V_{\rm vir}$ and $f_M=M_\star/M_{\rm vir}$. Under these assumptions, the evolution in circular velocity is

\begin{equation}
    \Delta \log_{10} V_{\rm c}(z)
    =
    \frac{1}{6}\log_{10}
    \left[\frac{\Delta_{\rm c}(z)}{\Delta_{\rm c}(0)}\right]
    +
    \frac{1}{3}\log_{10}
    \left[\frac{H(z)}{H_0}\right],
    \label{eq: halo scaling}
\end{equation}

where $\Delta_{\rm c}(z)$ is the virial overdensity of \citet{Bryan_Norman_1998} for a flat $\Lambda$CDM cosmology with $H_0=70$~km\,s$^{-1}$\,Mpc$^{-1}$ and $\Omega_{\rm m}=0.3$. \rev{At the median redshift of the golden sample ($z=1.13$), this scaling predicts circular velocities 0.13~dex higher than locally at fixed stellar mass.}

\begin{figure}
    \centering
    \includegraphics[width=0.99\linewidth]{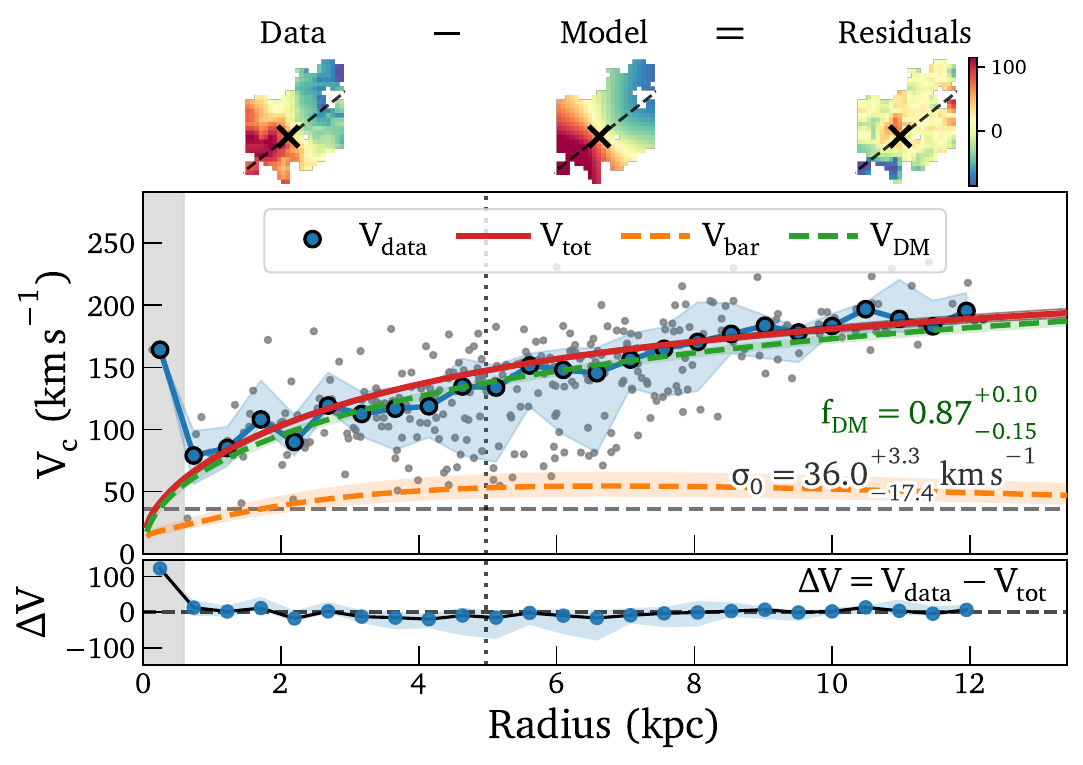}
   \caption{Example of the kinematic forward modelling of galaxy \texttt{10910}. \textbf{Top:} observed, model and residual velocity fields; the dashed line marks the kinematic major axis and the cross the fixed kinematic centre. \textbf{Middle:} circular velocity inferred from the data, corrected for beam smearing and pressure support (Section~\ref{subsection: Priors and modelling assumptions}), for individual pixels (grey points) and in radial bins (blue circles, with the 16th--84th percentile range shaded), compared with the total (red), baryonic (orange dashed) and dark matter (green dashed) model components. The grey region marks the innermost radii affected by beam smearing, the dotted line $R_{\rm e}$ and the horizontal dashed line $\sigma_0$; the inset gives $f_{\rm DM}(R_{\rm e})$. \textbf{Bottom:} residuals $\Delta V = V_{\rm data}-V_{\rm tot}$. The other galaxies are shown in the supplementary material.}
    \label{fig: Rotation curve 10910}
\end{figure}

\rev{We perform the primary comparison at $2R_{\rm e}$, the radius directly reported by \citet{Mancera_Pina_2026}. For \citet{Jeanneau_2026}, we evaluate their models at the same radius, while for MSA-3D we evaluate our models at $2R_{\rm e}$ for the 21 golden galaxies whose data reach this radius. For each galaxy, we compute the stellar-mass residual relative to the local relation evolved to its own redshift,}

\begin{equation}
    \Delta\log M_\star =
    \log M_{\star,{\rm obs}} - \log M_{\star,{\rm ref}}(V_{{\rm c},i},z_i).
\end{equation}

\rev{The MSA-3D galaxies lie $0.30\pm0.04$~dex below the self-similar reference in stellar mass, equivalent to $+0.07$~dex in velocity, whereas the samples of \citet{Jeanneau_2026} and \citet{Mancera_Pina_2026} lie $0.14\pm0.06$ and $0.47\pm0.06$~dex above it, respectively ($-0.04$ and $-0.12$~dex in velocity; Fig.~\ref{fig: TFR halo scaling}). Repeating the comparison at $1.8R_{\rm e}$, the radius adopted by \citet{Jeanneau_2026}, gives residuals of $-0.23\pm0.05$ and $+0.18\pm0.06$~dex for MSA-3D and \citet{Jeanneau_2026}, respectively, while at $2.2R_{\rm s}$ MSA-3D lies $0.23$~dex below the reference. The MSA-3D offset therefore changes little with the adopted radius (from $-0.23$ to $-0.30$~dex).}

\rev{The three samples thus show markedly different offsets from the self-similar expectation: MSA-3D lies below the reference, \citet{Jeanneau_2026} lies closest to it, and \citet{Mancera_Pina_2026} shows the largest positive offset, consistent with their reported departure from it. Although the self-similar scaling therefore provides a useful reference for the expected evolution, the three measurements do not yield a common normalisation relative to it. They span $\simeq0.8$~dex in stellar mass, or $\simeq0.2$~dex in circular velocity, substantially exceeding their statistical uncertainties.}

\rev{Interestingly, the ordering of the offsets broadly follows the effective spatial resolution of these three samples: the seeing-limited sample of \citet{Mancera_Pina_2026} lies furthest above the reference, the lensed sample of \citet{Jeanneau_2026} lies closer to it, and the higher-resolution MSA-3D sample lies below it. Our degradation experiment demonstrates that spatial resolution can shift the recovered velocities in the direction required to produce this ordering: degrading the MSA-3D models to seeing-limited resolution and extracting the kinematics without a beam-smearing correction lowers $V_{\rm c}$ by 0.04~dex, or by 0.11~dex when the intrinsic dispersion is used in the pressure-support correction. This simplified experiment is distinct from the full three-dimensional forward modelling used for the MSA-3D measurements, which explicitly accounts for beam smearing, as does the modelling of \citet{Mancera_Pina_2026}. It therefore demonstrates the direction and potential importance of resolution effects, rather than quantifying the residual bias expected in either forward-modelled analysis. The correspondence nevertheless suggests that differences in spatial resolution may contribute to the observed spread in sTFR normalisation. Resolution alone, however, cannot account for the ordering: the seeing-limited, forward-modelled KMOS$^{\rm 3D}$ disks of \citet{Ubler_2017}, which share our 3D-HST stellar masses, lie only 0.23~dex from MSA-3D in stellar mass (Section~\ref{section: Tully--Fisher}), whereas the \citet{Mancera_Pina_2026} sample, drawn largely from the same seeing-limited surveys, lies 0.77~dex away. Most of the difference between these two seeing-limited samples must therefore arise from their stellar-mass calibration, velocity modelling, pressure-support corrections and sample selection.}

\rev{Differences in stellar-mass calibration may also contribute. The local reference and both comparison samples use \texttt{BAGPIPES} masses, which are typically 0.1--0.2~dex higher than the 3D-HST catalogue masses adopted here. Placing MSA-3D on this mass scale would therefore reduce its offset from $-0.30$~dex to approximately $-0.20$ to $-0.10$~dex. Spatial resolution and stellar-mass calibration alone, however, cannot be assumed to explain the differences, since the studies also differ in sample selection, rotation-curve modelling, and pressure-support corrections. Moreover, \citet{Mancera_Pina_2026} show that the kinematic systematics they explore are unlikely to remove the departure they infer. These comparisons, therefore, highlight that the inferred evolution of the sTFR relative to simple halo-scaling expectations remains sensitive to differences in both the observations and their modelling.}

\section{Dark matter fractions}
\label{section: Dark matter fractions}

We quantify the contribution of dark matter in the inner regions of the MSA-3D galaxies using the dark matter fraction at the disk effective radius, $f_{\rm DM}(R_{\rm e}) = V_{\rm DM}^2(R_{\rm e})/V_{\rm c}^2(R_{\rm e})$, \rev{derived from each posterior sample of the dynamical model (Section~\ref{subsection: Mass model}), with $R_{\rm e}$ fixed to the photometric disk decomposition. This definition is commonly adopted in forward-modelling studies at similar redshifts (e.g., \citealp{Genzel_2017, Genzel_2020, Price_2021, Nestor_2023}).}

The dynamical modelling simultaneously constrains the baryonic and dark matter contributions to the circular velocity curve, allowing their relative contributions to be traced as a function of radius. Figure~\ref{fig: Rotation curve 10910} illustrates this decomposition for galaxy \texttt{10910}, showing the individual baryonic and dark matter components and their combined contribution to the total circular velocity.

\begin{figure}
    \centering
    \includegraphics[width=0.99\linewidth]{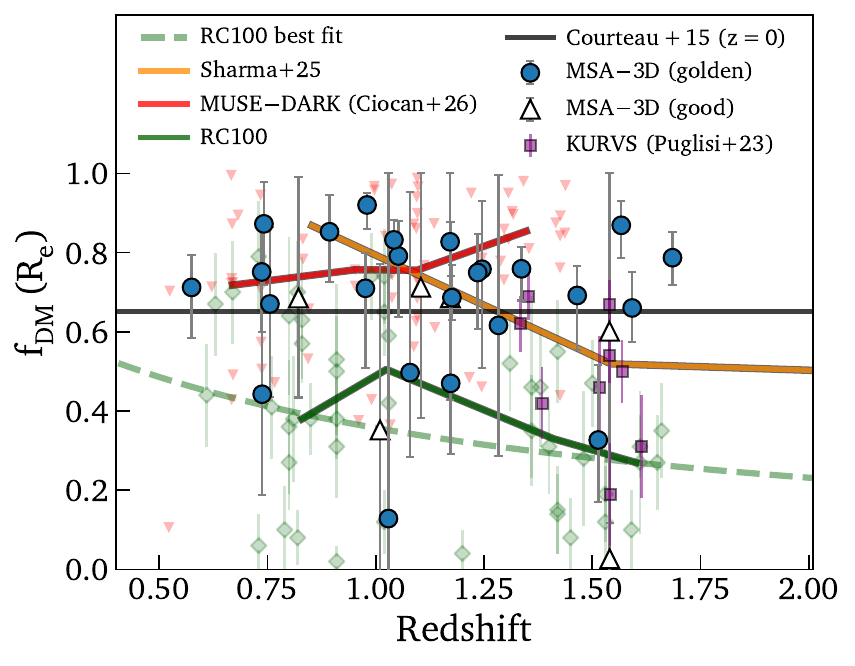}
    \caption{Dark matter fraction within the disk effective radius, $f_{\rm DM}(R_{\rm e})$, as a function of redshift. Filled circles: golden sample; open triangles: good sample; error bars are total uncertainties, truncated at $f_{\rm DM}=1$. Green diamonds and solid green line: RC100 galaxies \citep{Nestor_2023} and their running median; dashed green line: the best-fit redshift evolution of the full RC100 sample. Small red downward triangles and red line: MUSE-DARK galaxies \citep{Ciocan_2026}, converted to our definition of $f_{\rm DM}(R_{\rm e})$, and their running median. Purple squares: KURVS galaxies \citep{Puglisi_2023}. Orange line: published redshift-bin medians of \citet{Sharma_2025} (all stellar masses). Horizontal black line: the $z=0$ value of \citet{Courteau_2015}. RC100, MUSE-DARK and KURVS galaxies are restricted to the redshift and stellar-mass range of our common comparison window ($0.5\le z\le1.7$, $9\le\log_{10} (M_\star/M_\odot)\le11$); running medians use bins of $\Delta z=0.25$ with at least six galaxies.}
    \label{fig: fDM vs z}
\end{figure}

\rev{The golden sample has a median dark matter fraction of $f_{\rm DM}(R_{\rm e})=0.73$ (68\% bootstrap interval 0.70--0.76; Fig.~\ref{fig: fDM vs z}), with 68\% of the galaxies at $f_{\rm DM}(R_{\rm e})=0.49$--0.84 and individual values ranging from $\approx0.1$ to 0.9. The individual uncertainties are, however, large and differ substantially between galaxies; for example, the calibrated interval of \texttt{7561} spans the full physical range. To estimate the intrinsic scatter, we model $\log f_{\rm DM}$ as a Gaussian population truncated at $f_{\rm DM}=1$ and fit it by maximum likelihood using the individual asymmetric uncertainties. This gives an intrinsic scatter of 0.03~dex (68\% upper limit 0.04~dex; $<0.07$~dex at 95\% confidence), or 0.05~dex using only the statistical uncertainties, unchanged when \texttt{7561} is excluded. The observed spread is therefore comparable to that expected from the measurement uncertainties, and the present data do not require a large intrinsic variation in $f_{\rm DM}(R_{\rm e})$ among the golden galaxies.}

\begin{figure}
    \centering
    \includegraphics[width=0.98\linewidth]{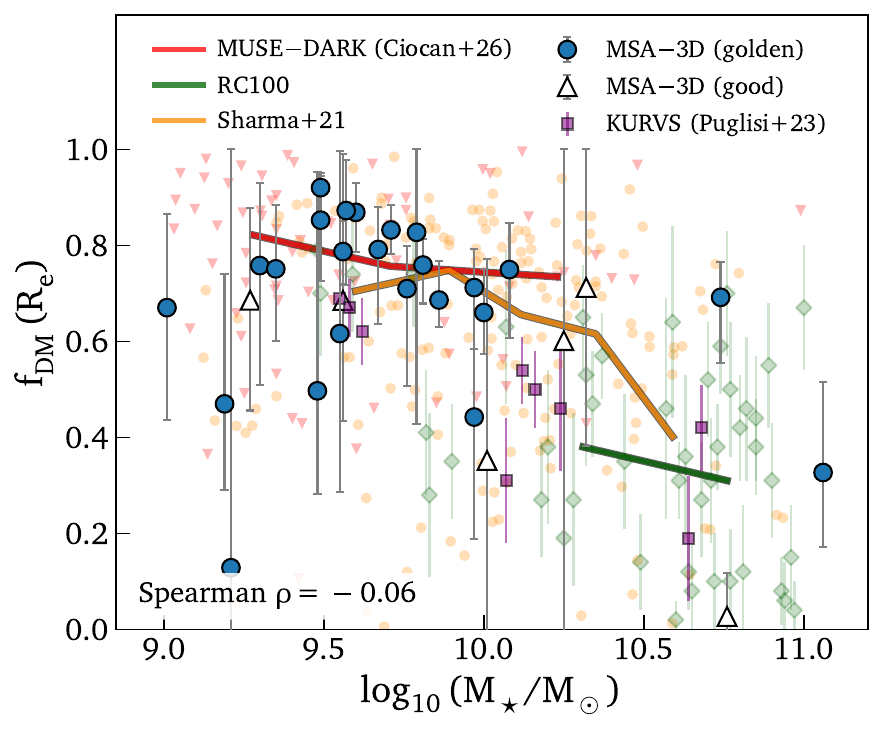}
    \caption{$f_{\rm DM}(R_{\rm e})$ as a function of stellar mass. MSA-3D symbols as in Fig.~\ref{fig: fDM vs z}. Green diamonds and green line: RC100 galaxies \citep{Nestor_2023} and their running median. Small red downward triangles and red line: MUSE-DARK galaxies \citep{Ciocan_2026}, converted to our definition of $f_{\rm DM}(R_{\rm e})$, and their running median. Small orange circles and orange line: individual KROSS galaxies at $z\sim0.8$--1 of \citet{Sharma_2021}, read from their Fig.~5 (170 of 225 galaxies), and their published binned values (50 galaxies per stellar-mass bin). Purple squares: KURVS galaxies \citep{Puglisi_2023}. RC100, MUSE-DARK and KURVS galaxies are restricted to $0.5\le z\le1.7$ and $9\le\log_{10} (M_\star/M_\odot)\le11$, the common comparison window; running medians use 0.5~dex bins with at least six galaxies. The Spearman coefficient refers to the golden sample.}
    \label{fig: fDM vs Mstar}
\end{figure}

\begin{figure}
    \centering
    \includegraphics[width=0.98\linewidth]{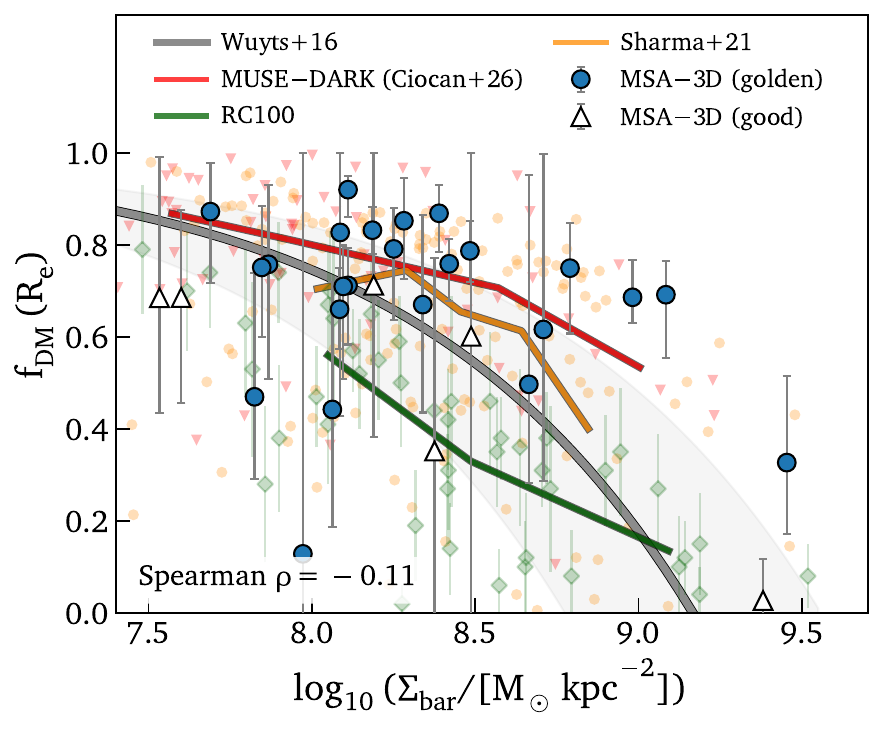}
    \caption{$f_{\rm DM}(R_{\rm e})$ as a function of the baryonic surface density within $R_{\rm e}$. MSA-3D symbols as in Fig.~\ref{fig: fDM vs z}. Grey curve and band: the relation of \citet[][Eq.~6]{Wuyts_2016}, $\log f_{\rm bar}=-0.34+0.51(\log\Sigma_{\rm bar}-8.5)$, converted with $f_{\rm DM}=1-f_{\rm bar}$, $\pm0.2$~dex. Green diamonds and green line: RC100 galaxies \citep{Nestor_2023}, with $\Sigma_{\rm bar}$ from their dynamically inferred baryonic masses, and their running median. Small red downward triangles and red line: MUSE-DARK galaxies \citep{Ciocan_2026}, with $\Sigma_{\rm bar}$ computed with our definition from their stellar masses, star formation rates and sizes, and their running median. Small orange circles and orange line: individual galaxies (read from their Fig.~5; 158 of 225) and binned values of \citet{Sharma_2021}, with their definition of $\Sigma_{\rm bar}$. RC100 and MUSE-DARK galaxies are restricted to $0.5\le z\le1.7$ and $9\le\log_{10} (M_\star/M_\odot)\le11$; running medians use 0.5~dex bins with at least six galaxies. The Spearman coefficient refers to the golden sample.}
    \label{fig: fDM vs Sigma bar}
\end{figure}

\rev{We compare our measurements with other kinematic studies at similar epochs in Figs.~\ref{fig: fDM vs z}--~\ref{fig: fDM vs Sigma bar}. For comparison samples with measurements of individual galaxies, we restrict the analysis to galaxies within a common window that matches the redshift and stellar-mass range of our golden sample ($0.5\le z\le1.7$ and $9\le\log_{10} (M_\star/M_\odot)\le11$; only \texttt{9960}, at $\log_{10} M_\star=11.06$, lies marginally above it). These include the RC100 galaxies \citep{Nestor_2023}, the KURVS galaxies at $z\sim1.5$ \citep{Puglisi_2023}, and the MUSE-DARK galaxies of \citet{Ciocan_2026}. The latter were modelled from deep MUSE data using \textsc{GalPaK3D} (\citealp{Galpak}), a parametric 3D forward-modelling tool incorporating a physical disk--halo decomposition, and provide a particularly relevant comparison to MSA-3D in redshift and stellar mass. Their published dark matter fractions are defined using enclosed masses, which we convert to our circular-velocity definition of $f_{\rm DM}(R_{\rm e})$, lowering their values by a median of 0.04. The KROSS galaxies of \citet{Sharma_2021} lie within our adopted range by construction. We additionally show the published redshift-bin medians of \citet{Sharma_2025}, the best-fitting redshift evolution of the full RC100 sample, and the local reference value of \citet{Courteau_2015}.}

\rev{As shown in Fig.~\ref{fig: fDM vs z}, even within this common redshift and stellar-mass range, the inferred dark matter fractions differ substantially between studies. The matched RC100, KURVS, and MUSE-DARK samples have median $f_{\rm DM}(R_{\rm e})$ values of 0.38, 0.50, and 0.78, respectively, compared with 0.73 for MSA-3D. Although they share the same window, the four samples populate it differently: the median stellar masses are $\log(M_\star/M_\odot)=10.6$, 10.1, 9.4 and 9.6 for RC100, KURVS, MUSE-DARK and MSA-3D, respectively, and the median dark matter fractions decrease in the same order as the stellar masses increase. MUSE-DARK is therefore the comparison sample closest to MSA-3D in both stellar mass and redshift (median $z=1.09$, compared with 1.13 for our golden sample), and we detect no significant difference between its dark matter fractions and ours (Mann--Whitney $p=0.09$), whereas the more massive RC100 and KURVS samples have significantly lower values ($p<10^{-6}$ and $p=0.006$). All 24 golden-sample galaxies except one lie above the best-fitting RC100 relation, while they are evenly distributed around the \citet{Sharma_2025} medians at their redshifts (11 of 24 lie below). The high median value of MSA-3D thus reflects, at least in part, its low stellar masses rather than a discrepancy with previous work. The remaining inter-sample differences likely reflect a combination of sample selection and methodology, including differences in kinematic tracer and spatial resolution, pressure-support corrections, baryonic mass modelling, and the adopted halo profile. Despite these differences in normalisation, none of the matched samples with individual measurements shows strong evolution over the redshift interval probed here: the matched RC100 galaxies show a mild decrease in $f_{\rm DM}(R_{\rm e})$ with redshift, MUSE-DARK a mild increase, and MSA-3D no significant trend. Taken together, these studies show a broad range of inner dark matter fractions at $z\sim1$, with substantial dark matter contributions being common while baryon-dominated systems are also present.}

\rev{Our inferred dark matter fractions are, however, sensitive to the adopted baryonic mass scale, because the baryonic and dark matter contributions are partially degenerate over the radii probed by the kinematics (Section~\ref{subsection: Fit quality and uncertainties}). Since this sensitivity is not included in the quoted uncertainties, the population statistics above are conditional on the adopted mass scale. In particular, the dynamical fits systematically prefer lower baryonic masses than the catalogue-based estimates: across the golden sample, the fitted baryonic masses lie a median of 0.31~dex below the catalogue stellar plus scaling-relation gas masses. Raising the baryonic-mass prior by 0.5~dex (Section~\ref{subsection: Fit quality and uncertainties}) substantially changes the inferred dark matter fractions, reducing the golden-sample median to $f_{\rm DM}(R_{\rm e})=0.48$ (68\% bootstrap interval 0.40--0.54) and substantially broadening the distribution (intrinsic scatter in $\log f_{\rm DM}$ of $\approx0.55$~dex, compared with 0.03~dex for the fiducial fits), as three galaxies move to $f_{\rm DM}(R_{\rm e})\lesssim0.1$. The MSA-3D median would then be comparable to that of the matched KURVS galaxies, between those of RC100 and MUSE-DARK.}

\rev{The kinematics, however, disfavour this higher mass scale: the shifted-prior fits are worse for 17 of the 24 golden galaxies and better for none (median $\Delta\chi^2_{\rm b}=+14$, from the maximum-likelihood solutions of the full fits; Table~\ref{tab: sensitivity fit quality}), while 18 converge to the lower bound of the shifted prior, corresponding to the nominal catalogue stellar plus gas mass. Conversely, for the ten galaxies whose baryonic mass reaches the lower bound of the fiducial prior (nine in the golden sample: \texttt{2824}, \texttt{3399}, \texttt{6199}, \texttt{8365}, \texttt{8942}, \texttt{11225}, \texttt{11843}, \texttt{12015} and \texttt{29470}; listed as upper limits on $\log M_{\rm bar}$ in Table~\ref{tab: stat sys}, and visible as posteriors truncated at the lower edge in the supplementary corner plots), $f_{\rm DM}(R_{\rm e})$ is itself prior-limited and could be substantially lower if their baryonic masses are closer to the catalogue-plus-gas values; complementary estimates based on the stellar-mass maps, which adopt this mass scale, indeed give higher baryonic fractions for these galaxies (Appendix~\ref{section: Baryonic rotation curve modelling}).}

\rev{The tension therefore reflects a broader mismatch between the baryonic mass distribution implied by the catalogue estimates and that preferred by the inner kinematics, rather than being restricted to the prior-bound galaxies. The galaxies that reach the fiducial prior bound are not those with the highest inferred gas fractions; they tend instead to have higher baryonic surface densities (median $\log\Sigma_{\rm bar}=8.42$, compared with 8.10 for the rest of the golden sample), and for eight of the nine the fitted baryonic mass lies below the catalogue stellar mass alone (by a median of 0.15~dex). This suggests that in the most compact systems the adopted mass normalisation and disk structure together over-predict the inner rotation. Possible contributors include systematic uncertainties in the catalogue stellar masses or stellar initial mass function, in the extrapolated molecular gas masses, and in the assumed vertical and radial structure of the baryons. The absolute normalisation of $f_{\rm DM}(R_{\rm e})$, and hence its comparison with other surveys, should therefore be regarded as uncertain at the level of $\approx0.2$ in $f_{\rm DM}(R_{\rm e})$ for galaxies whose baryonic mass is constrained by the data, and up to $\approx0.4$ for those at the prior bound, while the ranking of individual galaxies is largely preserved (Spearman $\rho=0.76$ between the two sets of values).}

\rev{Other modelling assumptions have a smaller impact. Adopting the self-gravitating thin-disk pressure-support correction used by RC100 increases $f_{\rm DM}(R_{\rm e})$ by a median of only 0.03 (Section~\ref{subsection: Priors and modelling assumptions}) and therefore cannot explain the difference between MSA-3D and RC100. The remaining modelling choices, including halo concentration, bulge size, beam (PSF) model, line-spread function, atomic gas, and kinematic centre, are incorporated into the systematic uncertainties; none of these alternatives both improves the fits and changes $f_{\rm DM}(R_{\rm e})$ appreciably (Table~\ref{tab: sensitivity fit quality}) and none of them alters the conclusions above.}

\rev{Figure~\ref{fig: fDM vs Mstar} shows $f_{\rm DM}(R_{\rm e})$ as a function of stellar mass. Within the golden sample, the correlation is weak and not significant (Spearman $\rho=-0.06$, $p=0.77$). The matched MUSE-DARK galaxies, which span a similar stellar-mass range, likewise show no significant trend ($\rho=-0.13$) and have slightly higher dark matter fractions than MSA-3D at fixed stellar mass, by $\approx0.04$ on average. The individual KROSS galaxies of \citet{Sharma_2021} broadly overlap with ours, with a median $f_{\rm DM}(R_{\rm e})=0.68$ below $10^{10}\,M_\odot$, compared with 0.75 for the MSA-3D galaxies in the same mass range. Their binned values decrease toward higher stellar masses, where the matched RC100 galaxies also exhibit lower dark matter fractions. Overall, the individual samples provide little evidence for a strong dependence of $f_{\rm DM}(R_{\rm e})$ on stellar mass over the mass range probed by MSA-3D.}

\rev{We next examine the baryonic surface density within $R_{\rm e}$, which provides a measure of the central baryonic mass concentration. We use the catalogue stellar masses (Table~\ref{tab: basic parameters}), the molecular gas masses from the \citet{Tacconi_2018} scaling relation adopted in Section~\ref{subsection: Priors and modelling assumptions}, and the atomic gas expected within $R_{\rm e}$ from the same prescription. Assuming that the stars and molecular gas follow the exponential disk, for which $R_{\rm e}$ is the half-mass radius, we define $\Sigma_{\rm bar}=[0.5\,(M_\star+M_{\rm mol})+M_{\rm HI}(<R_{\rm e})]/(\pi R_{\rm e}^2)$. This neglects the small bulge contributions (Section~\ref{subsection - Characterisation of geometrical parameters}), which would slightly increase $\Sigma_{\rm bar}$ for the few galaxies with significant bulges. For the MUSE-DARK galaxies, we compute $\Sigma_{\rm bar}$ in the same way using their stellar masses, star formation rates, and sizes.}

\rev{As shown in Fig.~\ref{fig: fDM vs Sigma bar}, MSA-3D shows no significant correlation between $f_{\rm DM}(R_{\rm e})$ and $\Sigma_{\rm bar}$ ($\rho=-0.11$, $p=0.6$), although an anticorrelation is expected if more centrally concentrated baryons contribute a larger fraction of the inner gravitational potential (e.g., \citealp{Wuyts_2016, Nestor_2023}). The larger comparison samples show this behaviour more clearly: $f_{\rm DM}(R_{\rm e})$ anticorrelates with $\Sigma_{\rm bar}$ for the matched MUSE-DARK galaxies ($\rho=-0.41$, $p=5\times10^{-4}$) and for the KROSS galaxies of \citet{Sharma_2021}. In each of the comparison samples, the correlation with $\Sigma_{\rm bar}$ is stronger than that with stellar mass. Part of this behaviour is expected, since $V_{\rm bar}^2\propto\Sigma_{\rm bar}R_{\rm e}$, and for samples where $\Sigma_{\rm bar}$ is derived from the same mass model as $f_{\rm DM}$ (RC100; \citealp{Sharma_2021}) the correlation is partly built in; it remains significant for MUSE-DARK, for which $\Sigma_{\rm bar}$ is computed independently of the fit. For MSA-3D, $\Sigma_{\rm bar}$ does not use the fitted baryonic mass, but it shares the catalogue and scaling-relation masses that centre the baryonic-mass prior, as well as $R_{\rm e}$, and is therefore not fully independent of the fit either. At fixed $\Sigma_{\rm bar}$, the MSA-3D galaxies lie close to the binned relation of \citet{Sharma_2021} and to the MUSE-DARK galaxies. Relative to the matched RC100 galaxies they lie $\approx0.3$ higher, but this offset depends on the definition of $\Sigma_{\rm bar}$: RC100 uses dynamically inferred baryonic masses, and computing $\Sigma_{\rm bar}$ for MSA-3D from our fitted baryonic masses, which are a median of 0.31~dex lower than the catalogue-based values, reduces it to $\approx0.2$. MSA-3D also lies above the empirical relation of \citet{Wuyts_2016}. On the catalogue-based definition, MSA-3D reaches central baryonic surface densities within $\approx0.2$~dex of those of RC100 despite being approximately an order of magnitude less massive and a factor of two smaller, and the MSA-3D galaxies are more compact than MUSE-DARK galaxies at similar stellar mass.}

\rev{Taken together, these comparisons suggest that the inner dark matter fraction at $z\sim1$ is more closely connected to central baryonic surface density than to stellar mass alone. Within MSA-3D, we find no significant trend. Given the size of the individual uncertainties and the limited dynamic range in $\Sigma_{\rm bar}$, however, our sample has limited power to detect such a trend. The remaining differences in normalisation between studies emphasise the importance of accounting for sample selection and methodology when comparing $f_{\rm DM}$ measurements.}

\subsection{Inner slope of the dark matter halo}
\label{subsection: Inner slope of the dark matter halo}

\rev{The generalised NFW halo allows us to test whether the kinematics favour an inner density slope different from the NFW value. Using the total uncertainties, five of the 24 golden galaxies (\texttt{2111}, \texttt{8365}, \texttt{9424}, \texttt{9636} and \texttt{29470}) have $\alpha<1$ at $2\sigma$, none has $\alpha>1$ at $2\sigma$, and the remainder are consistent with NFW; the sample median is $\alpha=0.82$. Injection--recovery tests with NFW mocks matched to each galaxy predict $\approx2.2$ such cases by chance (Appendix~\ref{section: injection recovery}); the probability of five or more is $\approx0.01$, so the excess is modest but significant for the sample as a whole. The false-positive rate differs strongly between galaxies: it is $\leq2\%$ for \texttt{2111}, \texttt{8365} and \texttt{9636}, whose shallow slopes are therefore robust individually, but high for \texttt{9424} and \texttt{29470}, in which non-circular residual structure can produce a spurious preference for $\alpha<1$. Because the mocks are generated and refitted with the same dynamical model, this test captures the effects of noise and residual structure at the observed amplitude and spatial correlation, but not systematic departures from the model, such as mismatches in the gas distribution, radial flows, or an inadequate pressure-support correction, which could also bias $\alpha$.}

\rev{A nested comparison with the matched NFW model gives a consistent picture: the approximately calibrated likelihood-ratio test (Section~\ref{subsection: Robustness}) favours the free-slope model at $p<0.05$ in 14 of the 24 galaxies. The larger number reflects the different question addressed by model comparison and does not require the marginalised $2\sigma$ interval on $\alpha$ to exclude unity. We therefore interpret $\alpha<1$ as an effective inner slope shallower than NFW over the radii probed, rather than as evidence for constant-density cores. The inferred slopes remain conditional on the adopted concentration, baryonic mass scale, and stellar-to-halo-mass prior. Freeing the concentration or shifting the baryonic-mass prior upward by $+0.5$~dex moves $\alpha$ to lower values (by a median of 0.13 for the latter), whereas the stellar-to-halo-mass prior pulls $\alpha$ upward through its degeneracy with $M_{\rm vir}$. The alternative concentrations also provide slightly better fits, while the higher baryonic mass scale fits the data worse (Table~\ref{tab: sensitivity fit quality}). Thus, among the modelling choices explored here, the main alternatives tend to strengthen rather than produce the inferred preference for $\alpha<1$.}

\begin{figure}
    \centering
        \includegraphics[width=0.98\linewidth]{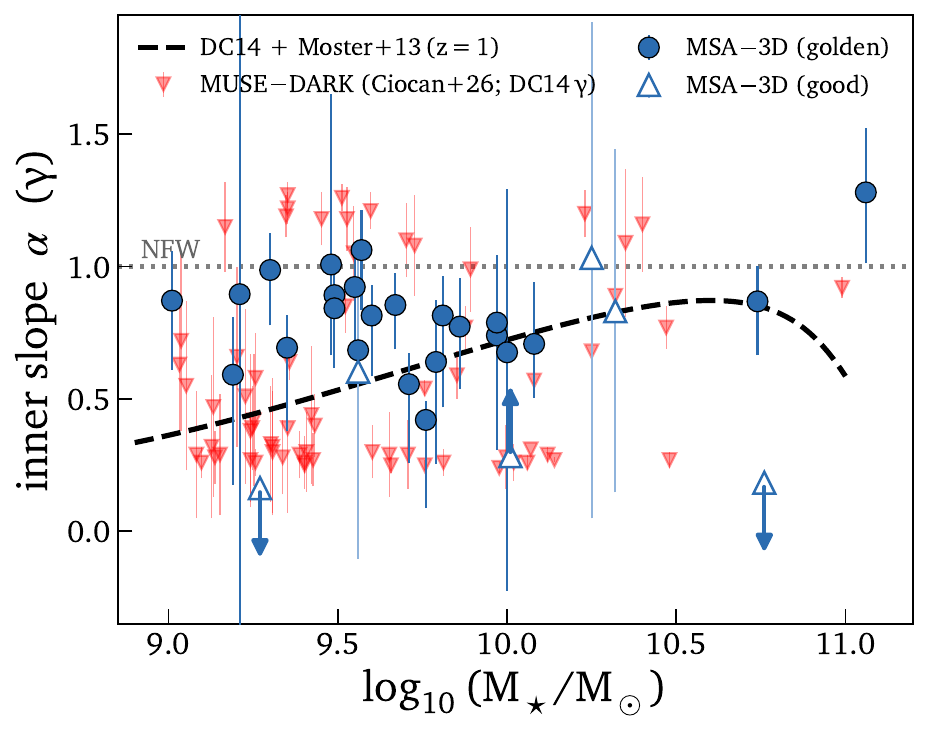}
    \caption{Inner slope of the dark matter density profile as a function of stellar mass. Filled circles and open triangles: gNFW inner slope $\alpha$ of the golden and good samples, with total uncertainties; arrows mark posterior limits. Small red downward triangles: inner slope $\gamma$ of the \citet{DiCintio_2014b} profile fitted to the MUSE-DARK galaxies \citep{Ciocan_2026} with $0.5\le z\le1.7$ and $9\le\log_{10} (M_\star/M_\odot)\le11$, which is set by the stellar-to-halo mass ratio. The dashed curve shows the \citet{DiCintio_2014b} prediction for the halo masses of \citet{Moster_2013} at $z=1$, and the dotted line the NFW slope.}
    \label{fig: alpha vs Mstar}
\end{figure}

\rev{The fitted virial masses lie systematically above the stellar-to-halo-mass relation, by a median of 1.5~dex for the golden sample and already by 0.9~dex for the matched NFW models, and the offset correlates with the deficit of fitted baryonic mass relative to the catalogue estimates (Spearman $\rho=-0.50$, $p=0.01$). With the concentration fixed, a larger $M_{\rm vir}$ is the only way for the model to increase the dark matter density within the radii probed, so the virial masses are extrapolations of the inner halo normalisation rather than measurements of the total halo mass. To test whether the shallow slopes depend on these large normalisations, we refitted all galaxies with $M_{\rm vir}$ restricted to within 1~dex of the stellar-to-halo-mass relation (Table~\ref{tab: sensitivity fit quality}). With the concentration fixed, the fits become significantly worse for 16 of the 24 golden galaxies (median $\Delta\chi^2_{\rm b}=+16$) and the slopes steepen to a median of $\alpha=0.98$. If the concentration is also allowed to vary, however, the fits are as good as the fiducial ones (median $\Delta\chi^2_{\rm b}=-0.5$), the concentrations increase by a median of 0.2~dex above the adopted relation, and ten golden galaxies have $\alpha<1$ at $2\sigma$ (five in the fiducial fits; median $\alpha=0.72$). In all cases $f_{\rm DM}(R_{\rm e})$ changes by less than 0.01. The kinematics therefore require a higher dark matter density within the inner few kiloparsecs than a halo of the expected mass and concentration provides, which can be achieved either through a larger normalisation or a higher concentration; the preference for $\alpha<1$ persists, and strengthens when the concentration is free and disappears only when both are fixed at their expected values, at the cost of significantly worse fits.}

\rev{The MUSE-DARK decompositions provide a comparison at the same epoch (Fig.~\ref{fig: alpha vs Mstar}). Using the \citet{DiCintio_2014b} profile, whose inner slope $\gamma$ is set by the stellar-to-halo mass ratio rather than fitted freely, \citet{Ciocan_2026} infer $\gamma<0.5$ for two thirds of their galaxies, most of which have $M_\star<10^{9.5}\,M_\odot$, where this profile predicts the shallowest slopes; within our redshift and stellar-mass range, their median is $\gamma=0.42$. Our fitted slopes are steeper, with a median $\alpha=0.82$ and only one golden galaxy having $\alpha<0.5$, and lie above the \citet{DiCintio_2014b} prediction evaluated at our stellar masses and redshifts using the halo masses of \citet{Moster_2013} (median 0.57). However, $\alpha$ and $\gamma$ are not directly equivalent because the two halo parameterisations differ in their transition and outer slopes. A more direct comparison is therefore provided by the logarithmic density slope evaluated at a common physical radius: at 1~kpc, the median slope is $-0.85$ for the MSA-3D golden galaxies and $-0.66$ for the matched MUSE-DARK galaxies. Both samples thus favour inner profiles shallower than NFW over the radii probed, with MSA-3D lying closer to the NFW form.}

\rev{Physically, effective slopes shallower than NFW but steeper than the \citet{DiCintio_2014b} prediction could indicate that feedback-driven expansion of the inner halo is less efficient in our galaxies than predicted at these stellar masses, or that it is partly offset by halo contraction in response to the baryons. For \texttt{9960}, the most massive galaxy and the one with the steepest slope in our golden sample ($\alpha=1.28^{+0.22}_{-0.27}$), Al~Farisy et al. (in prep.) find that, when the halo is adiabatically contracted by the baryons (as implemented in \texttt{DysmalPy} following \citealt{Burkert_2010}), the steep slope is reproduced by a halo with an initial slope close to NFW ($\alpha=0.91^{+0.21}_{-0.23}$); this model is preferred in leave-one-out cross-validation, while $f_{\rm DM}(R_{\rm e})$ is almost unchanged. For the sample as a whole, however, our data do not allow us to distinguish between these possibilities. Within the golden sample, $\alpha$ shows no significant correlation with stellar mass, star-formation rate, specific star-formation rate, size, baryonic surface density, or $f_{\rm DM}(R_{\rm e})$ (all $|\rho|\leq0.28$, $p\geq0.18$). In particular, we do not detect the stellar-mass dependence expected from the \citet{DiCintio_2014b} profile. However, across the mass range containing most of our sample, the predicted change is modest ($\gamma\approx0.4$ at $10^{9.2}\,M_\odot$ to $\approx0.7$ at $10^{10}\,M_\odot$) and comparable to the individual uncertainties on $\alpha$, so the absence of such a trend is not strongly constraining. In addition, the effective slope over the $\sim$1--3~kpc probed by the data is degenerate with the baryonic mass scale and distribution (Section~\ref{subsection: Fit quality and uncertainties}), and the innermost kiloparsec is only partially resolved at our spatial resolution. We therefore interpret the preference for $\alpha<1$ in a subset of the sample as evidence that at least some of the inner halo profiles are shallower than the collisionless NFW expectation, while the present data do not identify the baryonic process responsible.}

\section{Rotation-curve shapes}
\label{section: Rotation curve shapes}

\begin{figure*}
    \centering
    \includegraphics[width=\linewidth]{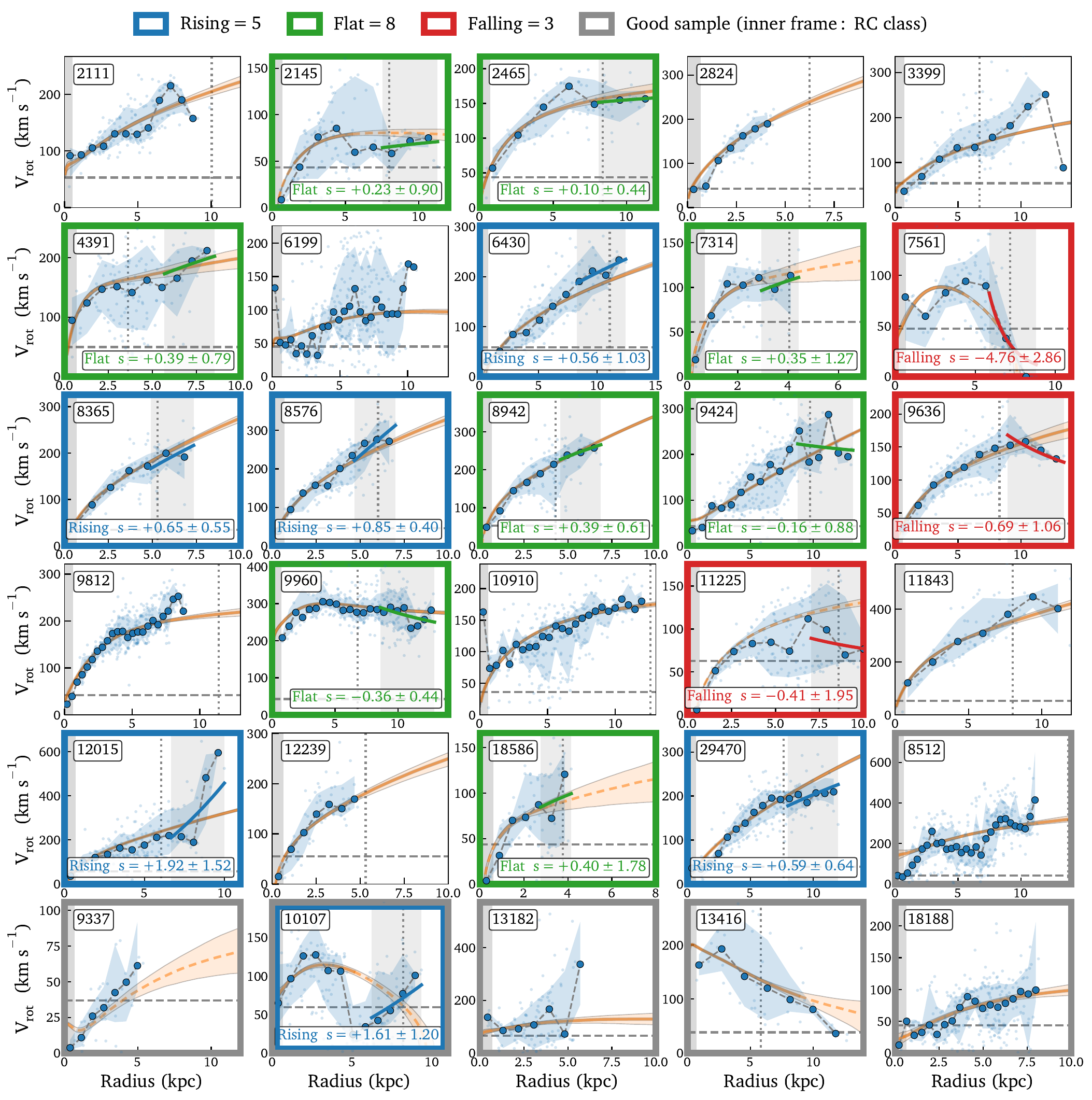}
    \caption{Rotation curves of the MSA-3D galaxies. In each panel, the blue points and grey dashed line show the median deprojected rotation velocity of the observed velocity field in radial bins (light points: individual pixels; shaded: 16th--84th percentile range per bin), and the orange curve the intrinsic rotation curve of the best-fitting \texttt{DysmalPy} model (dashed where the pressure-support term exceeds half of $V_{\rm c}^2$; band: 16th--84th percentile range of the posterior). For the classified galaxies, the grey band marks the outer window used for the classification and the coloured line the fitted outer slope $s$, whose value is given in the lower part of the panel. Panel frames give the class: rising (blue), flat (green) and falling (red); the numbers in the legend refer to the golden sample. Galaxies without a coloured frame either do not reach the dotted line or have too few pixels in the outer window. Galaxies of the good sample are shown with a grey outer frame, with an inner frame indicating their class. The dotted vertical line marks the coverage limit of the classification, $3\sqrt{2}\,R_{\rm s}\simeq2.5R_{\rm e}$, the horizontal dashed line the intrinsic dispersion $\sigma_0$, and the grey region at small radii the innermost pixel. Radii are in kpc; galaxy IDs are given in the upper left.}
    \label{fig: Rotation curves}
\end{figure*}

\subsection{Classification of the outer rotation curves} 
\label{subsection: RC classification}

The shape of galaxy rotation curves at cosmic noon has been the subject of significant debate over the past decade. Early integral-field studies of massive star-forming galaxies at $z\sim1$--$2$ reported rotation curves that rise rapidly and subsequently decline at large radii, a behaviour interpreted as evidence for baryon-dominated inner regions and relatively low dark matter fractions within the observed extent (e.g.\ \citealt{Genzel_2017, Lang_2017}). Subsequent analyses probing different mass ranges, spatial resolutions, and radial coverages have instead revealed a broader diversity of rotation-curve shapes, including flat and continuously rising profiles (\citealp{Tiley, Genzel_2020, Price_2021, Puglisi_2023, Sharma_2025}), more in line with the variety observed in local disk galaxies (e.g.\ \citealt{Lelli_2016, Sofue_2001, de_Blok_2008}).

To place the MSA-3D sample in the context of this debate, \rev{we quantify the outer shape of each rotation curve using a declared criterion applied to the observed velocities rather than to the model curves. We deproject the velocity of each pixel using the fitted geometry and fit $\ln V=s\ln r+c$ to the pixels in the outer third of the radial range covered by the data, with the inner edge never allowed inside $3R_{\rm s}$, corresponding to the inner boundary of the outer-shape window of \citet{Tiley} (grey bands in Fig.~\ref{fig: Rotation curves}). The fit uses the per-pixel uncertainties together with an intrinsic-scatter term that accounts for deviations from the fitted radial trend, including those induced by non-circular motions, and the uncertainty on the slope is increased by the square root of the number of pixels per resolution element to account for spatial correlations. A curve is classified as flat if $|s|<2\sigma_s$ and $|s|\le0.4$ (i.e. the velocity changes by less than $\sim30$\% per doubling of radius), and as rising ($s>0$) or falling ($s<0$) otherwise. We classify only galaxies whose data extend to at least $3\sqrt{2}\,R_{\rm s}\simeq2.5R_{\rm e}$, the logarithmic midpoint between $3R_{\rm s}$ and $6R_{\rm s}$, and that contain at least three resolution elements of pixels in the outer window. We also flag curves for which the model pressure-support term exceeds half of $V_{\rm c}^2$ within the window, because their observed shape can then be strongly affected by pressure support rather than directly tracing the shape of the circular-velocity curve.}

\rev{Of the 24 golden galaxies, 16 satisfy these criteria: five have rising, eight flat, and three falling outer rotation curves (Fig.~\ref{fig: Rotation curves}). Two further galaxies reach the required radius but have too few pixels in the outer window (\texttt{3399} and \texttt{11843}), while six do not extend far enough. Thus, among the classified galaxies, rising and flat curves substantially outnumber falling ones. Independent of the discrete classification, 11 of the 16 measured outer slopes are positive, with a median $s=+0.37$, although the distribution is not significantly offset from zero (Wilcoxon $p=0.25$). The classes are based on the best-fitting slopes and describe the typical outer shape of each curve; individually, only one slope (\texttt{8576}) differs from zero at $2\sigma$, and none of the three falling slopes does. In \texttt{7561}, whose best-fitting model is baryon-dominated ($f_{\rm DM}=0.13$, although its calibrated interval spans almost the full physical range), the circular velocity of the model declines by $\sim15$\% across the outer window, while pressure support causes the observed rotation velocity to decline much more steeply. In \texttt{9636} and \texttt{11225}, which are dark-matter dominated within $R_{\rm e}$, the observed declines are smaller than their uncertainties and the best-fitting model circular-velocity curves rise; these galaxies are classified as falling because their best-fitting outer slopes satisfy $|s|>0.4$. A falling $V_{\rm rot}$ is therefore not necessarily indicative of a falling $V_{\rm c}$.}

\rev{The classification of individual galaxies is sensitive to the precise definition of the outer window, but the population-level result is more robust. Varying its inner edge between $3R_{\rm s}$ and the outer half of the observed radial range changes the class of about half of the golden galaxies, yet rising and flat curves outnumber falling ones for every window tested. Requiring two rather than three resolution elements in the window classifies one additional galaxy (\texttt{3399}, flat with a poorly constrained slope), while lowering the radial-coverage threshold to $2R_{\rm e}$ adds one rising and one falling curve (\texttt{10910} and \texttt{2111}), neither with a slope significant at $2\sigma$. Individual slopes also carry substantial uncertainties (typically $\sigma_s\simeq0.9$). We therefore regard the classes as descriptive summaries that are most informative at the population level rather than as precise classifications of individual galaxies. We next examine whether these outer slopes are connected to other galaxy properties.}

\subsection{Connection between rotation-curve shape and galaxy properties}
\label{subsection: Connection between rotation-curve shape and galaxy properties}

\begin{figure}
    \centering
    \includegraphics[width=0.99\linewidth]{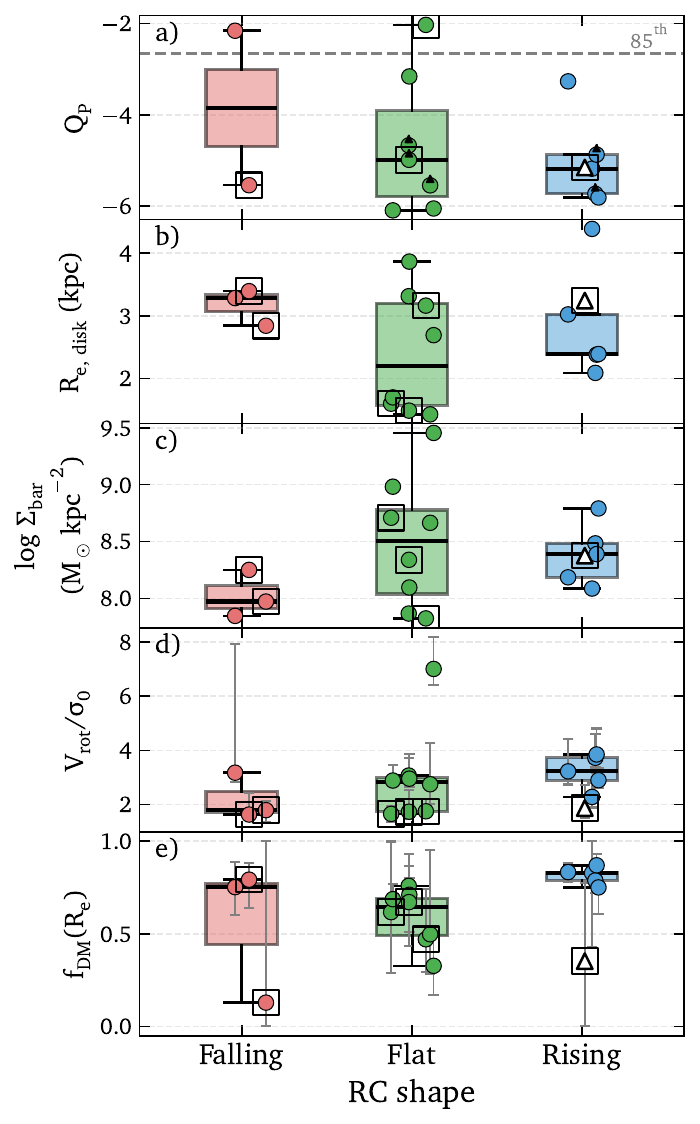}
    \caption{Structural, kinematic and environmental properties as a function of the outer rotation-curve class: (a) projected interaction strength $Q_P$, (b) disk effective radius, (c) baryonic surface density within $R_{\rm e}$, (d) $V_{\rm rot}/\sigma_0$ at $2.2R_{\rm s}$ and (e) $f_{\rm DM}(R_{\rm e})$. Filled circles: golden sample; open triangles: good sample (shown at the class given by the same criterion, not included in the statistics); open squares mark curves whose outer window is dominated by pressure support. Boxes summarise the golden sample (median, interquartile range, whiskers to $1.5\times$ the interquartile range). Only $f_{\rm DM}(R_{\rm e})$ differs nominally between the classes (Kruskal--Wallis $p=0.02$), not significant after accounting for the number of tests (Section~\ref{subsection: Connection between rotation-curve shape and galaxy properties}); for the other properties $p\geq0.23$. The dashed line in (a) marks the 85th-percentile threshold in $Q_P$.}
    \label{fig: rot curves vs params}
\end{figure}

If the outer shape of the rotation curve reflects the relative distributions of baryons and dark matter, it may be expected to depend on the inner dark matter fraction and the baryonic structure of the galaxy. Previous studies at $z\sim1$--$2$ found declining outer profiles predominantly among massive galaxies with high baryonic surface densities and low inner dark matter fractions \citep[e.g.,][]{Genzel_2017, Genzel_2020, Price_2021}, whereas larger samples extending to lower stellar masses found predominantly flat or rising curves \citep{Tiley}. \rev{We therefore test whether the outer slopes of the 16 classified golden galaxies depend on their structural, kinematic, or environmental properties, considering both the three discrete classes and the continuous slope $s$ (Fig.~\ref{fig: rot curves vs params}).}

\rev{We find no statistically robust dependence on any of the quantities examined. The inner dark matter fraction differs nominally between the three classes (Kruskal--Wallis $p=0.02$), with medians of 0.83, 0.64 and 0.75 for the rising, flat and falling curves, respectively, but this ordering is not the one expected if falling curves traced baryon-dominated systems, and with eight quantities tested it is not significant after accounting for the number of tests. The continuous outer slope correlates positively with $f_{\rm DM}(R_{\rm e})$, as expected if more dark-matter-dominated galaxies have more steeply rising outer curves, but not significantly ($\rho=0.49$, $p=0.06$). Disk size, baryonic surface density, stellar mass, rotation velocity, intrinsic dispersion, rotational support, and the inner halo slope do not differ significantly between the classes ($p\ge0.17$), and the continuous slope does not correlate significantly with any of them ($|\rho|\le0.40$, $p\ge0.13$). Thus, within MSA-3D, the outer rotation-curve shape shows only a weak, non-significant tendency with the inner dark matter fraction, and no dependence on the mean baryonic surface density is detected.}

\rev{The absence of significant correlations does not, however, imply the absence of an underlying physical connection. The typical uncertainty on the measured outer slopes is large ($\sigma_s\simeq0.9$), the curves generally extend to $\lesssim3R_{\rm e}$, and only 16 galaxies satisfy the classification criteria. Accounting for these limitations, a correlation of the strength expected from our mass models would be recovered at $p<0.05$ in only $\sim10$\% of samples of this size. The present data therefore have limited power to detect a moderate underlying relation. They nevertheless provide no evidence for the strong ordering of rotation-curve shape with baryonic structure reported for more massive galaxies at higher baryonic surface densities \citep{Genzel_2020}.}

\rev{The three galaxies with falling outer curves provide a useful illustration of this complexity. They have the lowest rotational support among the classified galaxies, with a median $V_{\rm rot}/\sigma_0=1.8$ at $2.2R_{\rm s}$ compared with $\simeq3$ for the rising and flat classes, although the difference is not significant given the small number of falling curves. Moreover, these three systems are physically heterogeneous (Sect.~\ref{subsection: RC classification}). In \texttt{7561}, whose dark matter fraction is unconstrained (its calibrated interval spans the full physical range), pressure support steepens an underlying decline in the model circular-velocity curve, whereas in the dark-matter-rich galaxies \texttt{9636} and \texttt{11225}, the observed declines are not significant and the best-fitting circular-velocity curves rise. A declining observed rotation curve therefore does not by itself imply a low dark matter fraction within $R_{\rm e}$. More generally, the shape of $V_{\rm rot}$, and any characteristic velocity measured from it, depends on pressure support as well as on the underlying mass distribution. This distinction is important when comparing kinematic scaling relations between samples that adopt different velocity definitions or probe different radial extents (Section~\ref{section: Tully--Fisher}).}

\rev{Interactions provide another possible source of perturbations to the outer kinematics. Of the five galaxies identified as likely interacting based on the projected interaction strength $Q_P$ (Appendix~\ref{section: interaction strengths}), only two have a classified outer shape: \texttt{2145} is flat and \texttt{9636} is falling. The $Q_P$ distributions do not differ significantly between the rotation-curve classes (Kruskal--Wallis $p=0.74$), but with only two classified likely interacting galaxies, the present sample provides little leverage on a possible connection between interactions and outer rotation-curve shape.}

\section{Conclusions}
\label{section: Conclusions}

We have presented a spatially resolved kinematic analysis of 30 star-forming galaxies at $0.5<z<1.7$ from the MSA-3D Cycle 1 survey, \rev{24 of which form our golden sample}. This sample provides new constraints on disk settling and inner mass assembly toward the end of the cosmic noon epoch, complementing existing ground-based IFU and AO studies with efficient, high-spatial-resolution kinematic measurements extending to lower stellar masses. Combining \textit{JWST}/NIRSpec ionised-gas kinematics, \textit{JWST}/NIRCam structural constraints, and \texttt{DysmalPy} forward modelling, we characterise their rotational support, stellar Tully-Fisher relation, inner dark matter fractions, and rotation-curve shapes. Our main findings are:


\textbf{Disk classification and rotational support:} The MSA-3D kinematic sample is \rev{predominantly composed of rotation-supported galaxies. Most systems (24/30) exhibit disk-like morphologies in the \textit{JWST} imaging, while the remaining six also show ordered velocity gradients. At $2.2R_{\rm s}$, all galaxies have $V_{\rm rot}/\sigma_0>1$, with a median of 2.9 for the golden sample and 10 of its 24 galaxies reaching $V_{\rm rot}/\sigma_0>3$, highlighting the substantial contribution of ordered rotation to their kinematics. The intrinsic dispersions in the golden sample span $\sigma_0\simeq34$--63~km\,s$^{-1}$, and our rotational support agrees to within 5\% in the median with the independent analysis of the same galaxies by \citet{Ju_2026}. The line-spread function remains the dominant systematic uncertainty on $\sigma_0$, but tests with alternative instrumental resolutions do not alter these conclusions.}

\textbf{Stellar Tully--Fisher relation:} \rev{Using a common definition of the circular velocity at $2.2R_{\rm s}$ and the same pressure-support correction across samples, the golden sample follows a relation with slope $a=3.8\pm0.5$, consistent with the slopes typically measured at $z\sim1$. At a common slope, MSA-3D galaxies rotate faster at fixed stellar mass than the $z\sim1$ KMOS$^{\rm 3D}$ and KROSS samples and local SPARC galaxies, with offsets of 0.06--0.17~dex in velocity. A simplified seeing-degradation test lowers our recovered circular velocities by 0.04~dex, suggesting that spatial resolution could contribute to these differences. Relative to the local SPARC relation evolved with self-similar halo scaling, MSA-3D lies $0.30\pm0.04$~dex below in stellar mass (+0.07~dex in velocity), whereas the samples of \citet{Jeanneau_2026} and \citet{Mancera_Pina_2026} lie 0.14 and 0.47~dex above it; the ordering broadly tracks spatial resolution, but differences in stellar-mass calibration, sample selection, and methodology prevent attributing it to any single factor. The spread between these samples ($\simeq0.2$~dex in velocity) exceeds the 0.13 dex evolution in the Tully-Fisher zero-point predicted by self-similar scaling from $z\approx0$ to $z\approx1$, indicating that current inter-survey differences are comparable to or larger than the expected evolution.}

\textbf{Dark matter fractions:} \rev{Within the disk effective radius, 68\% of the golden galaxies have $f_{\rm DM}(R_{\rm e})=0.49$--0.84 (median 0.73). These high fractions are conditional on the baryonic mass scale favoured by the kinematics, which is $\approx0.3$~dex below the catalogue estimates. The individual uncertainties are large, and when they are taken into account, the intrinsic scatter is $\lesssim0.07$~dex (95\% confidence) for the fiducial mass model; we find no significant trend with redshift or stellar mass. Within a common redshift and stellar-mass range, comparison samples span substantially different dark matter fractions, with median $f_{\rm DM}(R_{\rm e})=0.38$, 0.50, and 0.78 for RC100, KURVS, and MUSE-DARK, respectively. These medians decrease with the median stellar mass of each sample ($\log M_\star\approx10.6$, 10.1 and 9.4, compared with 9.6 for MSA-3D), and MSA-3D agrees best with MUSE-DARK, the sample closest to ours in both stellar mass and redshift; the remaining differences likely reflect a combination of sample selection and methodology. Across the individual comparison samples, $f_{\rm DM}(R_{\rm e})$ is more closely connected to central baryonic surface density than to stellar mass alone; within MSA-3D, we find no significant trend with either. The inferred dark matter fractions are sensitive to the adopted baryonic mass scale: increasing the baryonic-mass prior by $+0.5$~dex lowers $f_{\rm DM}(R_{\rm e})$ by a median of 0.19 for galaxies whose baryonic mass is constrained by the kinematics, although the kinematics disfavour this higher mass scale, while the values for galaxies whose baryonic mass reaches the lower prior bound could be substantially lower. Other modelling choices are incorporated into the systematic uncertainties.}

\rev{\textbf{Inner slope of the dark matter halo:} The inner halo slopes are generally consistent with NFW (median $\alpha=0.82$),  with no galaxy significantly steeper than NFW and none favouring a constant-density core. Five of the 24 golden galaxies have $\alpha<1$ at $2\sigma$, compared with $\approx2.2$ expected from injection--recovery tests with NFW mocks matched to each galaxy (probability of five or more $\approx0.01$); under this NFW null test the excess is modest but significant for the sample, and three galaxies remain individually significant. These tests capture noise and residual structure but not a systematic model mismatch, so we interpret $\alpha$ as an effective inner slope over the radii probed. The preference for $\alpha<1$ is conservative with respect to the adopted concentration and persists when $M_{\rm vir}$ is restricted to within 1~dex of the stellar-to-halo-mass relation and the concentration is allowed to vary. Evaluated at 1~kpc, the profiles inferred under the respective halo models are steeper for MSA-3D than for the matched MUSE-DARK galaxies (median logarithmic slopes of $-0.85$ and $-0.66$), although this comparison depends on the different halo parametrisations. The slopes show no significant dependence on stellar mass or other galaxy properties, and we cannot attribute them to a specific baryonic process.}
    
\textbf{Rotation-curve shapes:} \rev{With a declared criterion based on the outer slope of the deprojected velocities, the 16 golden galaxies with sufficient radial coverage comprise five rising, eight flat and three falling rotation curves. The classes of individual galaxies depend on the adopted outer window, but rising and flat curves outnumber falling ones for every choice. The outer shape shows only a weak, non-significant tendency with the inner dark matter fraction, and no dependence on disk size, baryonic surface density, stellar mass or rotational support is detected, although the present data have limited power to detect a moderate dependence. The falling curves are physically heterogeneous: in one case, whose dark matter fraction is unconstrained, the decline of the model circular velocity is amplified by pressure support, while in the other two, dark-matter-rich galaxies the decline is not significant. A falling rotation curve at these masses therefore need not indicate a deficit of dark matter within $R_{\rm e}$, and any single characteristic velocity read from the outer rotation curve depends on the pressure-support correction as well as on the mass distribution.}

\textbf{MSA-3D efficiency:} A key outcome of this study is the demonstration of the efficiency of the slit-stepping strategy with the \textit{JWST}/NIRSpec MSA for resolved kinematic studies at high redshift. Measurements that previously relied on much larger investments of ground-based IFU observing time, including rotation-curve shapes, $V_{\rm rot}/\sigma_0$, dark matter fractions, and their connection to galaxy structure, can now be obtained for sizable samples with only $\sim30$ hours\footnote{The MSA-3D Cycle 1 sample required $\sim30$ hours for the 43 galaxies, of which 30 are suitable for this analysis.} of space-based observations. The combination of stable space-based spatial resolution, high sensitivity, and multiplexed slit-stepping enables systematic dynamical analyses across parameter spaces that were previously difficult to access.


Building on these results, and on the wealth of high-spatial-resolution \textit{JWST}/NIRCam imaging, we will extend this analysis to the $z\sim2$ and $z\sim3$ MSA-3D samples\footnote{Cycle 2 ($z\sim3$) has been observed, and Cycle 5 ($z\sim2$) has been accepted.}. Applying the same slit-stepping strategy and analysis framework to a substantially larger sample spanning $0.5\lesssim z\lesssim3$ will enable direct comparisons of rotation curves, velocity dispersions, and mass distributions across cosmic time. Together, these observations will provide a space-based benchmark for tracing the evolution of disk dynamics, mass assembly, and the baryon–dark matter interplay throughout cosmic noon.


\begin{acknowledgements}
      We thank Lilian Lee and Hannah Übler for their valuable suggestions. We thank the anonymous referee for their comments and suggestions, which helped improve the clarity and robustness of this work. J.M.E.S. and N.M.F.S. acknowledge financial support from the European Research Council (ERC) Advanced Grant under the European Union’s Horizon Europe research and innovation programme (grant agreement AdG GALPHYS, No. 101055023). Views and opinions expressed are however those of the authors only and do not necessarily reflect those of the European Union or the European Research Council Executive Agency. Neither the European Union nor the granting authority can be held responsible for them. T.T. is supported by the JSPS Grant-in-Aid for Research Activity Start-up (25K23392) and the JSPS Core-to-Core Program (JPJSCCA20210003). K.B.W. gratefully acknowledges support from Simons Foundation Award SFI-MPS-SSRFA-00012894. The specific observations analysed can be accessed via DOI: \href{https://doi.org/10.17909/s8wp-5w10}{10.17909/s8wp-5w10}. These observations are associated with program JWST-GO-2136. We acknowledge financial support from NASA through grant JWST-GO-2136.  This work made use of observations and catalogues from the 3D-HST Treasury Program (GO 12177 and 12328) with the NASA/ESA Hubble Space Telescope, which is operated by the Association of Universities for Research in Astronomy, Inc., under NASA contract NAS5-26555. We acknowledge the DAWN \JWST Archive at the Cosmic Dawn Centre funded by the Danish National Research Foundation. We further appreciate the open-source software packages used throughout this work, including \textsc{astropy} (\citealp{astropy}), \textsc{scipy} (\citealp{scipy}), \textsc{numpy} (\citealp{numpy}), CM\textsc{asher} (\citealp{cmasher}), \textsc{matplotlib} (\citealp{matplotlib}), \galfit (\citealp{Peng_2002, Peng_2010}), \textsc{trilogy} \citep{Coe_2012}.
\end{acknowledgements}

\bibliographystyle{aa}
\bibliography{references}

\appendix

\section{Sensitivity to the halo concentration}
\label{section: concentration impact}

\rev{The fiducial models adopt the redshift-dependent halo concentration $c=10.9\,(1+z)^{-0.83}$ from \citet{Genzel_2020}, based on \citet{Dutton_Maccio_2014} (Table~\ref{tab: dysmalpy parameters}). We assess the sensitivity to this assumption by refitting all galaxies with two alternative prescriptions: (i) allowing the concentration to vary with a log-normal prior centred on the fiducial value and a scatter of 0.11~dex, consistent with \citet{Dutton_Maccio_2014}, bounded to $1\leq c \leq20$; and (ii) defining the concentration relative to the radius where the logarithmic density slope is $-2$, as in standard NFW concentration--mass relations, implemented as $c\rightarrow c(2-\alpha)$ for the generalised NFW profile. All other modelling settings remain unchanged.}

\begin{figure*}
    \centering
    \includegraphics[width=0.75\linewidth]{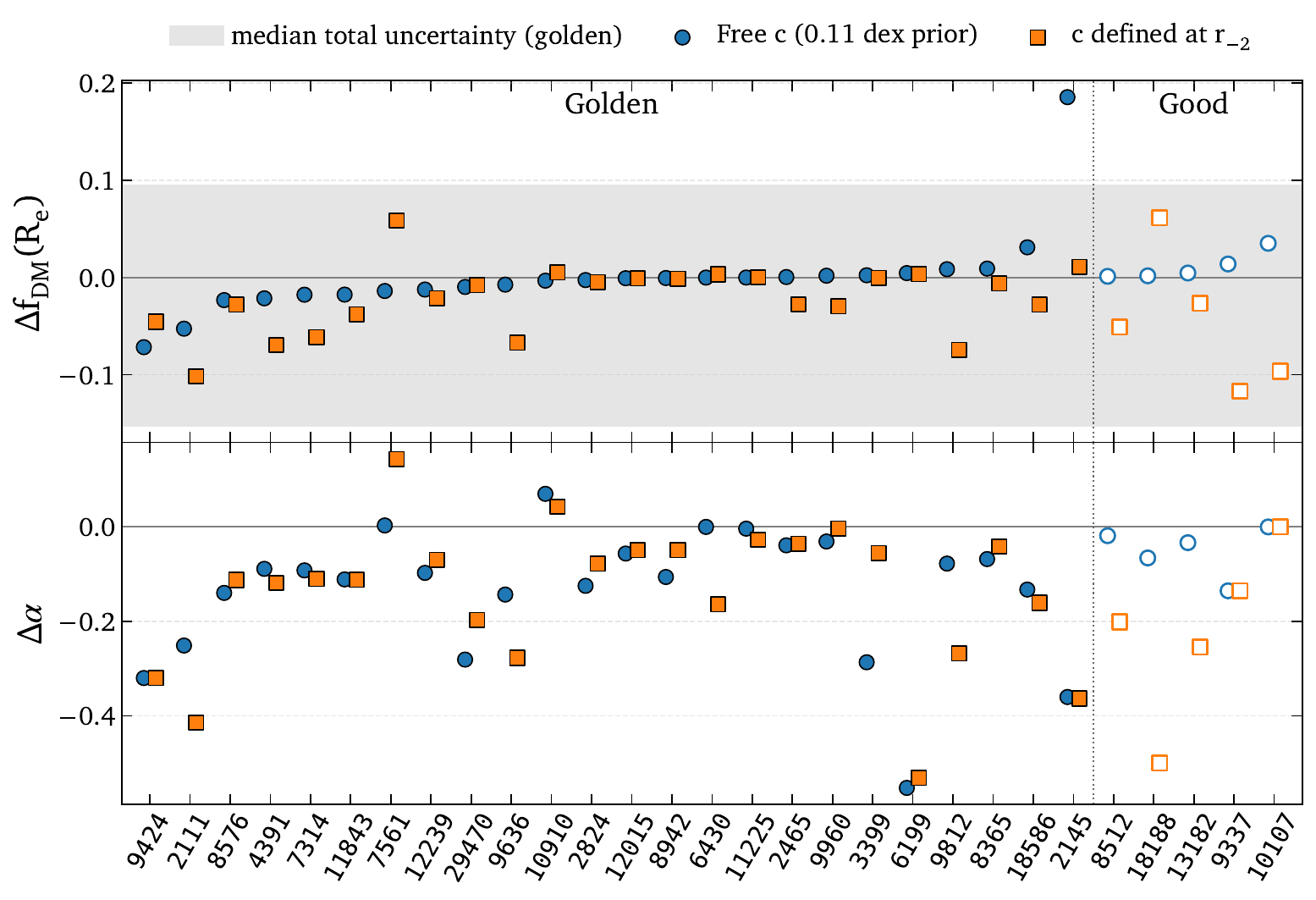}
    \caption{Sensitivity of the dark matter fractions (top) and inner halo slopes (bottom) to the concentration prescription. Symbols show the changes relative to the fiducial model when the concentration is allowed to vary with a 0.11~dex log-normal prior (circles) or is defined relative to the radius where the logarithmic density slope is $-2$ (squares). Filled and open symbols denote the golden and good samples, respectively. The grey band in the top panel indicates the median total uncertainty on $f_{\rm DM}(R_{\rm e})$ for the golden sample.}
    \label{fig: c sensitivity}
\end{figure*}

\rev{For the golden sample, the median changes in $f_{\rm DM}(R_{\rm e})$ are $-0.002$ and $-0.015$ for the two prescriptions, respectively, with median absolute changes of 0.009 and 0.024. The inferred inner slopes decrease by a median of 0.10 and 0.11, respectively (Fig.~\ref{fig: c sensitivity}). Both alternatives improve the fits for 17 of the 24 golden galaxies (median $\Delta\chi^2_{\rm b}=-4.0$ and $-3.7$; Table~\ref{tab: sensitivity fit quality}), the first at the cost of one additional free parameter. We retain the fiducial relation for consistency with \citet{Genzel_2020}; adopting either alternative would leave $f_{\rm DM}(R_{\rm e})$ unchanged and lower $\alpha$, strengthening rather than weakening our conclusions on the inner slope. These variations are included in the systematic uncertainty budget and dominate the downward systematic uncertainty on $\alpha$.}
    
\section{Impact of the bulge-to-total ratio and bulge size}
\label{section: B/T impact}

\begin{figure}
    \centering
    \includegraphics[width=0.95\linewidth]{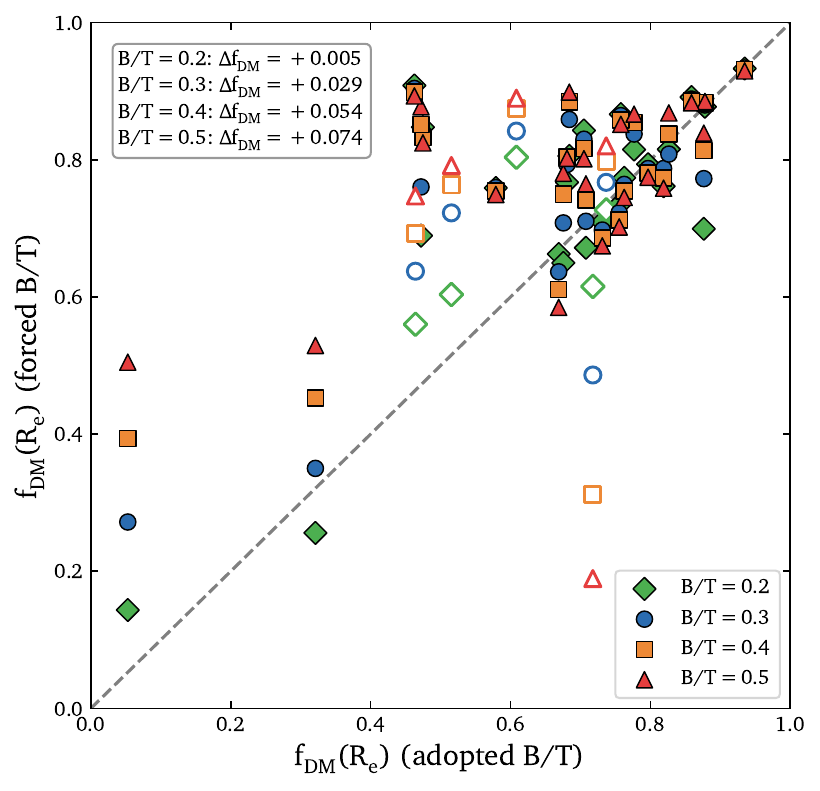}
    \caption{Sensitivity of $f_{\rm DM}(R_{\rm e})$ to the assumed bulge-to-total ratio. Values obtained from refits with fixed $B/T=0.2$, 0.3, 0.4 and 0.5 are compared with those of the fiducial models. Filled and open symbols denote the golden and good samples, respectively. The box indicates the median change for the golden sample, and the dashed line marks the one-to-one relation. Larger bulge fractions generally increase the inferred dark matter fractions, with the largest changes occurring in galaxies for which the forced bulge produces substantially worse fits.}
    \label{fig: bt_fdm_test}
\end{figure}

\rev{Most galaxies favour low bulge fractions, raising the possibility that an underestimated central stellar component could bias the inferred dark matter fractions. To test this, we refitted all galaxies with fixed bulge-to-total ratios of $B/T=0.2$, 0.3, 0.4 and 0.5, keeping all other modelling settings unchanged. Increasing $B/T$ raises, rather than lowers, $f_{\rm DM}(R_{\rm e})$ by a median of 0.01--0.07 for the golden sample (Fig.~\ref{fig: bt_fdm_test}), while worsening the fits ($\Delta\chi^2_{\rm b}>2$) for 14, 15, 17 and 18 of the 24 galaxies for $B/T=0.2,\,0.3,\,0.4$ and $0.5$, respectively (median $\Delta\chi^2_{\rm b}=+4$ to $+15$; Table~\ref{tab: sensitivity fit quality}). This occurs because a more massive central bulge steepens the inner rotation curve, requiring a reduction in the total baryonic mass to reproduce the observed kinematics. We also tested bulge effective radii of 0.5 and 2~kpc, finding median changes in $f_{\rm DM}(R_{\rm e})$ below 0.01 and negligible changes in fit quality, which are included in the systematic uncertainties. Finally, fixing $B/T$ to its photometric value worsens the fit ($\Delta\chi^2_{\rm b}>2$) for seven of the 11 galaxies in which this parameter is allowed to vary (Section~\ref{subsection: Mass model}), including five of the seven golden galaxies listed in Table~\ref{tab: sensitivity fit quality}. These tests indicate that the inferred dark matter fractions are not driven by the low bulge fractions preferred by the fiducial models.}

\section{Baryonic rotation curves from the stellar-mass maps}
\label{section: Baryonic rotation curve modelling}

\begin{figure}
    \centering
    \includegraphics[width=0.99\linewidth]{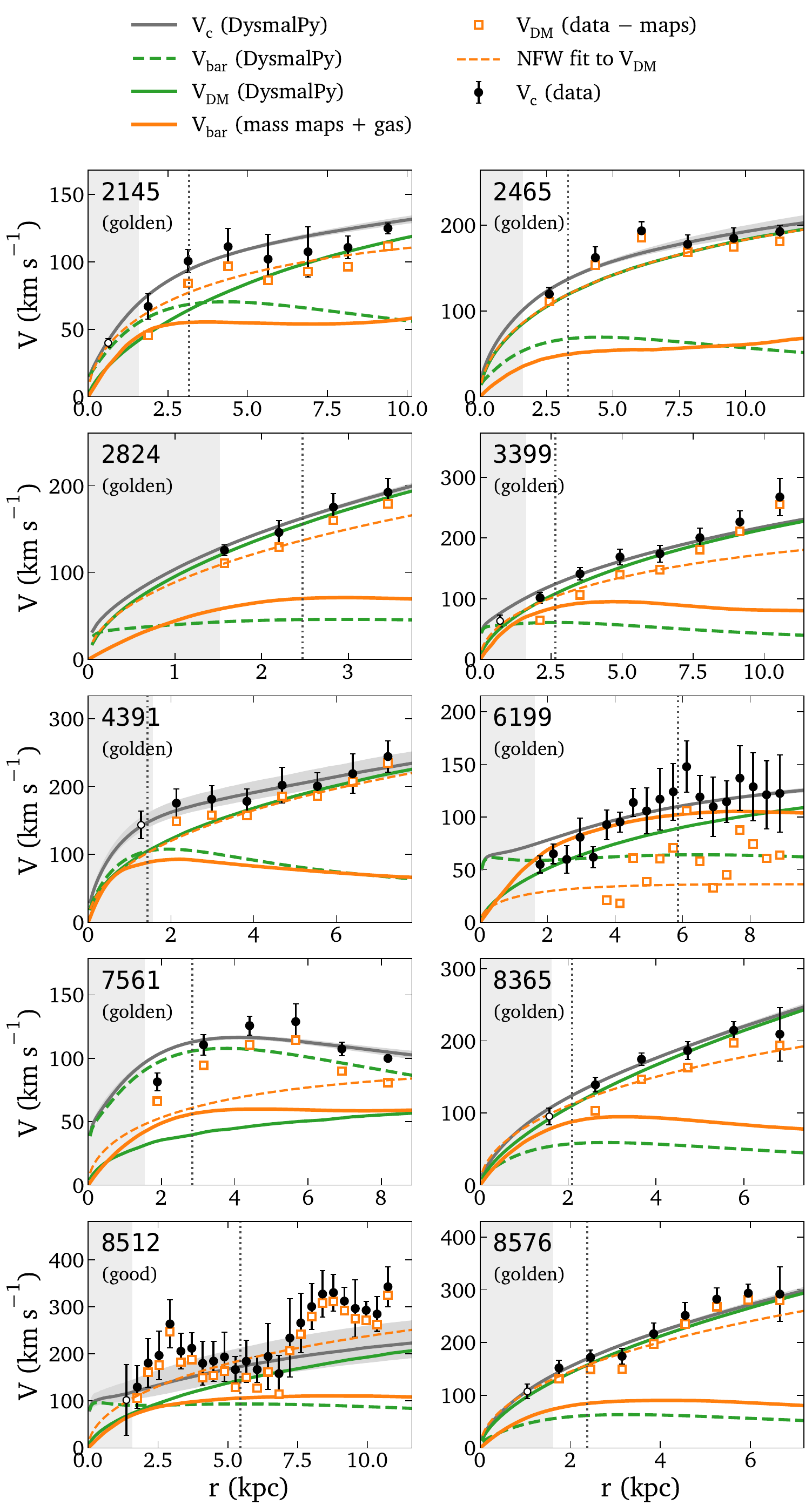}
    \caption{Baryonic rotation curves derived from the stellar-mass maps for the ten galaxies with reliable maps. Black points show the observed circular velocities corrected for inclination, beam smearing and pressure support; open points lie within the beam FWHM (grey band). The solid orange curve shows the baryonic circular velocity derived from the stellar-mass map and the estimated molecular and atomic gas distributions. Open squares indicate the inferred dark matter velocities, $V_{\rm DM}=(V_{\rm c,obs}^2-V_{\rm bar}^2)^{1/2}$, with the corresponding NFW fit shown by the dashed orange curve. The green dashed and solid curves show the baryonic and dark matter contributions from the forward modelling, respectively, while the grey curve and shaded region indicate the total circular velocity and its 16th--84th percentile range. The vertical dotted line marks $R_{\rm e}$.}
    \label{fig: Baryonic rot curves with SFR mosaic}
\end{figure}

\rev{For the ten galaxies with reliable spatially resolved stellar-mass maps (Section~\ref{subsection: MSA-3D data}), we derive an alternative estimate of the baryonic contribution to the circular velocity, without assuming a dark matter halo profile. We deproject each map using the adopted kinematic centre, position angle and inclination, and extract the azimuthally averaged stellar-mass surface-density profile, $\Sigma_\star(r)$. The profiles are normalised to the catalogue stellar masses, such that only the radial distribution is taken from the maps. We include molecular gas masses from the scaling relations of \citet{Tacconi_2018}, distributed like the stellar disk, and atomic gas masses from \citet{Chowdhury_2022}, adopting the radial profile of \citet{Wang_2016}. The circular velocity of each component is calculated following \citet{Casertano_1983}, with a thickness $h_z = 0.25R_s$, which reproduces the baryonic circular velocity of the $q = 0.25$ model disk to $\approx2\%$ beyond $R_{\rm e}$. We compare the resulting baryonic circular velocity, $V_{\rm bar}$, with the observed circular velocity, $V_{\rm c,obs}$, corrected for inclination, beam smearing and pressure support as in the supplementary figures. The baryonic fraction is then estimated as $f_{\rm bar}(R_{\rm e})=V_{\rm bar}^2/V_{\rm c,obs}^2$ (Fig.~\ref{fig: Baryonic rot curves with SFR mosaic}).}

\begin{figure}
    \centering
    \includegraphics[width=0.99\linewidth]{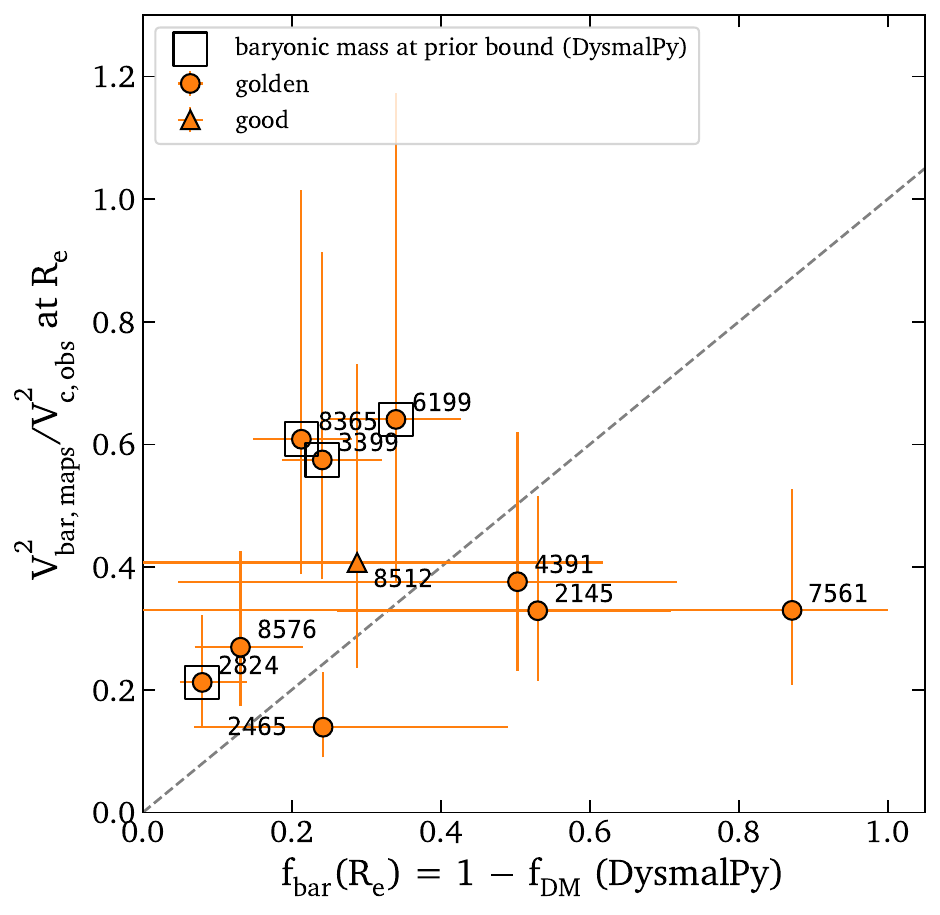}
    \caption{Comparison of the baryonic fractions at $R_{\rm e}$ derived from the stellar-mass maps and gas scaling relations with those inferred from the forward modelling, $1-f_{\rm DM}(R_{\rm e})$. Circles and triangles denote the golden and good samples, respectively, while squares identify galaxies whose forward-model baryonic masses reach the lower prior bound. Uncertainties are truncated at the physical range, and the dashed line indicates the one-to-one relation.}
    \label{fig: fDM dysmalpy vs fDM baryons}
\end{figure}

\rev{The comparison between the map-based and forward-modelled baryonic fractions is shown in Fig.~\ref{fig: fDM dysmalpy vs fDM baryons}. For the six galaxies whose baryonic masses are constrained by the kinematics, the two estimates are generally consistent, with a median difference of $-0.11$. For the remaining four galaxies, whose forward-model baryonic masses reach the lower prior bound, the map-based fractions are higher, by 0.13--0.40, corresponding to lower inferred dark matter fractions. The map-based estimates are not truncated at $f_{\rm bar}=1$. Because the map-based estimates adopt the catalogue stellar and scaling-relation gas masses, they test the spatial distribution of the baryons rather than independently establishing the absolute mass scale, and these offsets largely reflect the different mass normalisation of the two estimates (Section~\ref{subsection: Fit quality and uncertainties}). Moreover, molecular gas contributes 40--60\% of $V_{\rm bar}^2(R_{\rm e})$ in eight of the ten galaxies, making the comparison particularly sensitive to the assumed gas masses. We therefore cannot determine which baryonic mass scale is more accurate. Nevertheless, the comparison indicates that the dark matter fractions of some of the prior-limited galaxies could be lower than the adopted values.}

\rev{Finally, we use the map-based baryonic rotation curves to estimate the dark matter contribution as $V_{\rm DM}=(V_{\rm c,obs}^2-V_{\rm bar}^2)^{1/2}$. The inferred dark matter velocities are consistent with an NFW halo of the adopted concentration for four of the ten galaxies, while two prefer a shallower and four a steeper inner slope, so this test shows no systematic departure from NFW. However, this one-dimensional comparison excludes the innermost, beam-smeared regions and therefore provides weaker constraints on the inner halo slope than the forward modelling.}

\section{Interaction strengths}
\label{section: interaction strengths}

\begin{figure}
    \centering
    \includegraphics[width=0.99\linewidth]{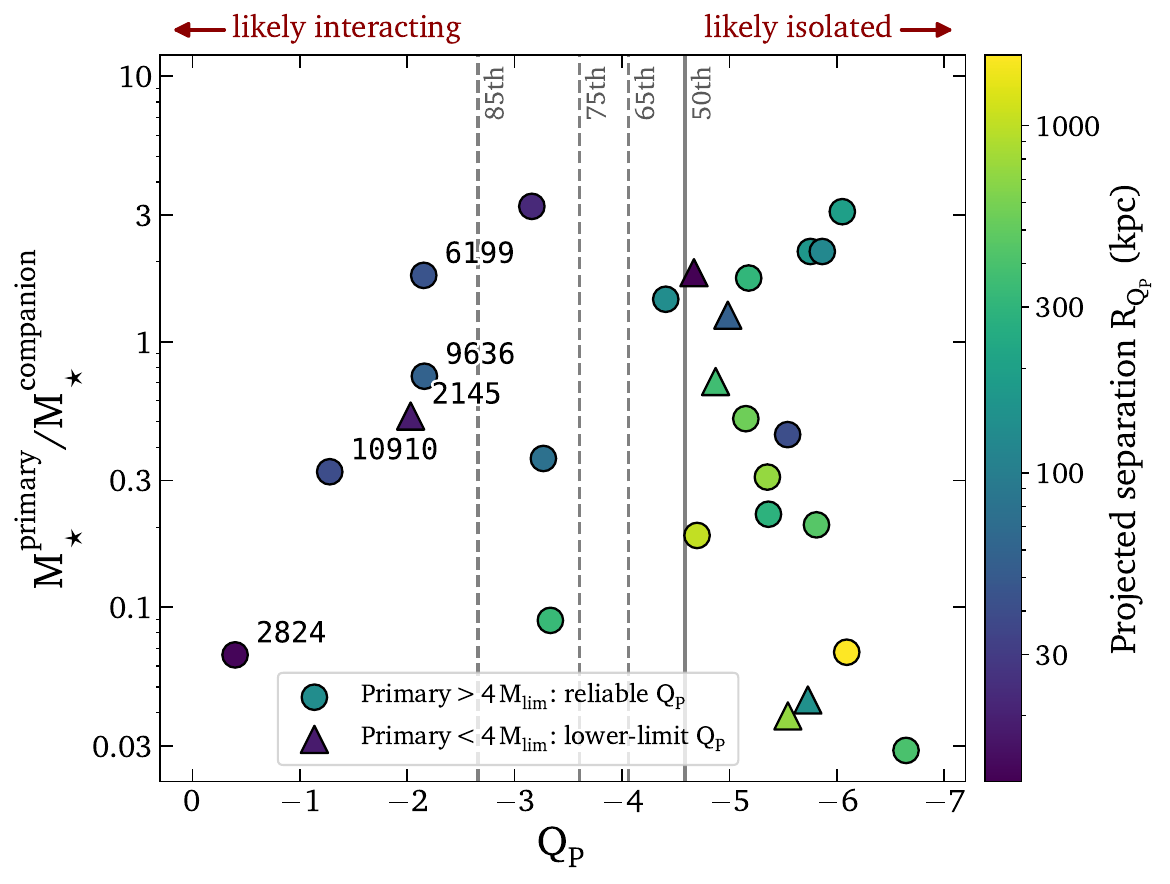}
    \caption{Primary-to-companion stellar-mass ratio as a function of the projected interaction strength, $Q_P$ \citep{Kanowski_2026}, for the 26 MSA-3D galaxies with available measurements. The companion is selected as the one contributing the largest individual interaction strength, and symbols are colour-coded by its projected separation, $R_{Q_P}$. Circles indicate primaries above four times the stellar-mass completeness limit, while triangles indicate lower-mass primaries for which $Q_P$ is a lower limit. Vertical lines mark the 50th (solid), 65th, 75th and 85th (dashed) percentiles of the mass-complete 3D-HST reference population. Galaxies above the 85th percentile are classified as likely interacting and are labelled.}
    \label{fig: interaction strength}
\end{figure}

\rev{Figure~\ref{fig: interaction strength} shows the projected interaction strength, $Q_P$, and the stellar-mass ratio between each MSA-3D galaxy and its dominant companion. Following \citet{Kanowski_2026}, we classify galaxies above the 85th percentile of the mass-complete 3D-HST reference population as likely interacting. The reference population and the companion search are restricted to galaxies above the 90\% stellar-mass completeness limit of 3D-HST, which increases from $\log(M_\star/M_\odot)\approx8.5$ at $z\approx0.65$ to $\approx8.8$ at $z=1$ and $\approx9.4$ at $z=1.7$ \citep{Kanowski_2026}. For each primary, $Q_P$ is computed from neighbours with at least a quarter of its stellar mass, so that 1:4 or more equal-mass interactions are selected uniformly regardless of primary mass. Five of the 26 galaxies with $Q_P$ measurements meet this criterion (\texttt{2145}, \texttt{2824}, \texttt{6199}, \texttt{9636} and \texttt{10910}), all belonging to the golden sample. Their dominant companions have projected separations of 14--58~kpc and span a wide range of stellar-mass ratios. The companions of \texttt{9636} and \texttt{10910} have spectroscopic redshifts, while those of \texttt{2145}, \texttt{2824} and \texttt{6199} have grism redshifts. For \texttt{2145}, $Q_P$ is a lower limit because the primary lies below four times the stellar-mass completeness limit. Seventeen galaxies fall below the median of the reference distribution and are considered likely isolated.}

\rev{We find no significant association between interaction strength and rotation-curve shape. Of the five likely interacting galaxies, only two have sufficient radial coverage for classification (Section~\ref{section: Rotation curve shapes}): \texttt{2145} has a flat outer rotation curve and \texttt{9636} a falling one. The $Q_P$ distributions do not differ significantly between rotation-curve classes (Kruskal--Wallis $p=0.74$; Fig.~\ref{fig: rot curves vs params}). Likewise, $Q_P$ shows no significant correlation with dark matter fraction, inner halo slope or rotational support (Spearman $|\rho|\le0.24$, $p\ge0.29$ for the golden sample).}

\begin{figure}
    \centering
    \includegraphics[width=0.99\linewidth]{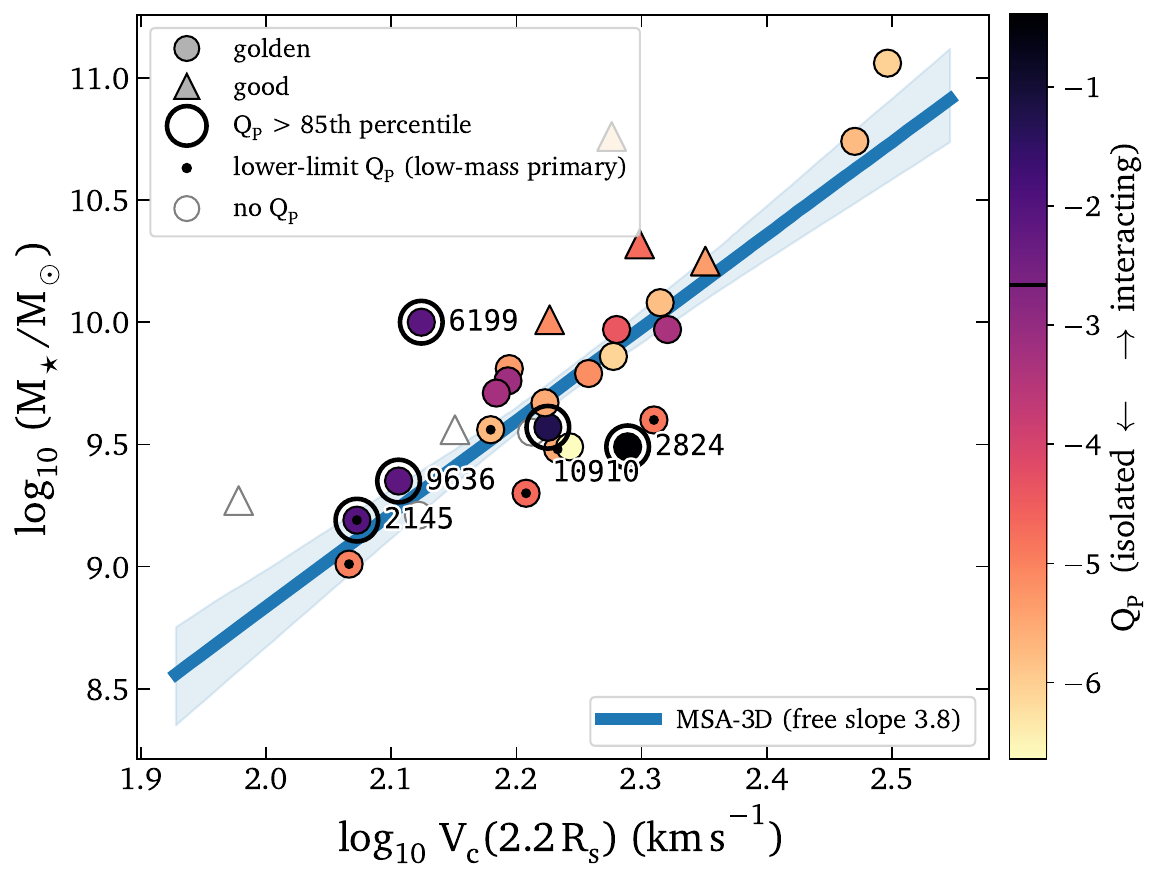}
    \caption{Stellar Tully--Fisher relation from Fig.~\ref{fig: Tully--Fisher}, colour-coded by the projected interaction strength, $Q_P$. Circles and triangles denote the golden and good samples, respectively. Large open circles highlight likely interacting galaxies ($Q_P$ above the 85th percentile, marked by the black line in the colour bar), while small black dots indicate lower limits on $Q_P$. Open grey symbols represent galaxies without $Q_P$ measurements. The blue line and shaded region show the free-slope fit to the golden sample.}
    \label{fig: Tully--Fisher QP}
\end{figure}

\rev{The likely interacting galaxies show no systematic displacement from the stellar Tully--Fisher relation of the remaining sample (Fig.~\ref{fig: Tully--Fisher QP}), and $Q_P$ is uncorrelated with velocity residuals at fixed stellar mass (Spearman $\rho=-0.01$). The two largest outliers, \texttt{6199} ($-0.18$~dex) and \texttt{2824} ($+0.12$~dex), are both classified as likely interacting but deviate in opposite directions. Interactions may therefore contribute to the scatter without producing a systematic offset. Galaxy \texttt{6199} has a companion at a projected separation of 45~kpc, although its grism-redshift uncertainty ($\sim10^3$~km\,s$^{-1}$) prevents us from establishing a physical association. The galaxy's velocity field remains regular and is well reproduced by the rotating-disk model. Its imaging inclination ($33^\circ$) satisfies our selection criterion, although its fitted inclination and baryonic mass reach their lower prior bounds. Excluding \texttt{6199} changes the golden-sample median $f_{\rm DM}(R_{\rm e})$ from 0.73 to 0.75. These conclusions remain limited by the small number of likely interacting galaxies and the projected nature of $Q_P$.}


\section{Intrinsic velocity dispersion and the line-spread function}
\label{section: sigma0 LSF}

\begin{figure*}
    \centering
    \includegraphics[width=0.98\linewidth]{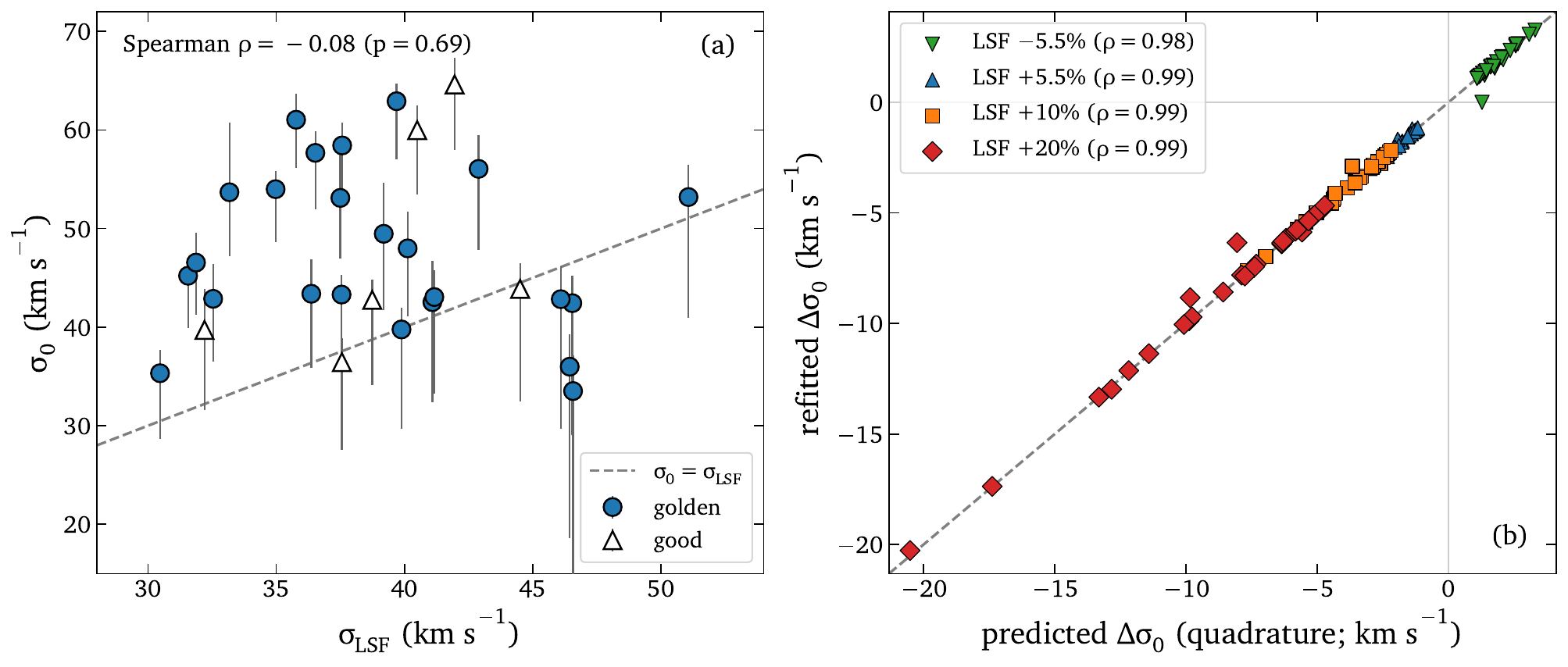}
    \caption{Sensitivity of the inferred intrinsic velocity dispersions to the instrumental resolution. (a) $\sigma_0$ as a function of the LSF width at the observed wavelength of each galaxy \citep{Shajib_2025}, including total uncertainties. Filled circles and open triangles denote the golden and good samples, respectively, and the dashed line indicates $\sigma_0=\sigma_{\rm LSF}$. (b) Changes in $\sigma_0$ after refitting with the LSF width varied by $-5.5\%$, $+5.5\%$, $+10\%$ and $+20\%$, compared with the first-order expectation from subtracting the instrumental broadening in quadrature, $\Delta\sigma_0\simeq-\sigma_{\rm LSF}\Delta\sigma_{\rm LSF}/\sigma_0$. The dashed line marks the one-to-one relation, and the Spearman correlation coefficients are indicated in the legend.}
    \label{fig: sigma0 LSF}
\end{figure*}

\rev{Figure~\ref{fig: sigma0 LSF} presents our tests of the sensitivity of the inferred intrinsic velocity dispersion to the instrumental line-spread function (LSF). Despite $\sigma_{\rm LSF}$ varying by a factor of 1.7 across the sample, $\sigma_0$ shows no correlation with the LSF width at the observed wavelength (Spearman $\rho=-0.08$, $p=0.69$; panel a), and seven galaxies have $\sigma_0<\sigma_{\rm LSF}$ without reaching the lower prior bound. We further refitted the galaxies after varying the LSF width by $\pm5.5\%$, corresponding to the $1\sigma$ uncertainty of the \citet{Shajib_2025} calibration, and broadening it by 10\% and 20\% to account for the effect of spatially extended emission. The resulting changes in $\sigma_0$ closely follow those expected from quadrature subtraction of the instrumental broadening (Spearman $\rho\geq0.97$; panel b). A 20\% broader LSF lowers $\sigma_0$ by a median of $\approx7$~km\,s$^{-1}$ and by up to 20~km\,s$^{-1}$ for galaxies with $\sigma_0\lesssim\sigma_{\rm LSF}$, although these broader LSFs are mildly disfavoured by the fits (median $\Delta\chi^2_{\rm b}=+4.0$; Table~\ref{tab: sensitivity fit quality}). Finally, injection--recovery tests with an input dispersion of $0.7\,\sigma_{\rm LSF}$ (median 29~km\,s$^{-1}$) recover $\sigma_0$ with a median bias of only $+1.2$~km\,s$^{-1}$ for the golden sample, with few refits reaching the prior bound (Appendix~\ref{section: injection recovery}). Together, these tests show that the inferred dispersions are not set by the instrumental resolution and can be constrained below $\sigma_{\rm LSF}$, while uncertainty in the adopted LSF remains the dominant systematic on their absolute values.}

\section{Uncertainty calibration}
\label{section: error budget}

\rev{We calibrate the posterior intervals using two ensembles of 400 synthetic realisations per galaxy, generated around the maximum-likelihood model. In the \emph{block} ensemble, the standardised velocity and dispersion residuals are jointly resampled in square two-dimensional blocks, with a size given by the larger of four pixels and the residual correlation length. The resampled residuals are then scaled by the error maps and added to the model maps. In the \emph{noise} ensemble, Gaussian noise consistent with the adopted likelihood is added instead. Each realisation is refitted using a maximum-a-posteriori optimiser with the same forward model, masks, error maps, likelihood, and priors as the fiducial fits. Because the maps are PSF-convolved and their residuals are spatially correlated, the absolute $\chi^2$ is not expected to equal the number of degrees of freedom \citep{Andrae_2010}, so we use $\Delta\chi^2_b$ only to compare models fitted to the same data.}

\rev{For each parameter, we define a calibration factor $f=\max(1,w_{\rm block}/w_{\rm noise})$, where $w$ is the 16th--84th percentile width of the corresponding distribution of refitted values. The posterior intervals are multiplied by this factor, with derived quantities recalculated for each realisation and calibrated in the same way. The median calibration factors for the golden sample are 1.9 for $\sigma_0$, 2.4 for $\alpha$, 2.4 for $f_{\rm DM}(R_{\rm e})$, and 2.8 for $V_{\rm rot}(2.2R_{\rm s})$. Parameters whose posteriors reach a prior bound are reported as limits. We adopt the sample-median calibration factor for parameters strongly correlated with another parameter at a prior bound, for parameters whose refits reach a bound, or when fewer than five independent residual blocks are available.}

\rev{Systematic uncertainties are estimated from shifts in the parameter medians obtained under the alternative modelling assumptions described in Section~\ref{subsection: Fit quality and uncertainties}. For each test, we refit both the alternative and fiducial models with the same MAP optimiser, data, masks, error maps, and likelihood. To compare their fit quality, we calculate $\Delta\chi_{\rm b}^2=\Delta\chi_{\rm data}^2/N_{\rm beam}=-2\Delta\ln\mathcal{L}_{\rm data}$, where $\Delta$ denotes the alternative model minus the re-optimised fiducial model and the likelihood is evaluated at the two MAP solutions. The priors affect the fitted parameters but are not included in $\Delta\chi_b^2$; this quantity therefore measures the relative fit to the maps and is not by itself a model-selection statistic.}

\rev{For each parameter, positive and negative systematic shifts are combined separately in quadrature and then added in quadrature to the calibrated statistical uncertainty on the corresponding side. For galaxy \texttt{13416}, whose alternative refits do not reproduce the fiducial solution, we adopt the sample-median systematic uncertainty. Table~\ref{tab: stat sys} reports the statistical and systematic contributions separately.}

\begin{table*}[ht]
\centering
\caption{Statistical and systematic uncertainties of the main fitted and derived parameters. The first and second uncertainties correspond to the calibrated statistical and systematic components, respectively, which are combined in quadrature on each side to obtain the total uncertainties reported in Table~\ref{tab: Fitted parameters}. $V_{\rm rot}$ and $V_{\rm rot}/\sigma_0$ are evaluated at $2.2R_{\rm s}$, and $\log M_{\rm bar}$ denotes the fitted baryonic mass. Values without uncertainties are reported as limits.}
\begin{tabular}{lcccccc}
\toprule
ID & $\sigma_0$ & $V_{\rm rot}(2.2R_{\rm s})$ & $V_{\rm rot}/\sigma_0$ & $f_{\rm DM}(R_{\rm e})$ & $\alpha$ & $\log M_{\rm bar}$ \\
 & (km\,s$^{-1}$) & (km\,s$^{-1}$) &  &  &  & ($M_\odot$) \\
\midrule
\multicolumn{7}{c}{\textbf{Golden sample}}\\
\hline
\texttt{2111} & $53.2\,^{+1.7}_{-1.7}\,^{+2.9}_{-12.1}$ & $154\,^{+4}_{-4}\,^{+12}_{-10}$ & $2.90\,^{+0.12}_{-0.10}\,^{+0.88}_{-0.25}$ & $0.71\,^{+0.06}_{-0.06}\,^{+0.05}_{-0.11}$ & $0.74\,^{+0.10}_{-0.09}\,^{+0.05}_{-0.42}$ & $10.05\,^{+0.10}_{-0.11}\,^{+0.16}_{-0.09}$ \\
\texttt{2145} & $43.3\,^{+1.0}_{-1.0}\,^{+1.8}_{-7.8}$ & $76\,^{+6}_{-6}\,^{+10}_{-5}$ & $1.75\,^{+0.14}_{-0.14}\,^{+0.44}_{-0.12}$ & $0.47\,^{+0.13}_{-0.12}\,^{+0.24}_{-0.13}$ & $0.59\,^{+0.18}_{-0.19}\,^{+0.12}_{-0.37}$ & $9.82\,^{+0.10}_{-0.11}\,^{+0.13}_{-0.22}$ \\
\texttt{2465} & $43.4\,^{+1.7}_{-1.7}\,^{+3.0}_{-7.3}$ & $133\,^{+4}_{-4}\,^{+5}_{-6}$ & $3.07\,^{+0.15}_{-0.15}\,^{+0.63}_{-0.30}$ & $0.76\,^{+0.15}_{-0.20}\,^{+0.09}_{-0.15}$ & $0.99\,^{+0.13}_{-0.18}\,^{+0.06}_{-0.11}$ & $9.82\,^{+0.33}_{-0.34}\,^{+0.22}_{-0.21}$ \\
\texttt{2824} & $42.5\,^{+2.2}_{-2.3}\,^{+3.6}_{-9.9}$ & $173\,^{+8}_{-8}\,^{+15}_{-14}$ & $4.06\,^{+0.37}_{-0.33}\,^{+1.26}_{-0.49}$ & $0.92\,^{+0.03}_{-0.05}\,^{+0.01}_{-0.04}$ & $0.89\,^{+0.09}_{-0.09}\,^{+0.12}_{-0.21}$ & $<9.43$ \\
\texttt{3399} & $54.0\,^{+1.1}_{-1.1}\,^{+1.5}_{-5.3}$ & $108\,^{+5}_{-5}\,^{+9}_{-8}$ & $2.00\,^{+0.09}_{-0.09}\,^{+0.27}_{-0.18}$ & $0.76\,^{+0.01}_{-0.01}\,^{+0.05}_{-0.08}$ & $0.82\,^{+0.13}_{-0.13}\,^{+0.08}_{-0.32}$ & $<9.61$ \\
\texttt{4391} & $49.5\,^{+1.6}_{-1.5}\,^{+5.0}_{-7.6}$ & $136\,^{+46}_{-28}\,^{+52}_{-19}$ & $2.75\,^{+0.77}_{-0.47}\,^{+1.30}_{-0.57}$ & $0.50\,^{+0.25}_{-0.20}\,^{+0.38}_{-0.08}$ & $1.01\,^{+0.31}_{-0.31}\,^{+0.49}_{-0.14}$$^{f}$ & $9.84\,^{+0.29}_{-0.32}\,^{+0.20}_{-0.39}$ \\
\texttt{6199} & $45.2\,^{+1.1}_{-1.1}\,^{+1.2}_{-5.1}$ & $93\,^{+17}_{-18}\,^{+9}_{-18}$ & $2.06\,^{+0.37}_{-0.38}\,^{+0.33}_{-0.40}$ & $0.66\,^{+0.02}_{-0.02}\,^{+0.09}_{-0.08}$ & $0.68\,^{+0.55}_{-0.55}\,^{+0.29}_{-0.68}$$^{f}$ & $<9.99$ \\
\texttt{6430} & $58.4\,^{+2.1}_{-2.1}\,^{+1.0}_{-5.9}$ & $133\,^{+19}_{-20}\,^{+15}_{-14}$ & $2.28\,^{+0.35}_{-0.32}\,^{+0.37}_{-0.22}$ & $0.83\,^{+0.06}_{-0.28}\,^{+0.17}_{-0.28}$$^{f}$ & $0.64\,^{+0.16}_{-0.23}\,^{+0.17}_{-0.31}$ & $9.82\,^{+0.54}_{-0.15}\,^{+0.27}_{-0.51}$ \\
\texttt{7314} & $61.0\,^{+0.7}_{-0.7}\,^{+2.6}_{-4.9}$ & $101\,^{+7}_{-7}\,^{+14}_{-15}$ & $1.65\,^{+0.11}_{-0.10}\,^{+0.26}_{-0.29}$ & $0.62\,^{+0.31}_{-0.32}\,^{+0.22}_{-0.09}$ & $0.92\,^{+0.15}_{-0.15}\,^{+0.07}_{-0.20}$ & $9.66\,^{+0.32}_{-0.40}\,^{+0.12}_{-0.31}$ \\
\texttt{7561} & $48.0\,^{+2.5}_{-2.6}\,^{+2.7}_{-6.4}$ & $86\,^{+10}_{-10}\,^{+6}_{-16}$ & $1.79\,^{+0.24}_{-0.24}\,^{+0.26}_{-0.39}$ & $0.13\,^{+0.87}_{-0.13}\,^{+0.51}_{-0.02}$$^{f}$ & $0.90\,^{+0.60}_{-0.90}\,^{+0.60}_{-0.00}$$^{f}$ & $10.14\,^{+0.09}_{-0.29}\,^{+0.06}_{-0.39}$ \\
\texttt{8365} & $35.3\,^{+1.6}_{-1.7}\,^{+1.8}_{-6.4}$ & $132\,^{+3}_{-3}\,^{+10}_{-9}$ & $3.73\,^{+0.19}_{-0.17}\,^{+0.86}_{-0.32}$ & $0.79\,^{+0.03}_{-0.06}\,^{+0.06}_{-0.04}$ & $0.68\,^{+0.07}_{-0.06}\,^{+0.10}_{-0.12}$ & $<9.53$ \\
\texttt{8576} & $46.5\,^{+1.4}_{-1.4}\,^{+2.7}_{-5.1}$ & $179\,^{+8}_{-9}\,^{+37}_{-25}$ & $3.85\,^{+0.24}_{-0.20}\,^{+0.93}_{-0.64}$ & $0.87\,^{+0.05}_{-0.07}\,^{+0.03}_{-0.04}$ & $0.82\,^{+0.10}_{-0.08}\,^{+0.06}_{-0.22}$ & $9.60\,^{+0.25}_{-0.21}\,^{+0.12}_{-0.13}$ \\
\texttt{8942} & $53.1\,^{+0.6}_{-0.6}\,^{+5.5}_{-6.2}$ & $153\,^{+7}_{-4}\,^{+25}_{-4}$ & $2.89\,^{+0.14}_{-0.08}\,^{+0.55}_{-0.19}$ & $0.69\,^{+0.01}_{-0.02}\,^{+0.08}_{-0.05}$ & $0.77\,^{+0.17}_{-0.17}\,^{+0.06}_{-0.16}$ & $<9.78$ \\
\texttt{9424} & $43.1\,^{+1.7}_{-1.3}\,^{+2.2}_{-9.7}$ & $127\,^{+5}_{-5}\,^{+11}_{-9}$ & $2.95\,^{+0.17}_{-0.15}\,^{+0.89}_{-0.25}$ & $0.71\,^{+0.05}_{-0.05}\,^{+0.07}_{-0.20}$ & $0.42\,^{+0.04}_{-0.04}\,^{+0.06}_{-0.33}$ & $9.84\,^{+0.03}_{-0.02}\,^{+0.32}_{-0.14}$ \\
\texttt{9636} & $33.5\,^{+1.4}_{-1.5}\,^{+3.3}_{-20.3}$ & $106\,^{+3}_{-3}\,^{+6}_{-6}$ & $3.17\,^{+0.17}_{-0.14}\,^{+4.75}_{-0.33}$ & $0.75\,^{+0.10}_{-0.13}\,^{+0.09}_{-0.08}$ & $0.69\,^{+0.12}_{-0.13}\,^{+0.04}_{-0.29}$ & $9.60\,^{+0.19}_{-0.20}\,^{+0.14}_{-0.20}$ \\
\texttt{9812} & $42.4\,^{+0.7}_{-0.7}\,^{+2.7}_{-13.3}$ & $189\,^{+5}_{-4}\,^{+34}_{-22}$ & $4.46\,^{+0.13}_{-0.12}\,^{+2.19}_{-0.59}$ & $0.44\,^{+0.23}_{-0.18}\,^{+0.16}_{-0.18}$ & $0.79\,^{+0.24}_{-0.25}\,^{+0.08}_{-0.29}$ & $10.59\,^{+0.14}_{-0.21}\,^{+0.18}_{-0.17}$ \\
\texttt{9960} & $42.9\,^{+2.3}_{-2.3}\,^{+2.7}_{-5.9}$ & $300\,^{+6}_{-5}\,^{+8}_{-7}$ & $7.01\,^{+0.34}_{-0.29}\,^{+1.13}_{-0.53}$ & $0.33\,^{+0.12}_{-0.11}\,^{+0.14}_{-0.11}$ & $1.28\,^{+0.18}_{-0.19}\,^{+0.16}_{-0.19}$ & $10.83\,^{+0.08}_{-0.09}\,^{+0.13}_{-0.15}$ \\
\texttt{10910} & $36.0\,^{+1.0}_{-1.0}\,^{+3.1}_{-17.4}$ & $150\,^{+3}_{-4}\,^{+12}_{-10}$ & $4.17\,^{+0.16}_{-0.16}\,^{+3.87}_{-0.43}$ & $0.87\,^{+0.09}_{-0.14}\,^{+0.05}_{-0.08}$ & $1.06\,^{+0.08}_{-0.10}\,^{+0.12}_{-0.21}$ & $9.79\,^{+0.37}_{-0.35}\,^{+0.21}_{-0.22}$ \\
\texttt{11225} & $62.9\,^{+1.2}_{-1.3}\,^{+1.3}_{-5.8}$ & $102\,^{+5}_{-4}\,^{+49}_{-5}$ & $1.63\,^{+0.06}_{-0.06}\,^{+0.79}_{-0.08}$ & $0.79\,^{+0.05}_{-0.10}\,^{+0.07}_{-0.12}$ & $0.85\,^{+0.12}_{-0.12}\,^{+0.02}_{-0.12}$ & $<9.80$ \\
\texttt{11843} & $53.7\,^{+3.8}_{-4.1}\,^{+6.0}_{-5.0}$ & $273\,^{+7}_{-8}\,^{+13}_{-15}$ & $5.09\,^{+0.39}_{-0.34}\,^{+0.58}_{-0.67}$ & $0.69\,^{+0.07}_{-0.12}\,^{+0.02}_{-0.07}$ & $0.87\,^{+0.13}_{-0.12}\,^{+0.04}_{-0.17}$ & $<10.52$ \\
\texttt{12015} & $57.7\,^{+1.7}_{-1.9}\,^{+1.3}_{-5.4}$ & $167\,^{+9}_{-9}\,^{+16}_{-13}$ & $2.90\,^{+0.21}_{-0.17}\,^{+0.40}_{-0.23}$ & $0.75\,^{+0.08}_{-0.13}\,^{+0.05}_{-0.05}$ & $0.71\,^{+0.11}_{-0.09}\,^{+0.21}_{-0.18}$ & $<9.89$ \\
\texttt{12239} & $56.1\,^{+2.2}_{-2.3}\,^{+2.6}_{-7.9}$ & $129\,^{+11}_{-11}\,^{+15}_{-12}$ & $2.31\,^{+0.23}_{-0.21}\,^{+0.45}_{-0.24}$ & $0.85\,^{+0.08}_{-0.12}\,^{+0.05}_{-0.04}$ & $0.84\,^{+0.12}_{-0.11}\,^{+0.08}_{-0.20}$ & $9.43\,^{+0.29}_{-0.25}\,^{+0.12}_{-0.19}$ \\
\texttt{18586} & $42.8\,^{+1.8}_{-1.9}\,^{+2.8}_{-13.0}$ & $74\,^{+5}_{-5}\,^{+11}_{-7}$ & $1.73\,^{+0.18}_{-0.16}\,^{+0.77}_{-0.21}$ & $0.67\,^{+0.13}_{-0.18}\,^{+0.14}_{-0.15}$ & $0.87\,^{+0.17}_{-0.15}\,^{+0.08}_{-0.21}$ & $9.26\,^{+0.23}_{-0.24}\,^{+0.19}_{-0.25}$ \\
\texttt{29470} & $39.8\,^{+0.9}_{-0.9}\,^{+2.0}_{-10.1}$ & $128\,^{+11}_{-8}\,^{+10}_{-9}$ & $3.22\,^{+0.46}_{-0.40}\,^{+1.10}_{-0.27}$ & $0.83\,^{+0.02}_{-0.03}\,^{+0.05}_{-0.04}$ & $0.56\,^{+0.05}_{-0.05}\,^{+0.11}_{-0.29}$ & $<9.53$ \\
\hline
\multicolumn{7}{c}{\textbf{Good sample}}\\
\hline
\texttt{8512} & $42.8\,^{+0.9}_{-0.9}\,^{+1.8}_{-8.6}$ & $177\,^{+144}_{-92}\,^{+76}_{-54}$ & $4.15\,^{+2.79}_{-1.77}\,^{+2.08}_{-1.29}$ & $0.71\,^{+0.25}_{-0.28}\,^{+0.15}_{-0.17}$ & $0.83\,^{+0.49}_{-0.56}\,^{+0.37}_{-0.39}$ & $10.27\,^{+0.37}_{-0.35}\,^{+0.12}_{-0.29}$ \\
\texttt{9337} & $36.5\,^{+1.3}_{-1.3}\,^{+2.1}_{-8.8}$ & $57\,^{+21}_{-24}\,^{+16}_{-42}$ & $1.55\,^{+0.47}_{-0.53}\,^{+0.43}_{-1.09}$ & $0.69\,^{+0.08}_{-0.10}\,^{+0.17}_{-0.21}$ & $<0.16$ & $<9.59$ \\
\texttt{10107} & $60.0\,^{+1.4}_{-1.4}\,^{+2.0}_{-6.4}$ & $112\,^{+12}_{-12}\,^{+50}_{-18}$ & $1.87\,^{+0.19}_{-0.20}\,^{+0.85}_{-0.30}$ & $0.35\,^{+0.33}_{-0.35}\,^{+0.26}_{-0.35}$$^{f}$ & $>0.28$ & $10.28\,^{+0.32}_{-0.30}\,^{+0.51}_{-0.33}$ \\
\texttt{13182} & $64.6\,^{+1.9}_{-2.0}\,^{+1.9}_{-6.4}$ & $179\,^{+173}_{-117}\,^{+111}_{-51}$ & $2.79\,^{+2.12}_{-1.55}\,^{+1.76}_{-0.77}$ & $0.60\,^{+0.40}_{-0.60}\,^{+0.32}_{-0.35}$$^{f}$ & $1.03\,^{+0.47}_{-0.65}\,^{+0.47}_{-0.74}$$^{f}$ & $10.39\,^{+0.47}_{-0.42}\,^{+0.15}_{-0.41}$ \\
\texttt{13416} & $39.7\,^{+3.3}_{-3.4}\,^{+2.6}_{-7.3}$ & $169\,^{+7}_{-6}\,^{+13}_{-12}$ & $4.27\,^{+0.36}_{-0.34}\,^{+0.79}_{-0.30}$ & $0.03\,^{+0.02}_{-0.01}\,^{+0.09}_{-0.03}$$^{f}$ & $<0.18$ & $10.48\,^{+0.04}_{-0.04}\,^{+0.13}_{-0.19}$ \\
\texttt{18188} & $43.9\,^{+1.2}_{-1.2}\,^{+2.3}_{-11.4}$ & $108\,^{+27}_{-14}\,^{+9}_{-4}$ & $2.46\,^{+0.66}_{-0.35}\,^{+0.79}_{-0.14}$ & $0.69\,^{+0.17}_{-0.25}\,^{+0.25}_{-0.05}$ & $0.60\,^{+0.43}_{-0.48}\,^{+0.18}_{-0.52}$ & $10.07\,^{+0.30}_{-0.30}\,^{+0.03}_{-0.40}$ \\
\bottomrule
\end{tabular}

\tablefoot{
\tablefoottext{a}{Sample-median statistical calibration factor (parameter correlated with, or refits reaching, a prior bound).}\\
\tablefoottext{b}{Sample-median statistical calibration factor (fewer than five independent residual blocks).}\\
\tablefoottext{c}{Sample-median systematic uncertainty (alternative refits unavailable).}\\
\tablefoottext{d}{Bootstrap width unstable between the two halves of the refits.}\\
\tablefoottext{e}{Statistical uncertainty estimated from the scatter of the injection--recovery tests.}\\
\tablefoottext{f}{Uncertainties truncated at the physical range of $f_{\rm DM}$ (0--1) or the prior range of $\alpha$ (0--1.5).}
}
\label{tab: stat sys}
\end{table*}

\begin{table*}[ht]
\centering
\setlength{\tabcolsep}{3.6pt}
\caption{Fit quality of the alternative models used in the systematic error budget and the robustness tests, relative to the re-optimised fiducial model, for the golden sample. $\Delta\chi^2_{\rm b}=\Delta\chi^2/N_{\rm beam}=-2\Delta\ln\mathcal{L}$ (positive values indicate worse fits); ``worse'' and ``better'' give the number of galaxies with $\Delta\chi^2_{\rm b}>2$ and $<-2$. The last two columns give the median changes in $f_{\rm DM}(R_{\rm e})$ and $\alpha$.}
\begin{tabular}{lrrrrrr}
\hline
Test & $N$ & median $\Delta\chi^2_{\rm b}$ & worse & better & $\Delta f_{\rm DM}(R_{\rm e})$ & $\Delta\alpha$ \\
\hline
PSF $0\farcs185\times0\farcs11$ (first principles) & 24 & $+3.1$ & 14 & 6 & $+0.008$ & $-0.04$ \\
PSF $0\farcs20\times0\farcs08$ & 24 & $+2.5$ & 13 & 9 & $+0.002$ & $-0.07$ \\
Concentration free (0.11~dex prior)$^{a}$ & 24 & $-4.0$ & 0 & 17 & $-0.002$ & $-0.10$ \\
Concentration, $r_{-2}$ convention & 24 & $-3.7$ & 2 & 17 & $-0.015$ & $-0.11$ \\
$R_{\rm e,bulge}=0.5$~kpc & 24 & $+0.7$ & 10 & 6 & $+0.006$ & $-0.00$ \\
$R_{\rm e,bulge}=2$~kpc & 24 & $-0.6$ & 6 & 10 & $-0.007$ & $+0.00$ \\
No atomic gas & 24 & $-0.3$ & 1 & 5 & $+0.013$ & $+0.01$ \\
LSF $\times0.945$ & 24 & $-0.6$ & 1 & 6 & $+0.003$ & $+0.00$ \\
LSF $\times1.055$ & 24 & $+1.0$ & 8 & 0 & $-0.003$ & $-0.00$ \\
LSF $\times1.10$ & 24 & $+1.8$ & 12 & 2 & $-0.006$ & $-0.01$ \\
LSF $\times1.20$ & 24 & $+4.0$ & 15 & 3 & $-0.014$ & $-0.02$ \\
Centre $+1\sigma$ ($x$) & 24 & $+12.4$ & 14 & 10 & $-0.003$ & $-0.02$ \\
Centre $-1\sigma$ ($x$) & 24 & $+13.0$ & 16 & 8 & $+0.011$ & $-0.00$ \\
Centre $+1\sigma$ ($y$) & 24 & $+16.5$ & 16 & 7 & $+0.002$ & $+0.00$ \\
Centre $-1\sigma$ ($y$) & 24 & $-1.0$ & 10 & 11 & $+0.006$ & $-0.00$ \\
Inclination, $b/a+0.05$$^{e}$ & 18 & $+2.0$ & 9 & 4 & $+0.008$ & $+0.03$ \\
Inclination, $b/a-0.05$$^{e}$ & 18 & $-0.5$ & 6 & 5 & $-0.005$ & $-0.01$ \\
$R_{\rm e,disk}\times0.9$ & 24 & $+0.3$ & 8 & 3 & $-0.012$ & $+0.00$ \\
$R_{\rm e,disk}\times1.1$ & 24 & $-0.2$ & 4 & 5 & $+0.011$ & $+0.00$ \\
$B/T=0.2$ fixed & 24 & $+4.0$ & 14 & 7 & $+0.005$ & $+0.00$ \\
$B/T=0.3$ fixed & 24 & $+4.4$ & 15 & 6 & $+0.029$ & $-0.00$ \\
$B/T=0.4$ fixed & 24 & $+10.2$ & 17 & 6 & $+0.054$ & $-0.01$ \\
$B/T=0.5$ fixed & 24 & $+14.9$ & 18 & 6 & $+0.074$ & $-0.02$ \\
$B/T$ fixed at the measured value$^{b}$ & 7 & $+20.2$ & 5 & 0 & $-0.000$ & $+0.00$ \\
$M_{\rm vir}$ within 1~dex of the \citet{Moster_2013} relation$^{d}$ & 24 & $+15.9$ & 16 & 0 & $+0.001$ & $+0.13$ \\
$M_{\rm vir}$ within 1~dex of the \citet{Moster_2013} relation, concentration free$^{a,d}$ & 24 & $-0.5$ & 8 & 8 & $+0.003$ & $-0.06$ \\
\hline
Baryonic-mass prior $+0.5$~dex$^{c}$ & 24 & $+13.8$ & 17 & 0 & $-0.201$ & $-0.13$ \\
Free kinematic centre$^{c}$ & 24 & $-105$ & 0 & 22 & $-0.008$ & $+0.10$ \\
\hline
\end{tabular}

\tablefoot{
\tablefoottext{a}{One additional free parameter.}\\
\tablefoottext{b}{Galaxies with free $B/T$ in the fiducial model.}\\
\tablefoottext{c}{From the maximum-likelihood solutions of the full nested-sampling fits rather than MAP refits; these tests are not included in the systematic uncertainties (Sections~\ref{subsection: Robustness} and \ref{section: Dark matter fractions}).}\\
\tablefoottext{d}{Robustness test of the halo normalisation (Section~\ref{subsection: Inner slope of the dark matter halo}); not included in the systematic uncertainties.}\\
\tablefoottext{e}{Galaxies with a fixed inclination; for the remaining galaxies this uncertainty is included in the posterior.}
}
\label{tab: sensitivity fit quality}
\end{table*}

\section{Injection--recovery tests}
\label{section: injection recovery}

\rev{To assess the reliability of the inferred inner halo slopes and intrinsic velocity dispersions, we performed injection--recovery tests for three input models per galaxy: $\alpha=1$ (NFW), $\alpha=0.5$, and $\sigma_0=0.7\,\sigma_{\rm LSF}$, keeping all other parameters fixed at their maximum-likelihood values (for the NFW null test, see below). For each model, we generated 100 block-resampled and 100 Gaussian-noise realisations, following Appendix~\ref{section: error budget}, together with one noise-free realisation, and refitted them using the same procedure. Each realisation is refitted with the maximum a posteriori optimiser used for the uncertainty calibration (Appendix~\ref{section: error budget}), and a realisation is counted as favouring $\alpha<1$ if its refitted slope plus twice the adopted upper uncertainty of the corresponding galaxy lies below unity. This approximates, but does not fully repeat, the posterior-based procedure applied to the data.}

\rev{Mocks built by setting $\alpha=1$ in the fiducial solution retain its large halo normalisation and therefore rotate more strongly than the observed galaxies (by a median of 0.2~dex in $V_{\rm rot}(R_{\rm e})$). For the null test, we therefore construct the NFW mocks from the maximum-likelihood solution of the matched NFW fit of each galaxy, which reproduces the observed velocity fields as well as the fiducial model, and refit them with the generalised NFW model. For the golden sample, 9\% of the block-resampled realisations (2.5\% of the Gaussian-noise realisations) are classified as having $\alpha<1$ at $2\sigma$, corresponding to $\approx2.2$ expected false detections among the 24 galaxies, compared with the five observed. Treating each galaxy as an independent trial with its own false-positive rate, the number of false detections follows a Poisson-binomial distribution, which gives a probability of 0.013 for five or more if all halos followed NFW profiles (a single pooled rate of 9\% would give 0.06). The false-positive rate varies strongly between galaxies: it reaches 24--90\% for \texttt{9424}, \texttt{6430}, \texttt{29470} and \texttt{2145}, which include the two galaxies flagged for large residuals (Section~\ref{subsection: Fit quality and uncertainties}), is 10--16\% for \texttt{3399} and \texttt{11225}, and is $\leq2\%$ for the remaining 18 galaxies, including \texttt{2111}, \texttt{8365} and \texttt{9636}. Non-circular residual structure can therefore bias the inferred slope low in individual galaxies: of the five detections, those in \texttt{2111}, \texttt{8365}, and \texttt{9636} remain individually robust, whereas those in \texttt{9424} and \texttt{29470} do not.}

\rev{The median recovered slope is 0.94, partly because the priors pull the noise-free solutions slightly below the input value (median 0.95). The block-resampled realisations include the observed residual structure, but since the mocks are built from the same model used in the fits, these tests do not assess the impact of systematic model mismatch. For an input slope of $\alpha=0.5$, the median recovered value is 0.63, with approximately half of the realisations significantly favouring $\alpha<1$ at $2\sigma$. The priors therefore pull the recovered slopes towards intermediate values, biasing shallow input slopes high and NFW slopes slightly low; the latter effect is included in the false-positive rate quoted above. Finally, intrinsic dispersions below the instrumental resolution (median input 29~km\,s$^{-1}$) are recovered with a median bias of only $+1.2$~km\,s$^{-1}$; the refits reach the prior bound in more than 10\% of the realisations for only four golden galaxies. The sample-median bias in $f_{\rm DM}(R_{\rm e})$ is within 0.05 in all three tests ($-0.04$, $+0.01$ and $-0.01$ for the original $\alpha=1$, $\alpha=0.5$ and low-$\sigma_0$ inputs), although individual galaxies show median biases of up to $\approx0.4$.}

\section{Fits with a free kinematic centre}
\label{section: free centre}

\rev{For the fiducial models, we compared four independent centre estimates measured in the frame of the kinematic maps: the centroid of the line-free continuum in the NIRSpec cube, the centroid in the reddest available NIRCam image, the photometric centre used in our previous analysis, and the solution from an initial free-centre kinematic fit. Based on their overall consistency with the light distribution and the symmetry of the velocity fields, we adopt the centroid in the reddest available NIRCam image (F444W or F356W) for the 26 galaxies with suitable coverage, and the centroid of the line-free continuum for the remaining four. The independent light-based centroid estimates typically differ by $0\farcs05$--$0\farcs07$.}

\rev{To assess the impact of fixing the kinematic centre, we refitted all galaxies, allowing their position to vary within $\pm0\farcs24$ of the adopted centre, while keeping all other modelling settings unchanged. The free-centre fits reach substantially higher likelihoods (median $\Delta\chi^2_{\rm b}=-105$ for the golden sample, based on the maximum-likelihood solutions of the full fits) and reduce the median velocity residuals from 18.0 to 15.4~km\,s$^{-1}$. We nevertheless retain the fixed centres for three reasons.}

\rev{First, the improved fits often require large centre displacements: in eleven galaxies, the preferred centre lies more than one PSF FWHM from the nearest light centroid, and in five it reaches the prior boundary. By contrast, perturbing the fixed centre by the centroid uncertainty degrades the fit in three of four directions (Table~\ref{tab: sensitivity fit quality}). This suggests that the free centre is not refining the gravitational centre, but instead allows the symmetric disk model to absorb asymmetric velocity structure from, e.g., lopsidedness, clumps, spiral arms, or non-circular motions. Consistently, the improvement correlates strongly with the excess-residual statistic used to flag poorly fitting galaxies in Section~\ref{subsection: Fit quality and uncertainties} (Spearman $\rho=0.84$, $p<10^{-3}$) and is among the largest for \texttt{6430} and \texttt{9424}, the two galaxies flagged there, whereas eight galaxies improve by less than $\Delta\chi^2_{\rm b}=30$. Second, the likelihood does not account for spatially correlated residuals, which broaden the recovered parameter distributions by factors of 2--3 (Appendix~\ref{section: error budget}); consequently, $\Delta\chi^2_{\rm b}$ overstates the significance of the improvement. Third, the free-centre posteriors are implausibly narrow, with a median uncertainty of $0\farcs014$, well below the $0\farcs05$--$0\farcs07$ differences among independent light-based centroid estimates. Their formal precision therefore likely reflects sensitivity to residual velocity structure rather than a constraint on the gravitational centre.}

\rev{The inferred dynamical parameters are only moderately affected. Relative to the free-centre fits, the fixed-centre models yield slightly higher dark matter fractions (median difference $+0.01$; median absolute difference 0.03) and shallower inner slopes (median difference $-0.10$), while $\sigma_0$ remains unchanged. Adopting the calibration factors and systematic uncertainties of the fiducial fits, four of the 24 golden galaxies have $\alpha<1$ at $2\sigma$ with a free centre (\texttt{2111}, \texttt{8365}, \texttt{9636} and \texttt{11225}; the first three are also identified in the fiducial fits) and none has $\alpha>1$, compared with five and none, respectively, for the fixed-centre models. The three galaxies whose shallow slopes are robust in the injection tests (Appendix~\ref{section: injection recovery}) are therefore also recovered with a free centre, whereas \texttt{9424} and \texttt{29470} are not. We therefore prefer the independently determined fixed centres as the more physically motivated and conservative choice, while including the uncertainty associated with the adopted centre in the systematic error budget (Section~\ref{subsection: Fit quality and uncertainties}).}

\newif\ifsupplement
\supplementtrue  

\ifsupplement

    \clearpage


\onecolumn

\setcounter{figure}{0}\renewcommand{\thefigure}{S\arabic{figure}}

\begin{center}


\vspace{1.2em}

{\Large\bfseries
Supplementary material: Diagnostic figures for individual galaxies
\par}

\vspace{1.2em}




\end{center}


\vspace{1.5em}


This supplementary material presents the full set of diagnostic figures for the galaxies analysed in the main paper. For each galaxy, the figures combine imaging, two-dimensional kinematic maps, one-dimensional kinematic profiles, the dynamical model and its posterior distributions, providing a comprehensive assessment of the dynamical modelling. Since all figures follow the same layout, the panels and plotting conventions are described here once for reference.

\textbf{Top row:} From left to right, the panels show the emission-line flux map (H$\alpha$, or [O\,\textsc{iii}]$\lambda5007$ for \texttt{13182}), the observed line-of-sight velocity field, the observed velocity-dispersion map, the reference image of the galaxy (Section~\ref{subsection: MSA-3D data}; the filter is indicated in the panel), the stellar-mass surface-density map derived from spatially resolved SED modelling, and the colour composite. The reference image and colour composite are shown when \textit{JWST} imaging is available, and the stellar-mass map when it could be derived. Only the ten stellar-mass maps listed in Section~\ref{subsection: MSA-3D data} are used quantitatively in the analysis. White contours trace the continuum morphology in the reference image and are overlaid on the H$\alpha$, velocity, velocity-dispersion and reference-image panels to facilitate comparison between the ionised-gas kinematics and the stellar distribution. The galaxy ID and sample classification (golden or good) are indicated in the first panel, while the filters assigned to the blue, green and red channels of the colour composite are listed on the image. The purple rectangle on the velocity and velocity-dispersion maps indicates the pseudo-slit along the kinematic major axis used to extract the line-of-sight velocity (LOSV) profiles. The white ellipse in the flux panel shows the FWHM of the PSF ($0\farcs19\times0\farcs16$).

\textbf{Bottom left:} The LOSV profiles show the observed line-of-sight velocity (blue circles) and LSF-corrected velocity dispersion (orange triangles) as a function of offset along the kinematic major axis. The horizontal dashed line indicates the mean outer-disk dispersion, $\langle\sigma_{\rm corr}\rangle$, used to centre the prior on the intrinsic dispersion in the dynamical modelling. The grey shaded region marks the central, PSF-dominated area excluded from this estimate, while the horizontal and vertical dotted lines indicate the systemic velocity and fixed kinematic centre, respectively.

\textbf{Bottom right:} The upper inset compares the observed velocity field with the best-fitting model and shows the corresponding residual map (data minus model). The black cross marks the fixed kinematic centre, and the dashed line indicates the kinematic major axis. The main panel shows the circular velocities inferred from the data after correcting for pressure support and beam smearing (Section~\ref{subsection: Priors and modelling assumptions}), both for individual pixels (grey points) and in radial bins (blue circles, with the 16th--84th percentile range shaded). These are compared with the total (solid red), baryonic (dashed orange) and dark matter (dashed green) circular velocity curves of the best-fitting model. The grey shaded region marks the innermost radii affected by beam smearing, the vertical dotted line indicates $R_{\rm e}$, and the horizontal dashed line indicates $\sigma_0$. The annotated values give the dark matter fraction within the disk effective radius, $f_{\rm DM}(R_{\rm e})$, and the intrinsic velocity dispersion, $\sigma_0$. The lower panel shows the velocity residuals, $\Delta V = V_{\rm data}-V_{\rm model}$, as a function of radius.

\textbf{Posterior panel:} The corner plot shows the posterior distributions of the fitted parameters and the derived dark matter fraction, $f_{\rm DM}(R_{\rm e})$, with the median and 16th--84th percentile interval reported above each histogram. The inset shows the position--velocity diagram extracted from the observed H$\alpha$ cube ([O\,\textsc{iii}] for \texttt{13182}) along the kinematic major axis, with contours from the maximum-likelihood model cube extracted in the same pseudo-slit and pixels as the data, and the corresponding model line-of-sight velocity curve overlaid.

%

\newcommand{\cornerkind}{full} 

%

\newcommand{\galpage}[2]{%
\begin{figure}[p!]
\centering
\setlength{\abovecaptionskip}{3pt}
\setlength{\belowcaptionskip}{0pt}

\includegraphics[
    width=0.92\textwidth,
    height=0.23\textheight,
    keepaspectratio
]{figures/maps/#1_v14.pdf}

\vspace{0.3em}

\begin{minipage}[t]{0.43\textwidth}
    \centering
    \includegraphics[
        width=\linewidth,
        height=0.24\textheight,
        keepaspectratio
    ]{figures/LOSV_diagrams/#1_v14.pdf}
\end{minipage}%
\hspace{-0.052\textwidth}%
\begin{minipage}[t]{0.59\textwidth}
    \centering
    \includegraphics[
        width=\linewidth,
        height=0.24\textheight,
        keepaspectratio
    ]{figures/rot_curves/#1_v14.pdf}
\end{minipage}

\vspace{0.3em}

\includegraphics[
    width=0.98\textwidth,
    height=0.46\textheight,
    keepaspectratio
]{figures/pv_diagrams/\cornerkind/#1_v14.pdf}

\caption{Diagnostic summary plots for galaxy~\texttt{#1} (#2 sample).}
\label{fig:supp-#1-panel}

\end{figure}

\FloatBarrier
}


\galpage{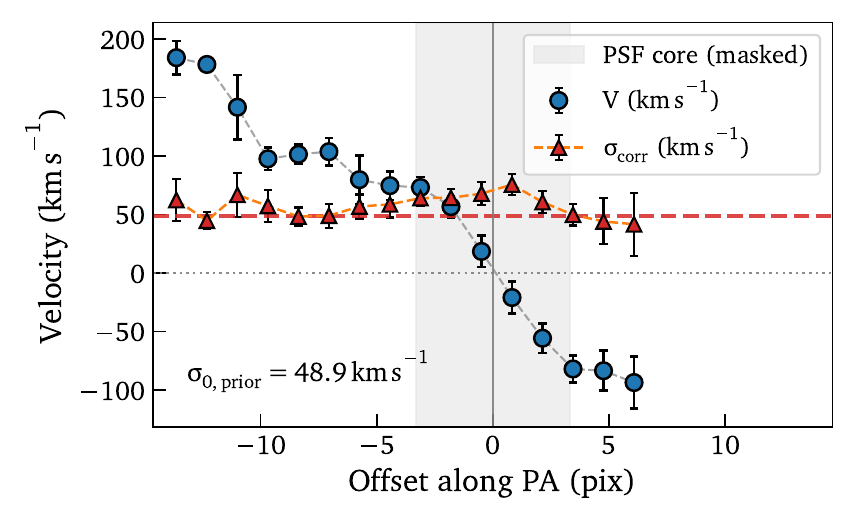}{golden}
\galpage{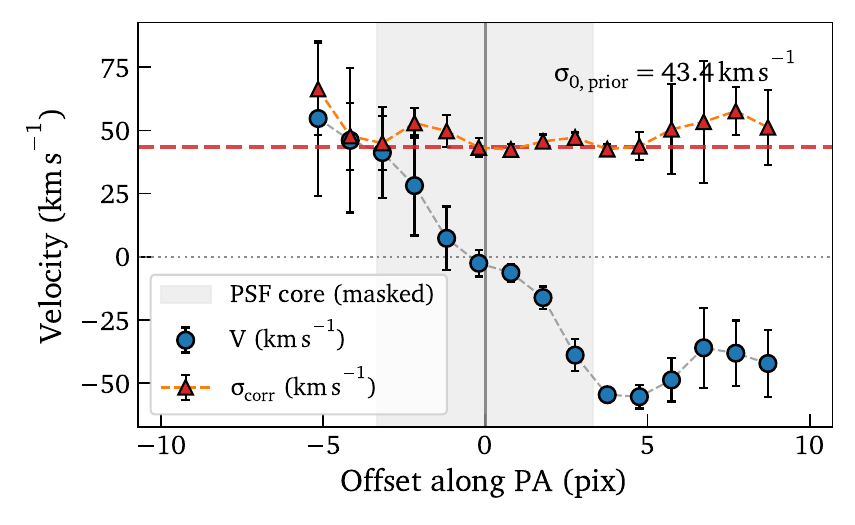}{golden}
\galpage{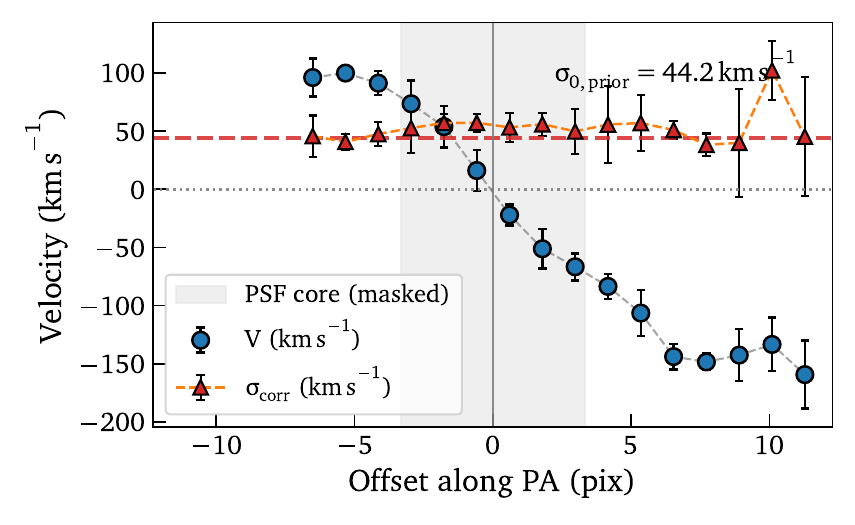}{golden}
\galpage{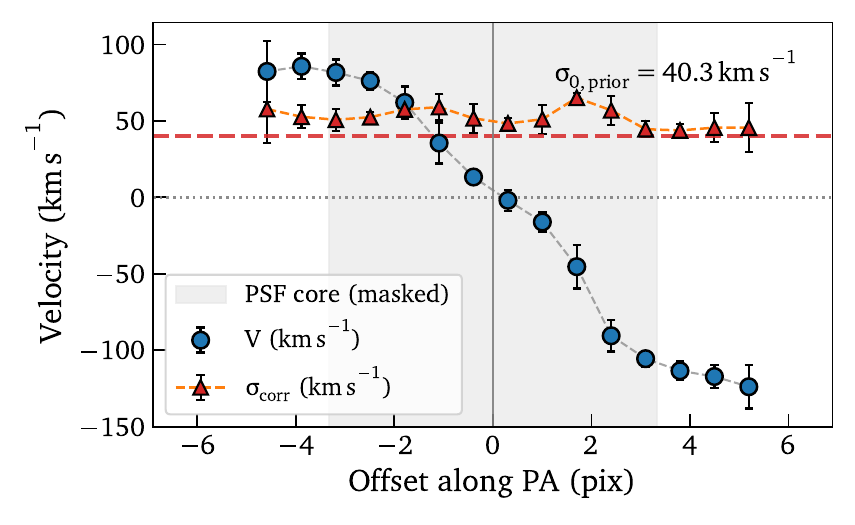}{golden}
\galpage{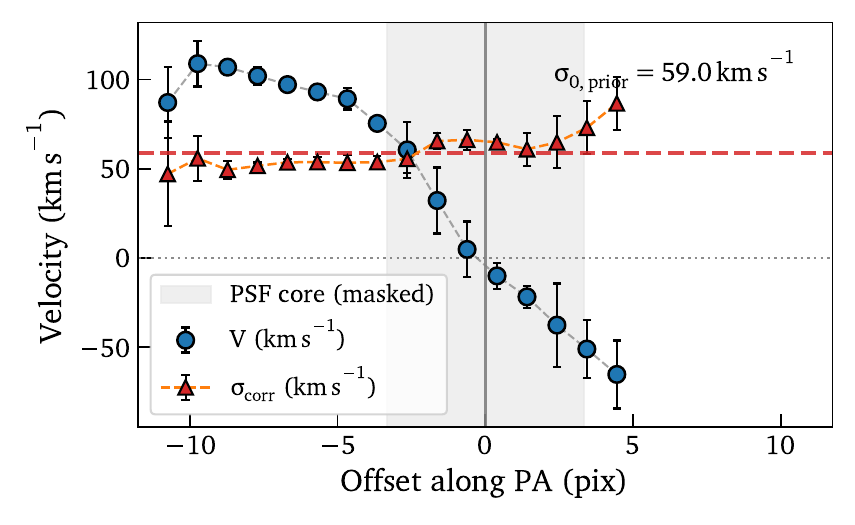}{golden}
\galpage{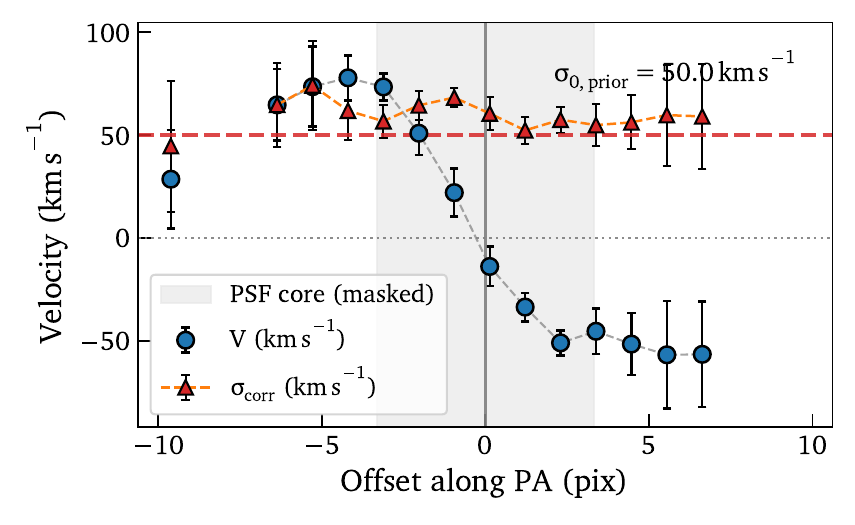}{golden}
\galpage{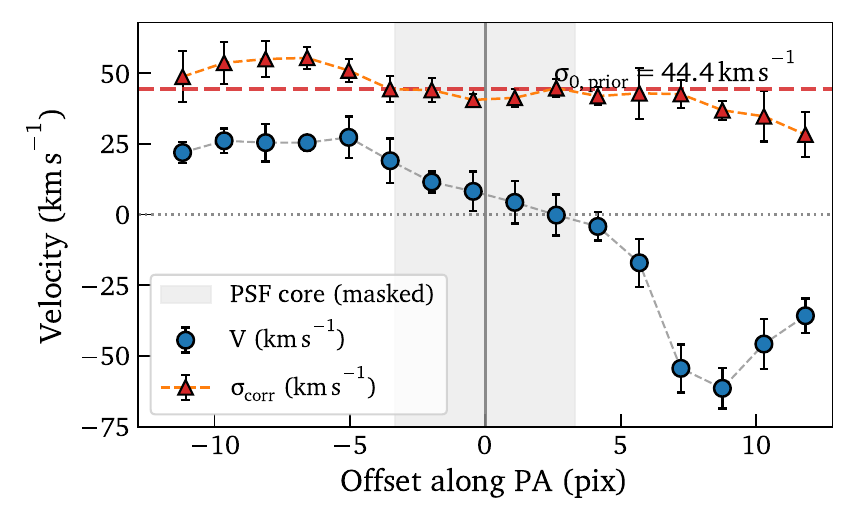}{golden}
\galpage{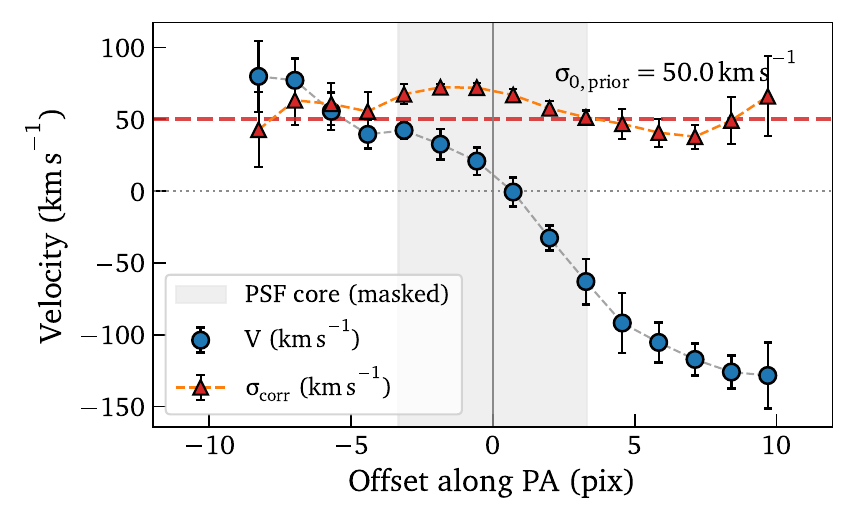}{golden}
\galpage{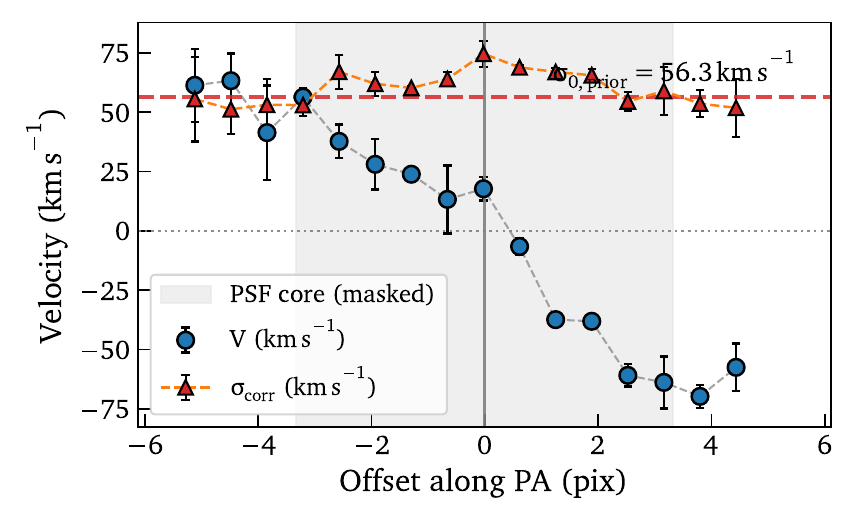}{golden}
\galpage{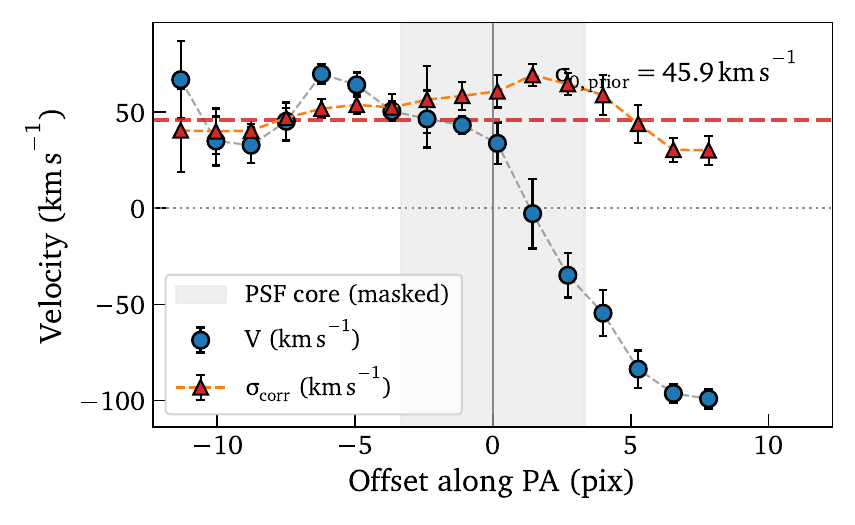}{golden}
\galpage{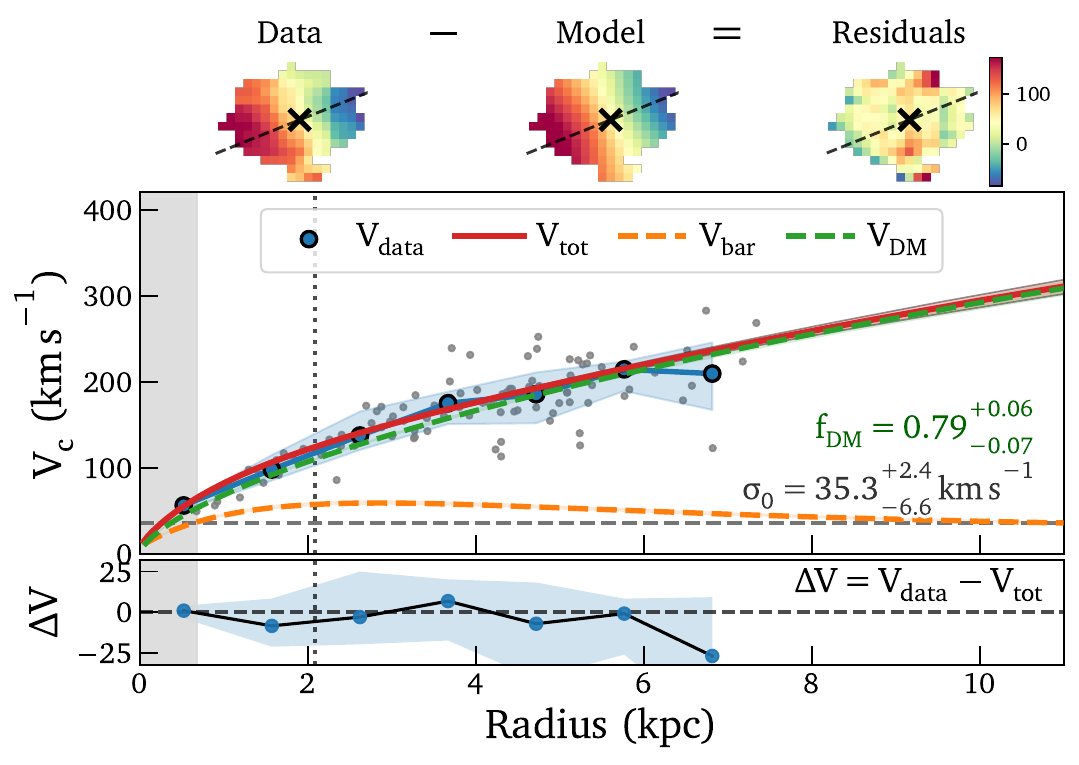}{golden}
\galpage{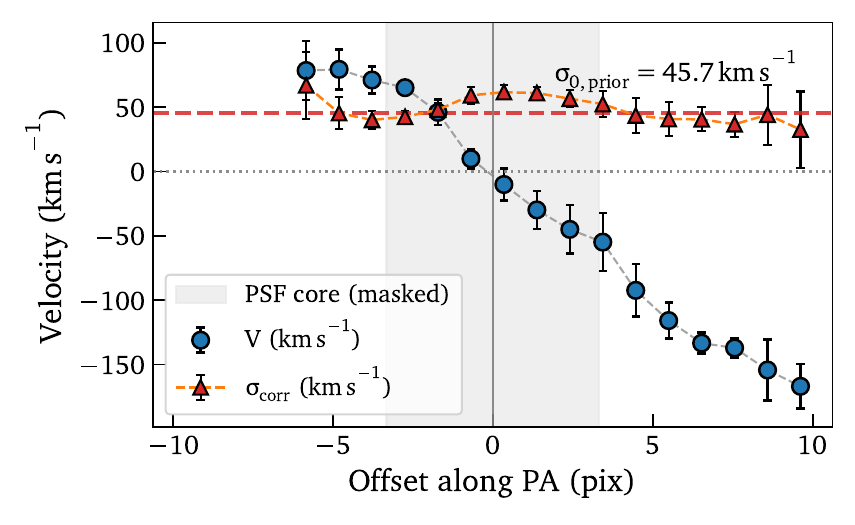}{golden}
\galpage{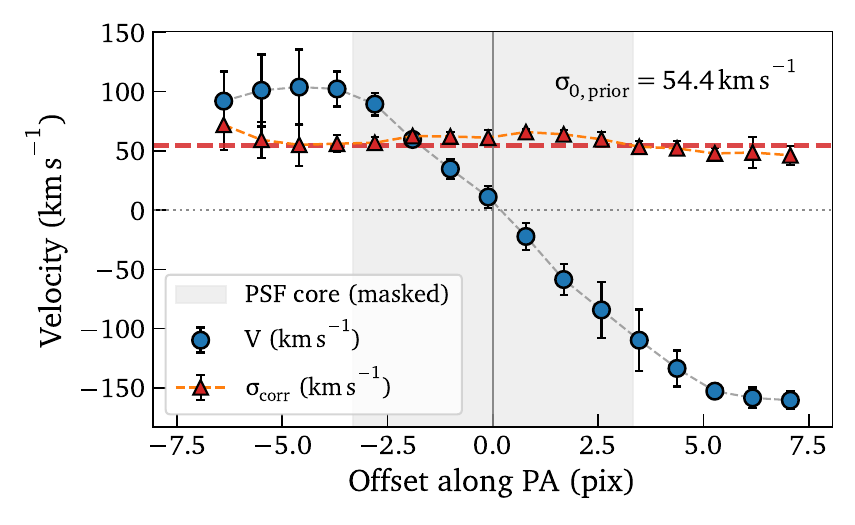}{golden}
\galpage{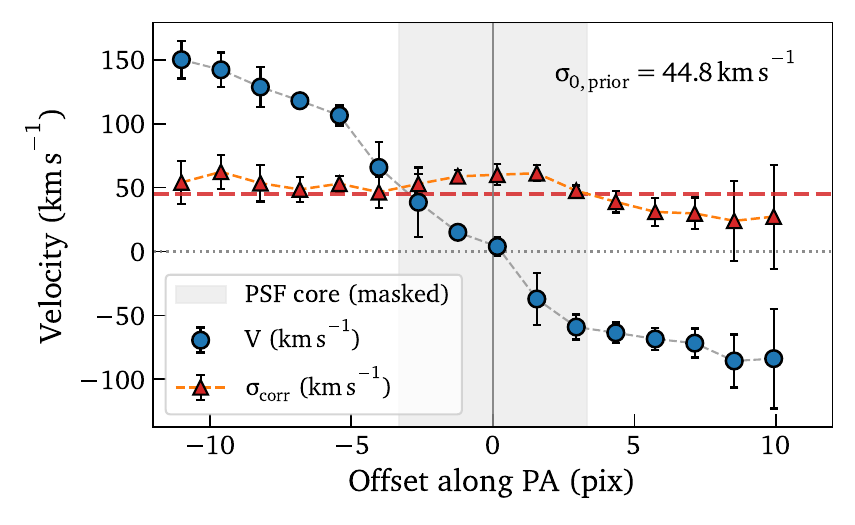}{golden}
\galpage{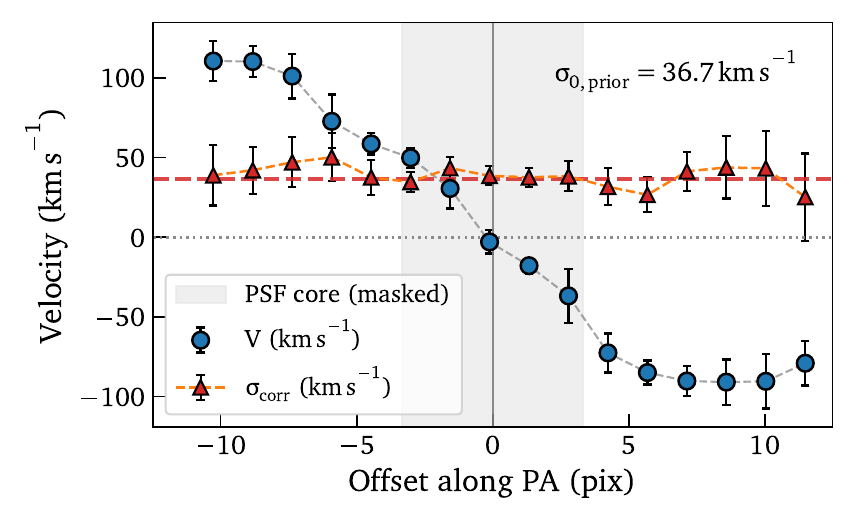}{golden}
\galpage{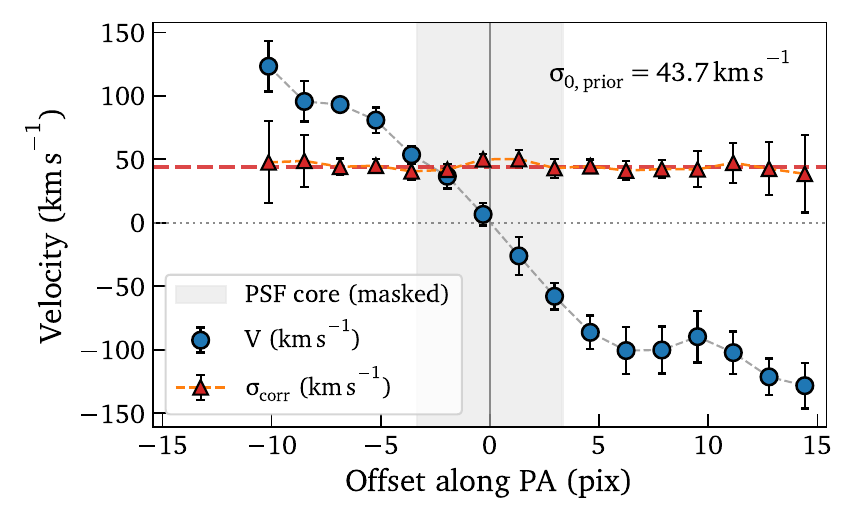}{golden}
\galpage{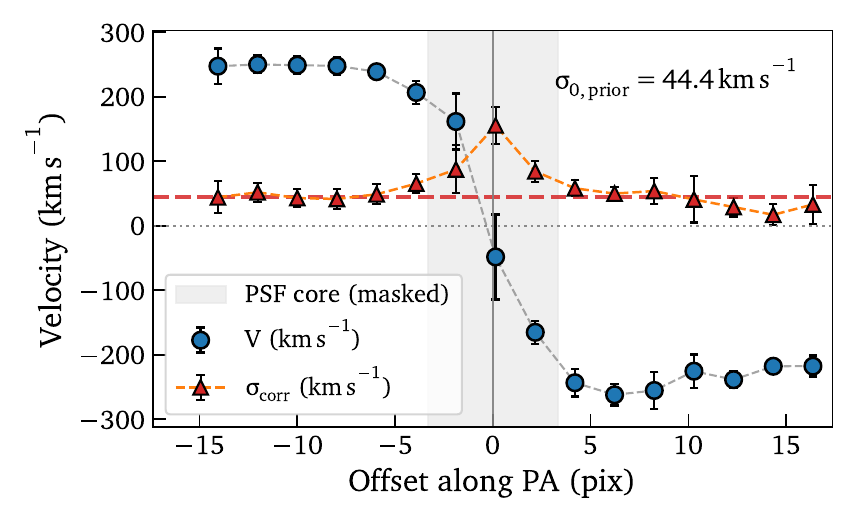}{golden}
\galpage{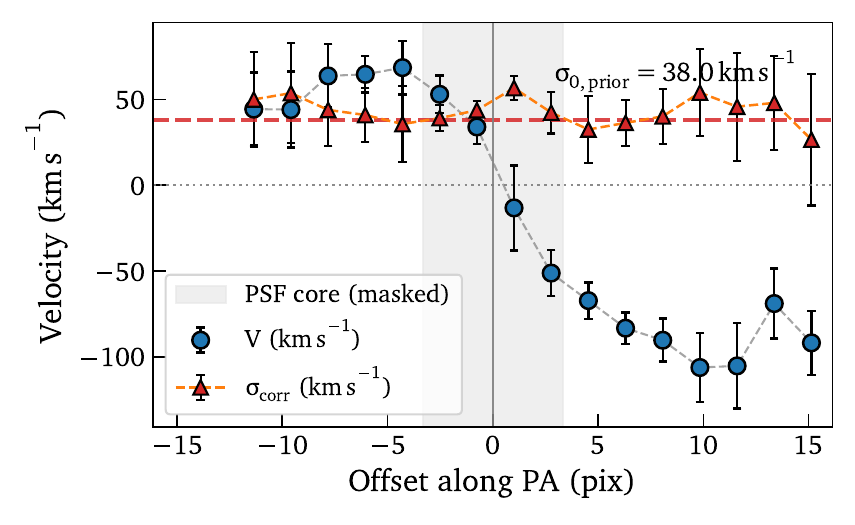}{golden}
\galpage{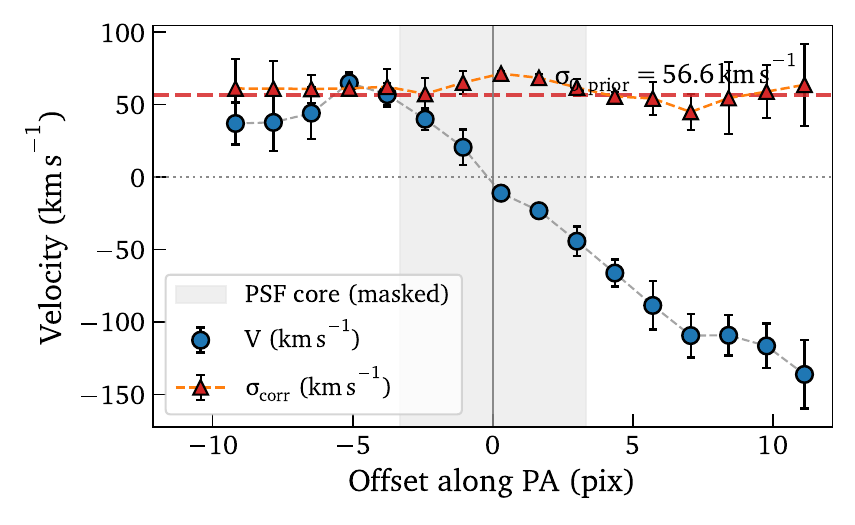}{golden}
\galpage{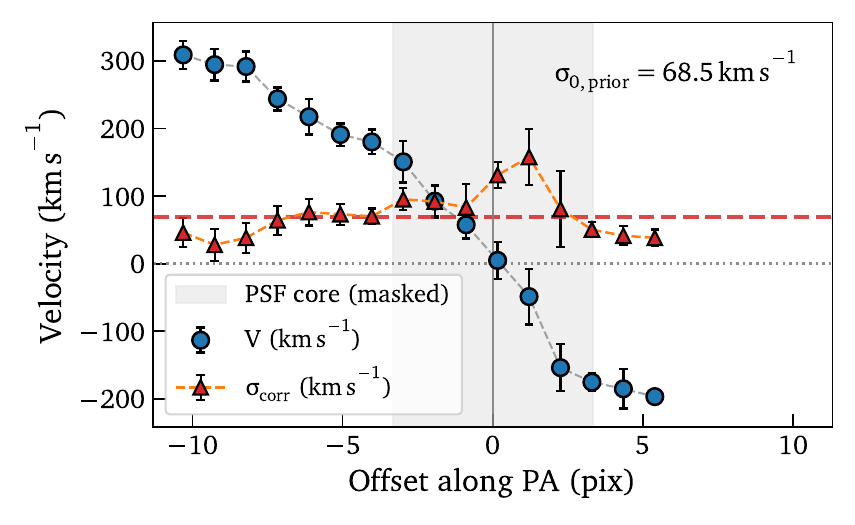}{golden}
\galpage{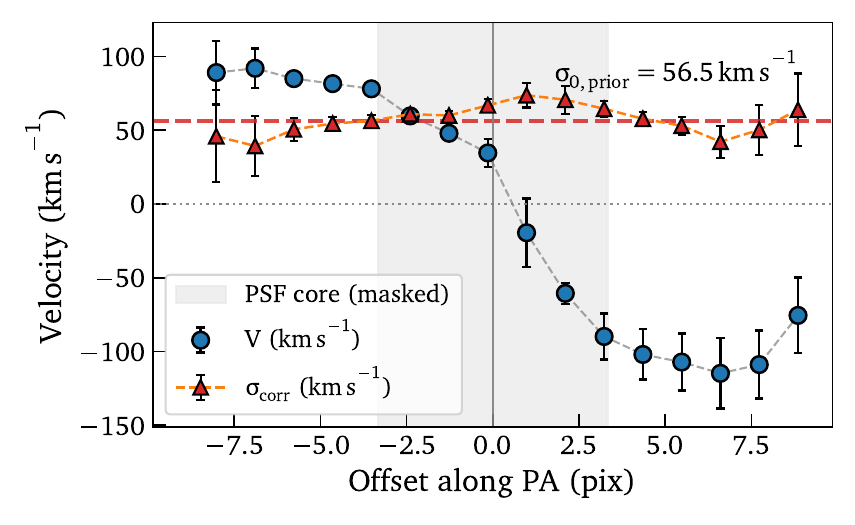}{golden}
\galpage{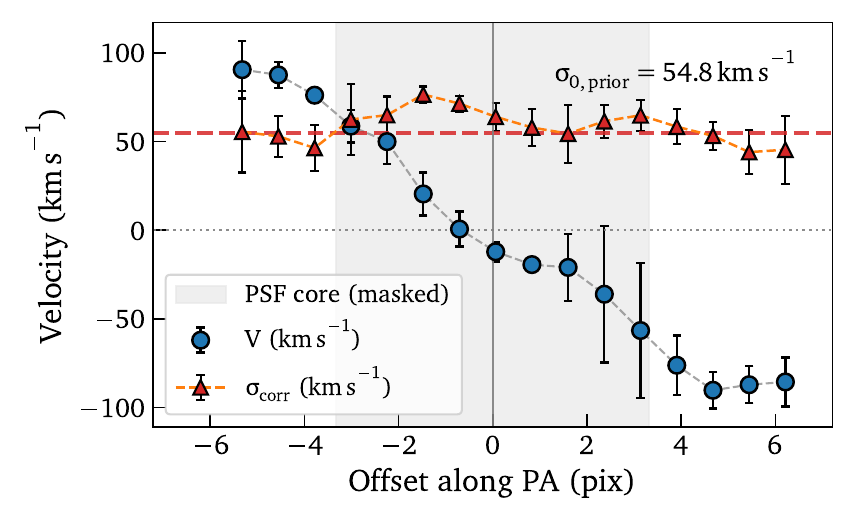}{golden}
\galpage{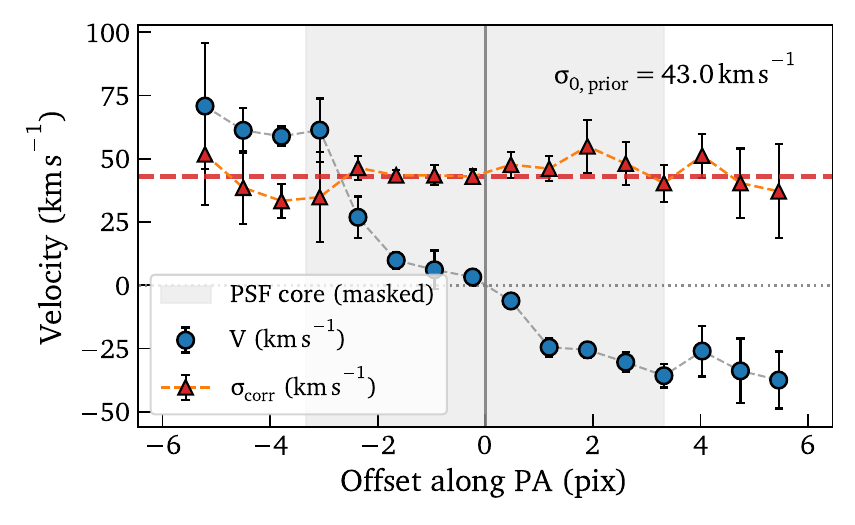}{golden}
\galpage{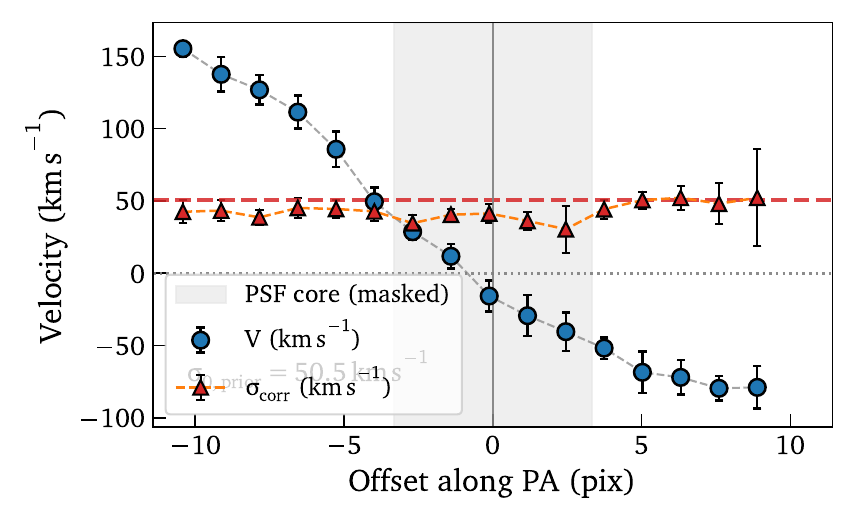}{golden}


\galpage{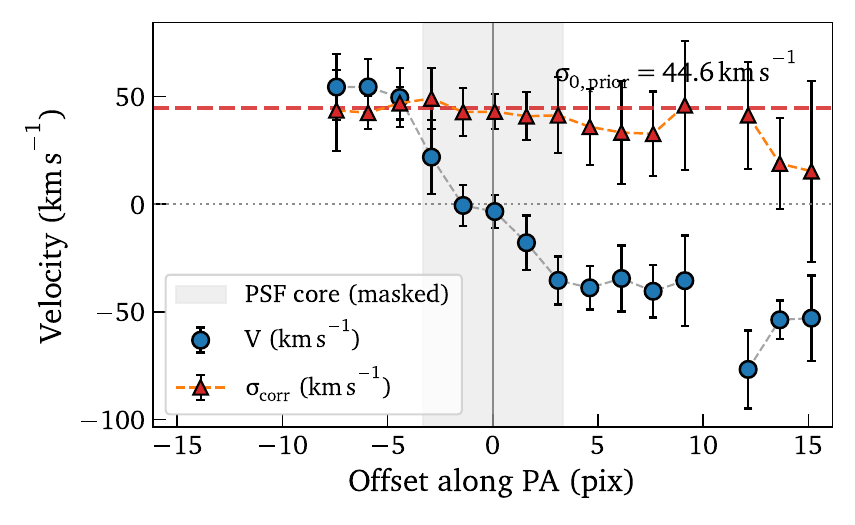}{good}
\galpage{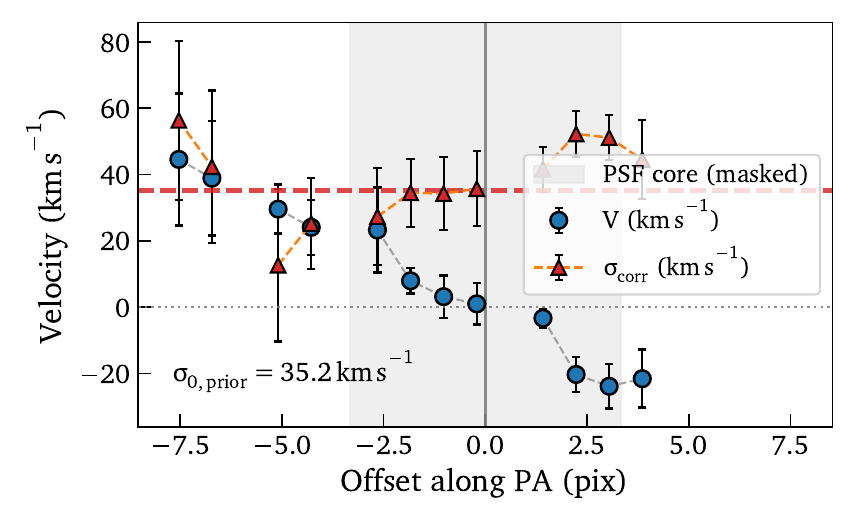}{good}
\galpage{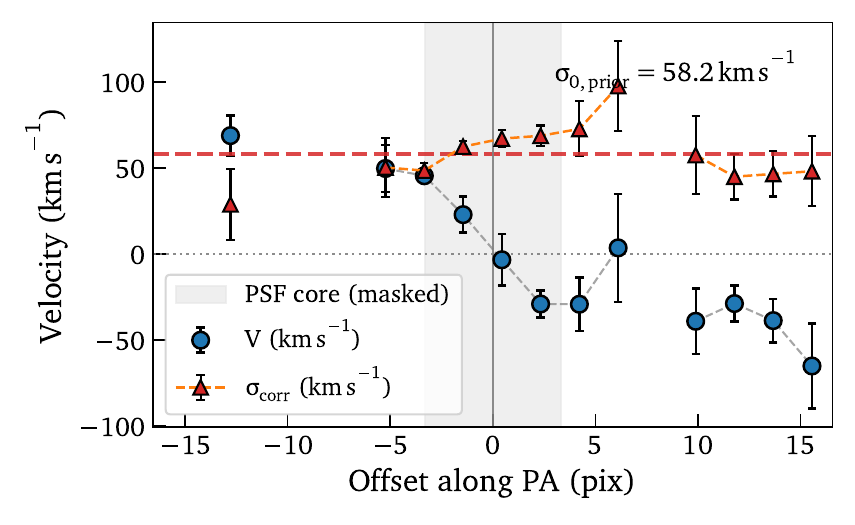}{good}
\galpage{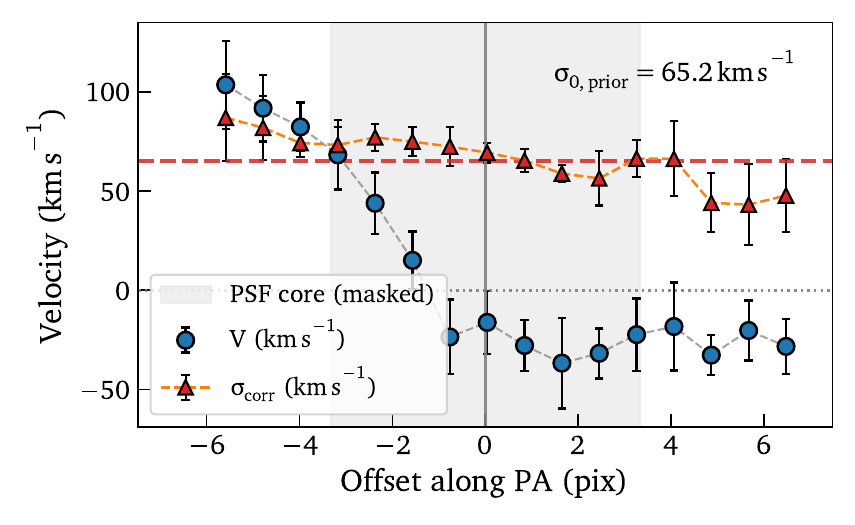}{good}
\galpage{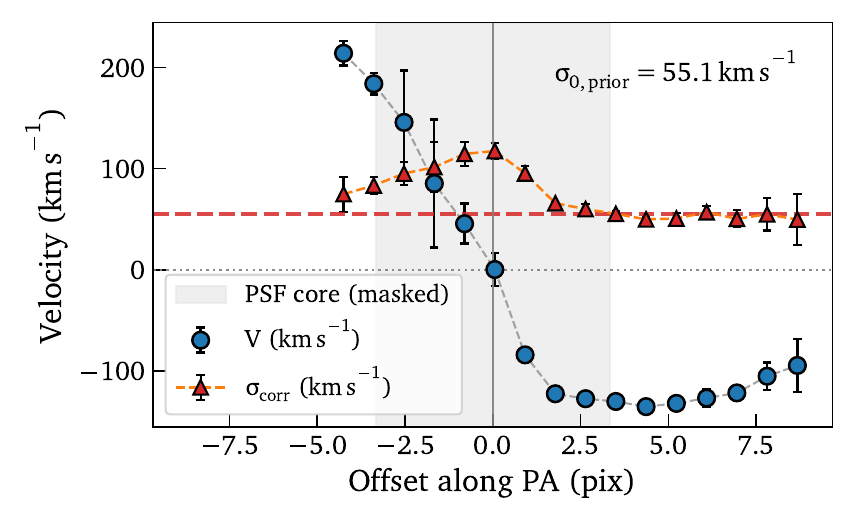}{good}
\galpage{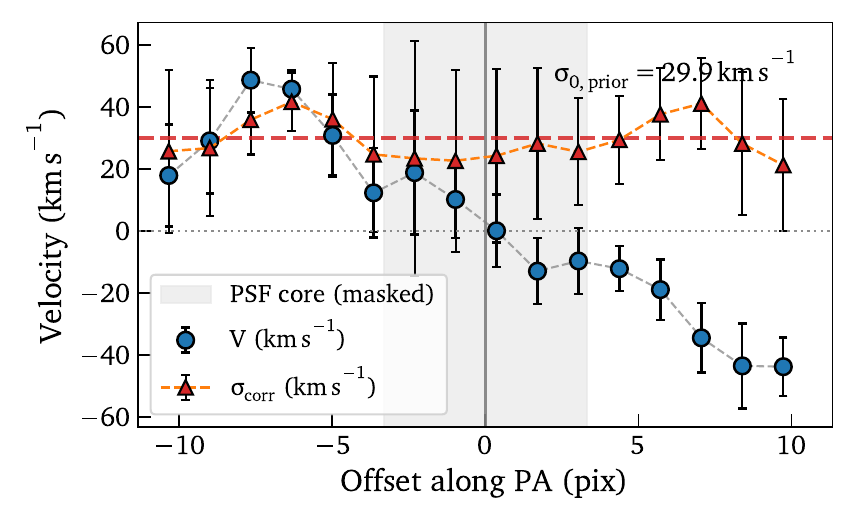}{good}

\clearpage

\fi

\end{document}